\documentclass[traditabstract]{aa} %
\usepackage[pdftex]{graphicx}
\usepackage{natbib}
\usepackage{txfonts}

\newcommand{\xmm}{XMM-{\em Newton}}
\newcommand{\chandra}{{\em Chandra}}

\newcommand{\lognlogs}{Log~$N$--Log~$S$}

\newcommand{\de}{{\rm d}}

\newcommand{\ergscmq}{erg s$^{-1}$ cm$^{-2}$}
\newcommand{\ergscmqdegq}{erg s$^{-1}$ cm$^{-2}$ deg$^{-2}$}

\newcommand{\e}[1]{\times 10^{#1}}

\newcommand{\cdfs}{\textit{Chandra} Deep Field South}
\newcommand{\obsid}{\textit{obsid}}
\newcommand{\pwxd}{{\tt PWXDetect}}
\newcommand{\emld}{{\tt EMLDetect}}

\newcommand{\srcsim}{src~sim$^{-1}$}
\newcommand{\prob}{\mathcal{P}}

\begin{document}
\title{The \xmm\ deep survey in the Chandra Deep Field South.\\
  III. Point source catalogue and number counts in the hard X-rays%
\thanks{Based on observations obtained with XMM-Newton, an ESA science mission with instruments and contributions directly funded by ESA Member States and NASA.}
}

   \author{P.~Ranalli
          \inst{1,2,3}
          \and
          A.~Comastri\inst{2}
          \and
          C.~Vignali\inst{3,2}
          \and
          F.~J.~Carrera\inst{4}
          \and
          N.~Cappelluti\inst{2}
          \and
          R.~Gilli\inst{2}
          \and
          S.~Puccetti\inst{6}
          \and
          W.~N.~Brandt\inst{7,8}
          \and
          H.~Brunner\inst{9}
          \and
          M.~Brusa\inst{3,2,9}
          \and
          I.~Georgantopoulos\inst{1,2}
          \and
          K.~Iwasawa \inst{5}
          \and
          V.~Mainieri\inst{10}
          }

   \institute{
     Institute for Astronomy, Astrophysics, Space Applications and
     Remote Sensing, National Observatory of
     Athens, Palaia Penteli, 15236 Athens, Greece\\
     \email{piero.ranalli@oabo.inaf.it} 
    \and
     INAF -- Osservatorio Astronomico di Bologna,
     via Ranzani 1, 40127 Bologna, Italy
    \and
     Universit\`a di Bologna, Dipartimento di Fisica e Astronomia,
     via Berti Pichat 6/2, 40127 Bologna, Italy
    \and
     Instituto de F\'{\i}sica de Cantabria (CSIC-UC), 39005 Santander, Spain
    \and
     ICREA and Institut de Ci\`encies del Cosmos (ICC), Universitat de
     Barcelona, (IEEC-UB), Mart\'\i\ y Franqu\`es 1, 08028 Barcelona,
     Spain
    \and
     ASI Science Data Center, via Galileo Galilei, 00044, Frascati,
     Italy
    \and
     Department of Astronomy and Astrophysics, Pennsylvania State
     University, University Park, PA 16802, USA
    \and
     Institute for Gravitation and the Cosmos, Pennsylvania State
     University, University Park, PA 16802, USA
    \and
     Max-Planck-Institut f\"ur extraterrestrische Physick, 85478
     Garching, Germany
    \and
     ESO, Karl-Schwarschild-Strasse 2, 85748, Garching bei München, Germany
}

   \date{Received 2013-1-31; accepted 2013-4-18}

  \abstract{
  Nuclear obscuration plays a key role in the initial phases of AGN
  growth, yet not many highly obscured AGN are currently
  known beyond the local Universe, and their search is an active topic of research.
  The \xmm\ survey in the \chandra\ Deep Field South (XMM-CDFS) aims at
  detecting and studying the spectral properties of a significant number of
  obscured and Compton-thick ($N_\mathrm{H}\gtrsim 10^{24}$ cm$^{-2}$) AGN.
  The large effective area of \xmm\ in the 2--10 and 5--10 keV bands,
  coupled with a 3.45 Ms nominal exposure time (2.82 and 2.45 Ms after
  lightcurve cleaning for MOS and PN respectively), allows us to build
  clean samples in both bands, and makes the XMM-CDFS the deepest
  \xmm\ survey currently published in the 5--10 keV band. The large
  multi-wavelength and spectroscopic coverage of the CDFS area allows
  for an immediate and abundant scientific return.
  In this paper, we present the data reduction of the XMM-CDFS
  observations, the method for source detection in the 2--10 and 5--10
  keV bands, and the resulting catalogues. A number of 339 and 137
  sources are listed in the above bands with flux limits of
  $6.6\e{-16}$ and $9.5\e{-16}$ \ergscmq, respectively. The flux
  limits at 50\% of the maximum sky coverage are $1.8\e{-15}$ and $4.0\e{-15}$
  \ergscmq, respectively.
  The catalogues
  have been cross-correlated with the \chandra\ ones: 315 and 130
  identifications have been found with a likelihood-ratio
  method, respectively. A number of 15 new sources, previously undetected by
  \chandra, is found; 5 of them lie in the 4~Ms area.
  Redshifts, either spectroscopic or photometric, are
  available for $\sim 92\%$ of the sources. The number counts in both bands
  are presented and compared to other works. The survey coverage has
  been calculated
  with the help of two extensive sets of simulations, one set per
  band. The simulations have been produced with a newly-developed
  simulator, written with the aim of the most careful reproduction of
  the background spatial properties. For this reason, we present a
  detailed decomposition of the \xmm\ background into its components:
  cosmic, particle, and residual soft protons. The three components have different
  spatial distributions. The importance of these
  three components depends on the band and on the camera; the particle
  background is the most important one (80--90\% of the background
  counts), followed by the soft protons (4--20\%).   %
  }


     \authorrunning{P. Ranalli et al.}
     \titlerunning{The XMM-CDFS survey: 2--10 and 5--10 keV catalogues}

   \maketitle

\section{Introduction}

The understanding that the X-ray background (XRB) is due to unresolved
emission by discrete extragalactic sources
\citep{giacconi87,schwartz1976,maccacaro1991}, and that obscured AGN
are a key ingredient to produce the observed shape of the XRB spectrum
\citep{sw89,comastri95} has put a strong focus on the role of
absorption as a driver of the observable properties. This idea has
shaped the original model of AGN unification \citep{antonucci1993,urrypadovani95}
and keeps its importance unaltered in the current modelling of AGN and
of their evolving populations \citep{treister2005,gilli07,treister09}.

Nuclear obscuration might be associated with the initial phases of
AGN emission \citep{page04,hopkins06,menci08}.  The idea
is that a large gas reservoir is available at high redshift to feed
(and obscure) an accreting supermassive black hole. The same gas
reservoir would also sustain high star formation rates in the host
galaxy, leading to the broad similarity between the cosmic
histories of accretion and star formation.

Deep X-ray surveys are a primary tool for the census of AGN and
the study of the properties of their populations. Since the launch of
the \textit{Einstein} observatory, the first imaging X-ray telescope,
many surveys have been perfomed with different combinations of area
and flux limit and in many locations on the sky.

The \chandra\ Deep Field South (CDFS) survey started with 1 Ms of
observations \citep{giacconi02} and was afterwards extended to 2 Ms
\citep{luo08} and finally to 4~Ms \citep[hereafter
X11]{xue-cdfs4Ms}. It currently reaches the deepest X-ray fluxes ever
probed, with a flux limit of $9.1\e{-18}$ \ergscmq\ in the soft
(0.5--2 keV) band. However, given the energy dependence of its
effective area, \chandra\ is much less sensitive in the hard and very
hard bands (2--10 keV and 5--10 keV, respectively) which are the most
important for obscured objects. The flux limit for 2--10 keV is
currently $5.5\e{-17}$ \ergscmq\ (X11). The only analysis for the
5--10 keV band was done for the initial 1 Ms survey \citep{rosati02}
and yielded a flux limit of $1.2\e{-15}$ \ergscmq; however, the 4~Ms
survey was analysed in the 4--8 keV band, moving the flux limit to
$6\e{-17}$ \ergscmq\ \citep{lehmer2012}. A larger area centred
  around the CDFS was surveyed with \chandra\ and with much shorter
  exposure times as the Extended CDFS (ECDFS; \citealt{lehmer05}).

\xmm\ is better suited to collect photons in the hard and very hard
bands, because at energies $\gtrsim 5$ keV the \xmm\ effective area
drops less sharply than \chandra's. However, it has a larger point
spread function (PSF), which makes it suffer more from source
confusion, and it has a higher instrumental background. While \xmm\
surveys cannot probe very faint fluxes as \chandra, they provide good
quality spectroscopy in the whole 0.5--8 keV band for a large number
of sources.  The large effective area of \xmm\ has been exploited in
several surveys, including the Lockman Hole (with 1 Ms of exposure,
\citealt{brunner08}), the ELAIS-S1 field \citep{puccetti06}, COSMOS
\citep{cappelluti07,xmm-cosmos}, and 2XMM \citep{mateos08}.

The \xmm\ survey in the CDFS (hereafter XMM-CDFS) has been started
with the main aim of finding obscured AGN, especially the
Compton-thick ones, and studying their properties. With a nominal
exposure of 3.45 Ms centred on a single point on the sky it is
currently the deepest \xmm\ observation ever performed. Moreover, it
has the advantage of having almost pan-chromatic coverage and a
large number of available optical spectra and redshifts.

The first results of the XMM-CDFS survey have been published in
\citet{comastri11-cdfs}. Different approaches to the selection and
study of obscured sources are being employed (\citealt{iwasawa2012};
\citealt{georgantopoulos2013}; spectral stacking (Falocco et al.,
submitted to A\&A) and spectral analysis of a flux limited sample (Comastri et
al., in prep.) are also ongoing. Extended sources will be discussed
in Finoguenov et al. (in prep.).

In the present paper, we present the hard and very hard catalogues,
and the \lognlogs\ in the two bands. A number of 339 and 137 sources
are detected, respectively, with a significance threshold 
roughly equivalent to $4.8\sigma$. Supplementary catalogues are
provided for sources with lower significance (%
$4\sigma$).

In Sect.~\ref{sec:data} we present the observation details and the
data reduction procedure. In Sect.~\ref{sec:catalogue} we describe the
source detection procedure and present the catalogues in the hard and
very hard bands.  The survey coverage and number counts are presented
in Sect.~\ref{sec:completeness}.  Simulations of mock XMM-CDFS fields
are instrumental in determining the coverage, and are presented in
Sect.~\ref{sec:simulations}. The simulations also used in
Sect.~\ref{sec:detmask} for a first estimate of the number of spurious
sources. We present in Sect.~\ref{sec:confusion} an analysis of the
source confusion in the hard band. In Sect.~\ref{sec:optical} we
identify the \chandra\ counterparts to the \xmm\ sources.  We
summarize our conclusions in Sect.~\ref{sec:conclusion}. Finally, in
Appendix~\ref{sec:simulator} we present the simulator that we have
developed.

\section{Observations and data reduction}
\label{sec:data}

The \cdfs\ was initially observed by \xmm\ in the years 2001--2002
(P.I.: J. Bergeron) with an exposure of 541 ks. It was observed again,
for the proposal which led to this series of papers, in the years
2008--2010 (P.I.: A. Comastri), to reach a nominal exposure of 3.45
Ms.

 Table~\ref{tbl:obsid} shows the observation dentification number
 (hereafter \obsid), date, pointing and exposure for all observations.
 The number of \obsid s is 33; considering the MOS1, MOS2 and
   PN cameras, the XMM-CDFS survey comprises a total of 99 event
 files.  We used the \xmm\ SAS software for our analysis (version 10
 for the initial processing and the catalogue; version 11 for the
 simulations; there have not been significant changes between the two
 versions which could affect this work), with the help of many
 custom-developed scripts to automate most of the data processing.

\begin{table*}[h]
\caption{\label{tbl:obsid}XMM-CDFS observation log.}
\centering
\begin{tabular}{llrrrrrrc}
\hline
  \multicolumn{1}{c}{OBS\_ID} &
  \multicolumn{1}{c}{DATE} &
  \multicolumn{1}{c}{RA} &
  \multicolumn{1}{c}{DEC} &
  \multicolumn{1}{c}{PA} &
  \multicolumn{1}{c}{MOS1 EXP} &
  \multicolumn{1}{c}{MOS2 EXP} &
  \multicolumn{1}{c}{PN EXP} & BORES.GR. \\
\hline
  0108060401 & 2001-07-27 & 53.0812 & -27.7929 &  58.74 &$^a$ 24.6&$^a$ 24.6&$^a$ 18.8 &A\\
  0108060501 & 2001-07-27 & 53.0919 & -27.7981 &  58.75 &     43.6&    44.6&  35.8    &A\\
  0108060601 & 2002-01-13 & 53.1457 & -27.8252 & 238.71 &$^a$ 52.6&$^a$ 52.6&  43.6   &B\\
  0108060701 & 2002-01-14 & 53.1406 & -27.8218 & 238.72 & 78.5 & 78.7 &  69.4         &B\\
  0108061801 & 2002-01-16 & 53.1450 & -27.8195 & 238.72 & 55.1 & 54.9 &  54.4         &B\\
  0108061901 & 2002-01-17 & 53.1456 & -27.8133 & 238.71 & 42.9 & 42.8 &  39.1         &B\\
  0108062101 & 2002-01-20 & 53.1511 & -27.8168 & 238.72 & 44.4 & 44.3 &  42.6         &B\\
  0108062301 & 2002-01-23 & 53.1462 & -27.8147 & 238.71 & 83.7 & 82.5 &  72.3         &B\\
  0555780101 & 2008-07-05 & 53.1476 & -27.7368 &  61.73 &106.6 &109.3 &  88.8         &C\\
  0555780201 & 2008-07-07 & 53.1488 & -27.7457 &  61.72 &116.7 &118.7 &$^a$ 91.5         &C\\
  0555780301 & 2008-07-09 & 53.1387 & -27.7448 &  61.73 &$^a$109.0&108.6 &  96.7         &C\\
  0555780401 & 2008-07-11 & 53.1385 & -27.7371 &  61.74 & 89.9 & 91.5 &  77.7         &C\\
  0555780501 & 2009-01-06 & 53.1325 & -27.8371 & 241.73 & 97.9 & 99.2 &  90.7         &D\\
  0555780601 & 2009-01-10 & 53.1343 & -27.8439 & 241.71 &104.9 &108.8 &  79.3         &D\\
  0555780701 & 2009-01-12 & 53.1322 & -27.8545 & 241.72 &101.5 &102.5 &  95.7         &D\\
  0555780801 & 2009-01-16 & 53.1233 & -27.8359 & 241.71 & 88.4 & 89.1 &  77.0         &D\\
  0555780901 & 2009-01-18 & 53.1230 & -27.8445 & 241.74 & 96.0 & 96.3 &  80.6         &D\\
  0555781001 & 2009-01-22 & 53.1234 & -27.8528 & 241.73 &106.4 &108.2 & 101.3         &D\\
  0555782301 & 2009-01-24 & 53.1235 & -27.8531 & 241.74 &106.5 &107.1 &  96.4         &D\\
  0604960101 & 2009-07-27 & 53.1486 & -27.7447 &  62.73 &101.5 &105.1 &  86.8         &C\\
  0604960201 & 2009-07-17 & 53.1404 & -27.7448 &  62.73 &104.7 &104.4 &$^a$ 76.9      &C\\
  0604960301 & 2009-07-05 & 53.1496 & -27.7550 &  62.72 &$^a$101.4&$^a$ 102.3&$^a$ 89.0 &C\\
  0604960401 & 2009-07-29 & 53.1393 & -27.7544 &  62.73 &119.2 &120.2 & 112.5         &C\\
  0604960501 & 2010-01-18 & 53.1307 & -27.8384 & 245.73 & 46.3 & 46.5 &  45.0         &D\\
  0604960601 & 2010-01-26 & 53.1312 & -27.8463 & 245.73 &$^a$ 106.9&$^a$ 106.8&  95.6  &D\\
  0604960701 & 2010-01-12 & 53.1224 & -27.8388 & 245.73 &$^a$  91.7&$^a$  92.6&$^a$ 51.1 &D\\
  0604960801 & 2010-02-05 & 53.1315 & -27.8569 & 245.73 & 90.3 & 90.3 &  82.3         &D\\
  0604960901 & 2010-02-11 & 53.1221 & -27.8573 & 245.72 &$^a$ 84.7&$^a$ 84.8&  75.4    &D\\
  0604961001 & 2010-02-13 & 53.1211 & -27.8476 & 245.72 & 91.4 & 93.7 &  86.1         &D\\
  0604961101 & 2010-01-04 & 53.1316 & -27.8301 & 245.72 &104.2 &105.9 & 100.6         &D\\
  0604961201 & 2010-01-08 & 53.1197 & -27.8294 & 245.71 &114.0 &115.0 &$^a$ 96.9      &D\\
  0604961301 & 2010-01-19 & 53.1307 & -27.8379 & 245.71 &$^a$ 11.2&$^a$ 11.9&$^a$ 5.8  &D\\
  0604961801 & 2010-02-17 & 53.1213 & -27.8474 & 245.73 & 94.9 & 95.5 &  88.8         &D\\
\hline\end{tabular}

\tablefoot{The columns show: the \obsid\ number; the date of
  observation start; the pointing coordinates (RA and DEC in deg and
  in J2000 coordinates; position
  angle in deg); the exposure times in ks for the MOS1, MOS2 and PN cameras after
  background filtering; and the
  boresight group (see Fig.~\ref{fig:expmap}). $^a$:
  highly flared observations for which the $3\sigma$-clipping failed
  (see text).
}
\end{table*}

\begin{figure}
  \centering
  \includegraphics[width=\columnwidth,bb=18 160 567 712]{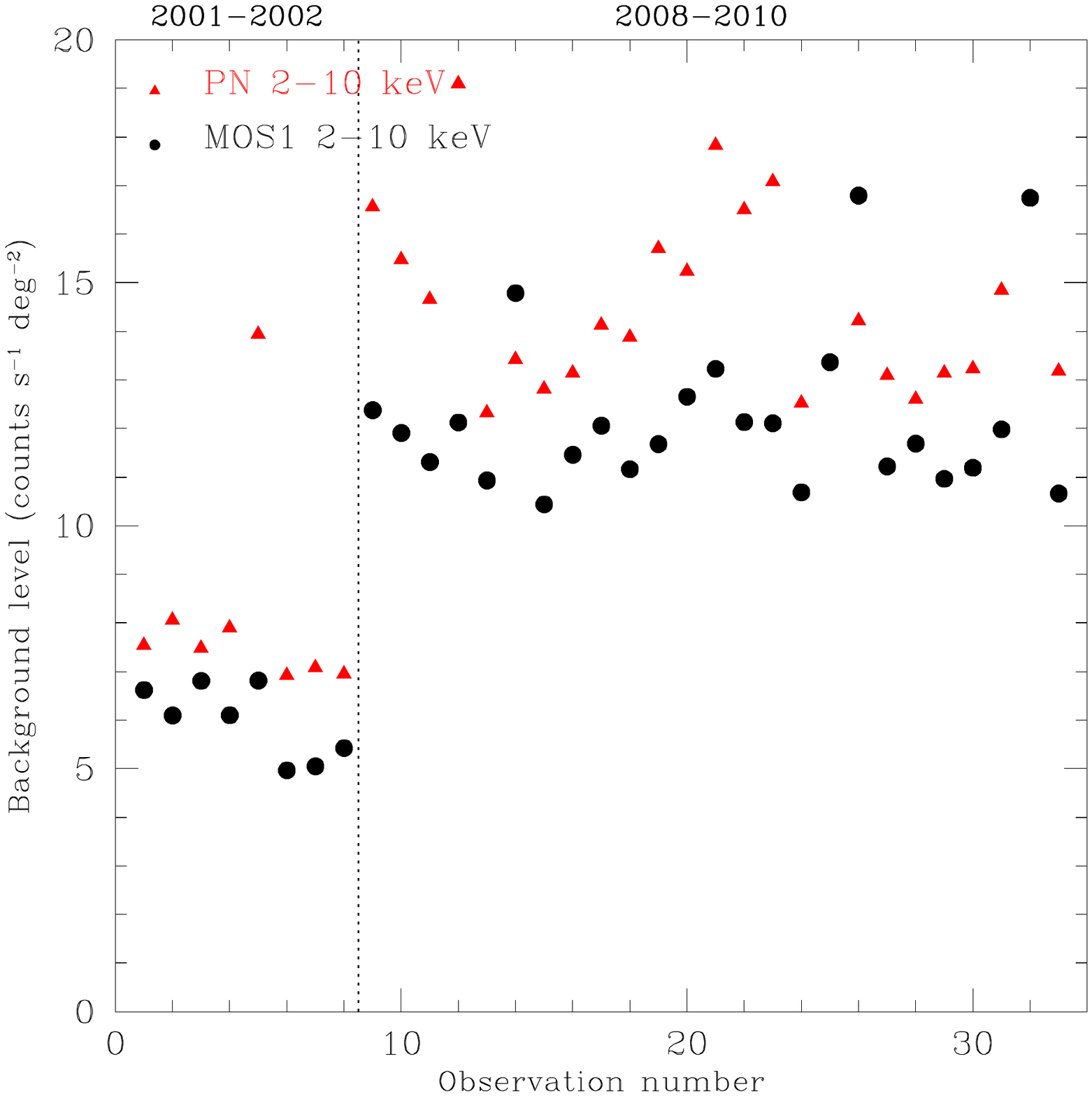}
  \caption{Total surface brightness in the 2--10 keV band for the 33
    \obsid s of the XMM-CDFS, after excluding th epositions around the
    30 brightest sources. The increase in background intensity between
    2002 and 2008 is evident. The \obsid s are numbered from 1 to 33
    according to observation date; the vertical dotted line marks the
    separation between the 2001--2002 and 2008--2010
    observations. Filled triangles: PN; filled circles: MOS1. The MOS2
    camera has values very similar to the MOS1.}
  \label{fig:sigmainbkg}
\end{figure}

\subsection{Background flares and quiescent level}
\label{sec:flares}

\begin{figure*}
  \centering
  \includegraphics[width=.49\textwidth,bb=43 157 563 695,clip]{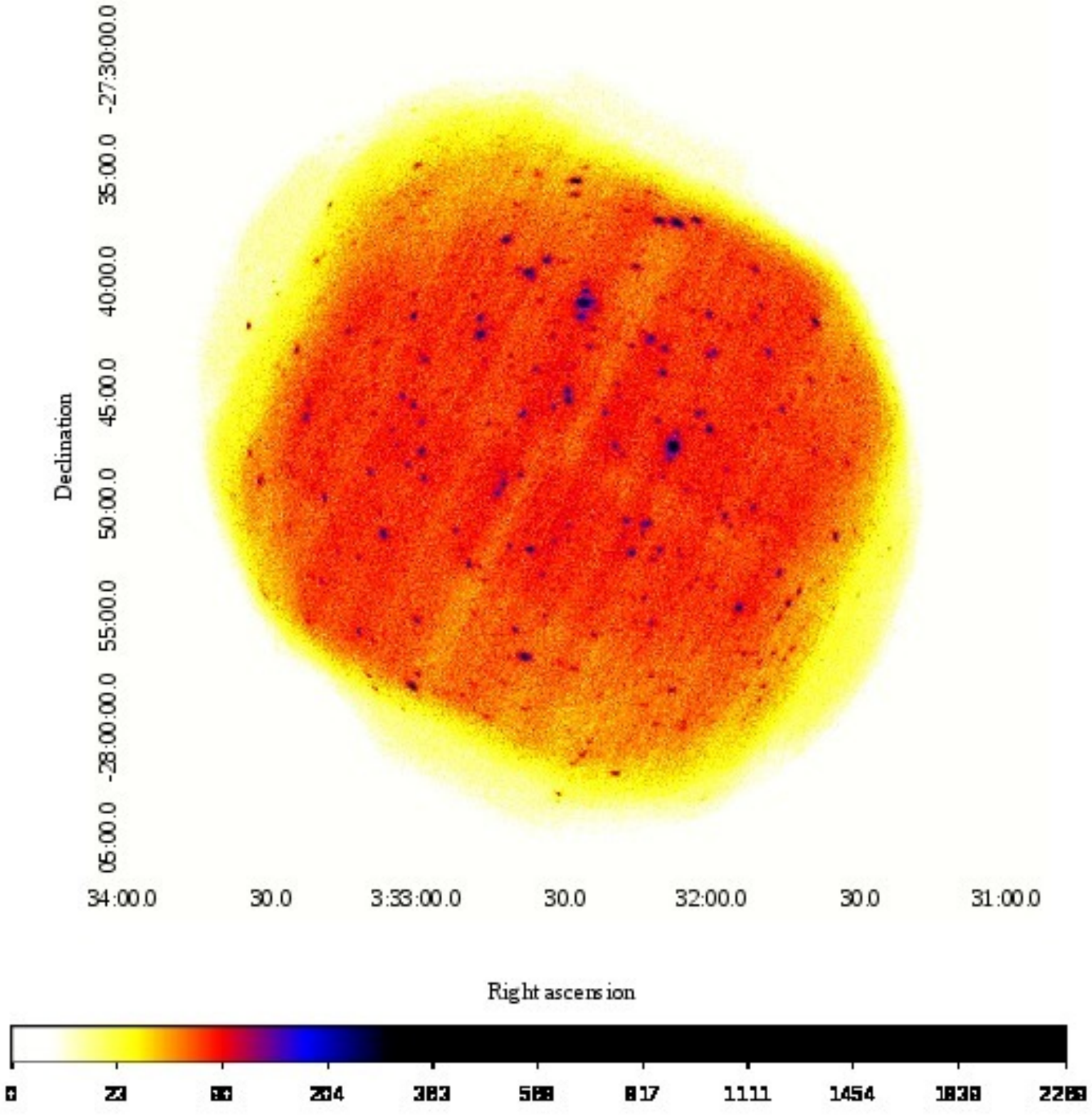}
  \includegraphics[width=.49\textwidth,bb=43 157 563 695,clip]{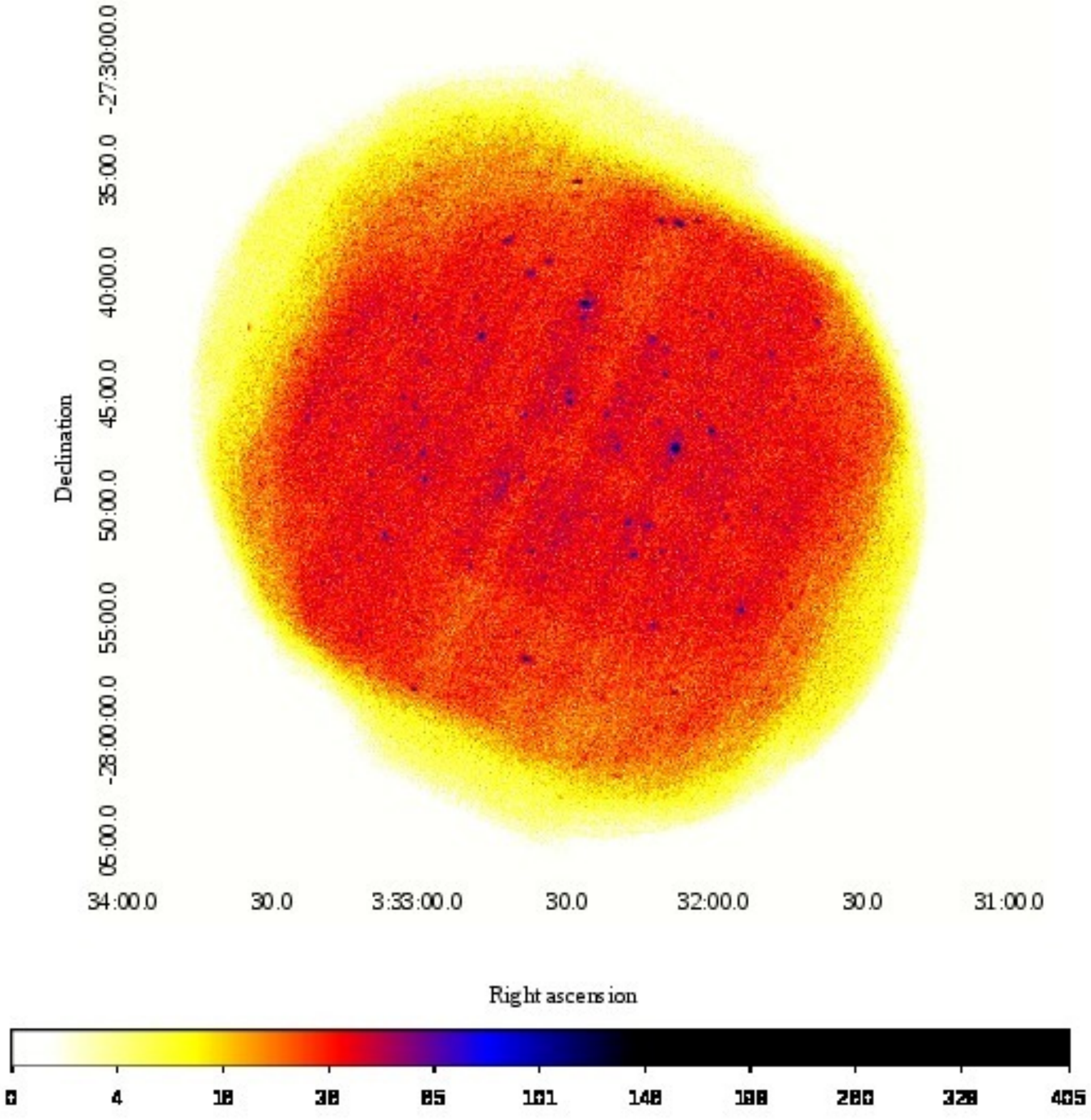}
  \caption{Images of the XMM-CDFS in the 2--10 (left) and 5--10
    keV (right) bands.  The colour wedges show the total counts
     as the sum of data from the MOS1, MOS2 and PN cameras.}
  \label{fig:allcts}
\end{figure*}

Each one of the 99 event files was screened individually for
background flares. Full-field light curves were generated in the
10--13.5 keV interval. A 3$\sigma$-clipping procedure was applied to
the light curves to identify and reject the high-background
periods. This procedure worked well most times, though it failed for
some event files which were severely flared. In such cases, we adopted
the nominal count rate thresholds given by the \xmm\
documentation\footnote{http://xmm.esac.esa.int/sas/current/documentation/threads/ EPIC\_filterbackground.shtml} to identify and reject the
high-background periods; these observations are marked in Table~\ref{tbl:obsid}.

A relevant and unexpected feature is an increase by a factor of $\sim
2$ in the quiescent background level in the 2008--2010 observations
with respect to the years 2001--2002 (Fig.~\ref{fig:sigmainbkg}). The
increase is seen in all three cameras and is due to the instrumental
(``particle'') component of the background (see
Sect.~\ref{sec:simulations}).  The reason for this increase is not
clear, though one possibility is that it is related to the Solar
cycle.

\begin{figure*}
  \centering
  \includegraphics[width=.75\textwidth,bb=61 205 543 640]{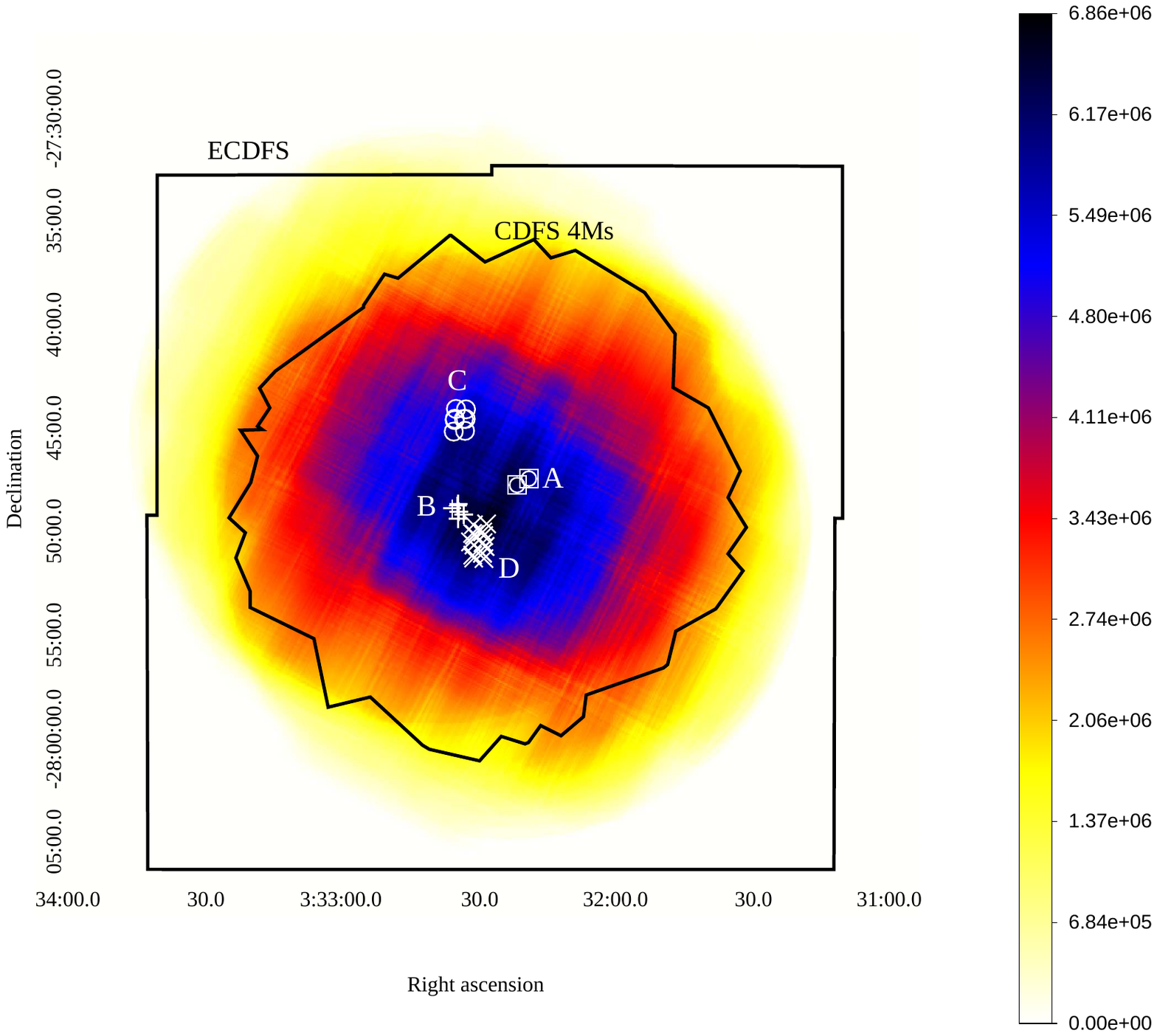}\\
  \includegraphics[width=.75\textwidth,bb=61 205 543 640]{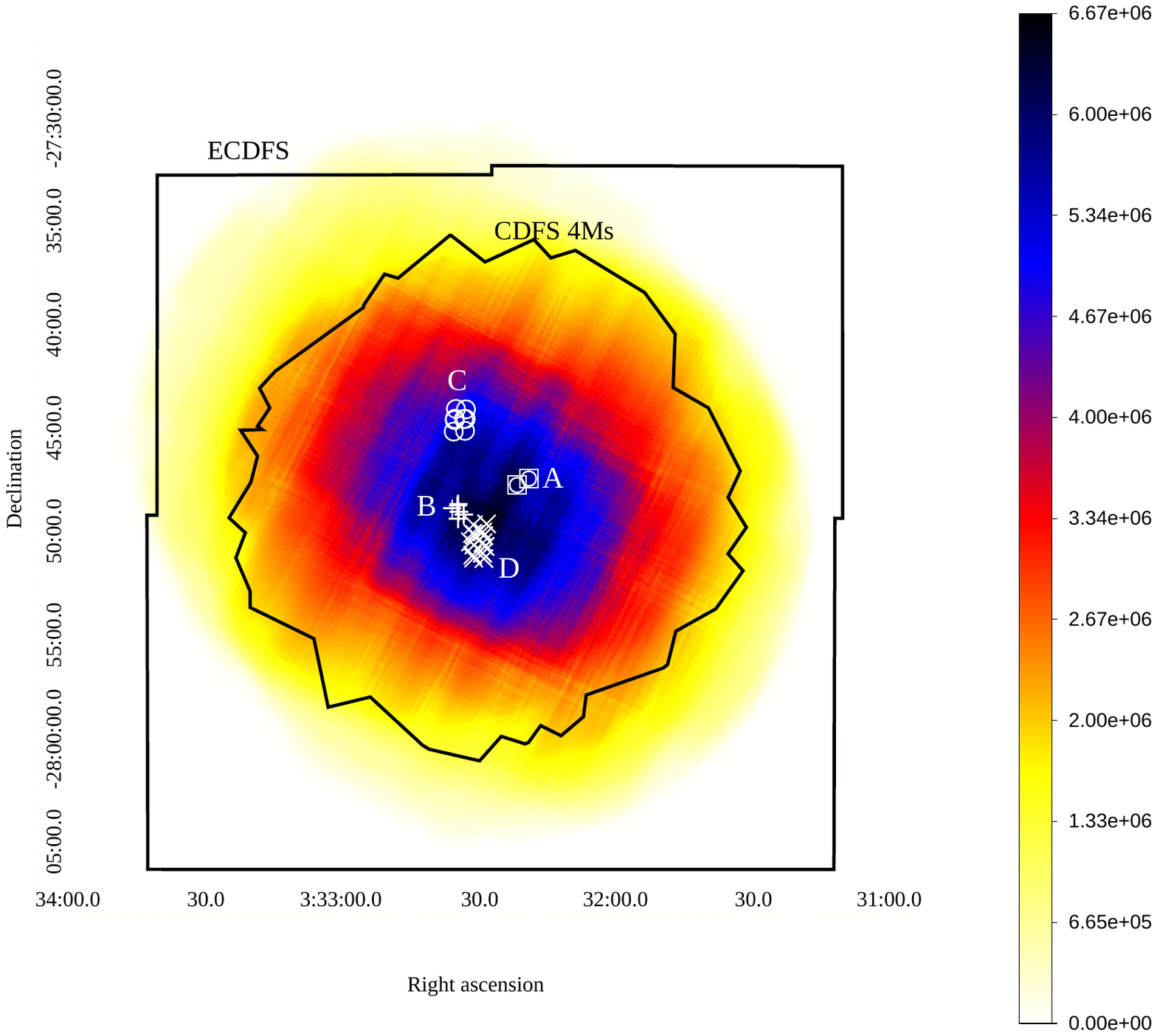}
  \caption{Exposure maps of the XMM-CDFS in the 2--10 (left) and 5--10
    keV (right) bands. The \chandra\ 4~Ms and ECDS areas, and the
    pointing positions and groups are superimposed with different
    symbols and letter marks according to the observing season (see
    Table~\ref{tbl:obsid}).  The colour wedges show the exposure time
    in s as the sum of the MOS1, MOS2 and PN exposures.}
  \label{fig:expmap}
\end{figure*}

The images in the 2--10 and 5--10 keV bands, showing the total counts
from the MOS and PN cameras, are presented in Fig.~\ref{fig:allcts}.
The total exposure, after the high-background period filtering, is
thus 2.82 and 2.45 Ms for the two MOS and the PN, respectively. The
exposure maps for the 2--10 and 5--10 keV bands are shown in
Fig.~\ref{fig:expmap}.

\subsection{Copper lines complex}
\label{sec:Cu_lines}

For the PN camera, the strongest background feature in the 2-10 keV
interval is a complex of lines (Cu K$\alpha$, Ni K$\alpha$, Zn
K$\alpha$, the Cu being the strongest one) around 8 keV, due to
fluorescence from the mirror holding structure, and which alone can
make $\sim 30\%$ of the total counts in the 2-10 keV band. This
complex has a well-defined spatial pattern with a central region where
it is virtually absent\footnote{\xmm\ Users Handbook, Issue 2.10,
  Sect. 3.3.7.2; http://xmm.esac.esa.int/
  external/ xmm\_user\_support/ documentation/ uhb/ epicintbkgd.html}.

Throughout this paper, in PN data we have excluded the energy intervals
pertaining to this complex.  We have chosen different trade-offs for
the 2--10 and 5--10 keV bands. For the 2--10 keV band, we have
excluded the Cu complex (energy intervals 7.2--7.6 and 7.8--8.2 keV)
across all the field, to favour a uniform coverage. For the 5--10
keV band, given the smaller energy band and the lower signal/noise
ratio, we have been more aggressive in excluding the Cu complex
(energy intervals 7.35--7.60, 7.84--8.28 and 8.54--9.00 keV), but we
have done so only in the outer detector regions where the line is
present. Our aim for 5--10 keV has been to retain the photons
around 8 keV in the centre of the FOV at the expense of a less uniform
coverage of the field.

\subsection{Pointings and astrometry}
\label{sec:astrometry}

The XMM-CDFS observations have been taken with some differences in the
pointing direction, in order to have a
field coverage as uniform as possible with the PN detector. Because of the
field visibility, the observations have been performed in the months
of January, February and July. The orbit of \xmm\ modulates the
position angle, so that the satellite rotates along its optical
axis by about 180 deg every six months. Therefore, when looked at in
detector coordinates, observations performed in summer appear
``upside-down'' with respect to the winter ones. While the MOS
detectors are centred on the respective telescope optical axes, the PN
is shifted by $2\arcmin$. The pointing coordinates have thus to be
shifted by twice that amount in order to position the PN on the
same sky region in the two seasons. This, however, has the effect of
making the MOS detectors offset with respect to the nominal CDFS
coordinates.

Minor shifts of $\lesssim 1\arcmin$ were added to all observations, to
smooth the chip gaps and to avoid the possibility that any source could
fall in a gap for a major part of the survey.

On top of these patterns, there are smaller random errors ($\sim
1\arcsec$, with 4 obsids in which they were $\sim 4\arcsec$) in the
pointing direction. These have been estimated by running a source
detection on the inner $5\arcmin$ of each obsid, and cross-correlating
the results with the \chandra\ catalogue (X11) using a modified
version of \chandra's routine {\tt align\_evt}. The event and attitude
files have been corrected for these shifts. All coordinates have been
aligned to those from X11 (which was itself aligned to VLA radio frame).

The \xmm\ PSF has a $\sim 5\arcsec$ FWHM on-axis, which degrades at
large off-axis angles\footnote{The \xmm\ Users
  HandBook. http://xmm.esac.esa.int/ external/ xmm\_user\_support/
  documentation/uhb/XMM\_UHB.html}.  We measured the observed 2--10
keV FWHM of the 50 brightest sources for the total 3Ms image and we
found that the median FWHM, across the whole field, is $8.5\arcsec$.

\subsection{Colour image}

\begin{figure*}
 \sidecaption %
  \includegraphics[width=12cm,bb=10 10 518 466]{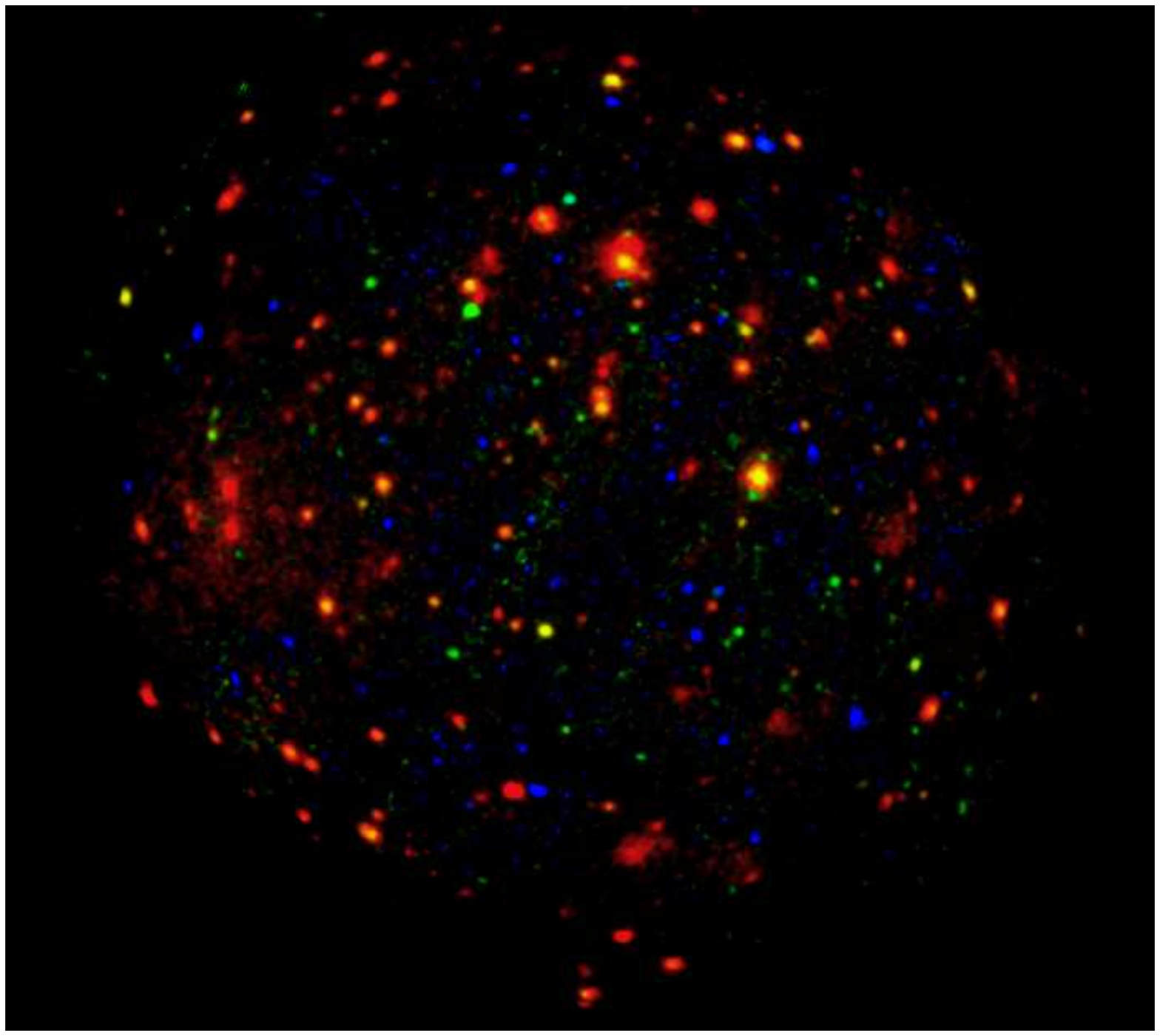}
  \caption{Colour image of the XMM-CDFS. Red: 0.4--1 keV; green: 1--2
    keV; blue: 2--8 keV. The colour scaling is non-linear, and is
done first by using a quadratic polynomial to suppress the
background, and then by following the \citet{lupton04} procedure.
}
  \label{fig:truecolour}
\end{figure*}

A colour image of the XMM-CDFS has been derived from the cleaned,
astrometry-aligned data. Three images in the 0.4-1, 1-2 and 2-8 keV
energy intervals have been produced. To better enhance the sources, the background
has been suppressed up to different thresholds in the three bands. %
Finally, the non-linear scaling and colour-mixing procedure by
\citet{lupton04}, whose main advantage is that bright
sources are not ``whited-out'', has been applied producing the colour image shown in
Fig.~\ref{fig:truecolour}.

\section{Catalogue}
\label{sec:catalogue}

The 2-10 and 5-10 keV catalogues have been produced independently. A
two-stage process has been used: the first step is a detection with a
wavelet code (\pwxd), followed by maximum likelihood fitting (\emld)
of the sources. This detection method can be regarded as a variant of
the standard \xmm\ detection process, which uses a sliding box
detection algorithm for the first stage instead of the wavelet (see
e.g.\ \citealt{brunner08} and \citealt{xmm-cosmos} for applications to
the Lockman Hole and COSMOS surveys, respectively).  The advantage of
using a wavelet-based tool is that it produces more accurate source
positions \citep{puccetti09}, and that it is especially suited for
crowded fields, as happens in a few small areas of the CDFS. Another
difference with respect to the Lockman Hole and COSMOS applications is
that we chose a higher threshold (i.e., more conservative) for the
first stage than for the second (see below); doing so means that the
selection is mostly done at the first stage, and that the second is
used mainly for the counts estimate.

First, we ran the \textit{Palermo Wavelet XMM Detect} (\pwxd) software
to identify source candidates. \pwxd\ is a version of the {\tt
  PWDetect} tool \citep{damiani97}, specially tuned for \xmm\ data. It
can combine data from the three EPIC detectors and from different
observations, even taken with different pointings, provided that the
pointing boresights are close enough that a single model for the
off-axis angle dependence of the PSF can be used (in practice, this
amounts to a few arcmins). We used this feature to perform the
detection on all obsids and cameras. We chose a significance threshold
for source detection at the 4$\sigma$ level,
corresponding\footnote{The correspondence between probability and
  ``number of sigmas'' only applies for Gaussian probability
  distributions. Our use should just be regarded as a handy mnemonic.}
to a probability of $6.3\e{-5}$ for a random fluctuation to be
detected as a source (``false-positive'').  The first stage source
lists include 411 and 196 sources for the 2-10 and 5-10 keV bands,
respectively. The source identification numbers (ID210 for 2--10 keV
and ID510 for 5--10 keV) were assigned at this stage.

For the second stage we used \emld, a standard tool from the \xmm\ SAS
software. Originally developed for ROSAT
\citep{cruddace88,hasinger93}, the current version of
\emld\footnote{The XMM SAS reference manual for \emld;
  http://xmm.esac.esa.int/sas/11.0.0/doc/emldetect/node3.html}
includes many improvements in terms of PSF models, ability to
simultaneously fit data from different cameras and pointings, deblend
sources, and fit extended sources\footnote{While the CDFS field
  containes some extended sources, none of them was present in the
  \pwxd\ output.}.

\emld\ gives more accurate estimates of net source counts, with lower
systematic and random uncertainties \citep{puccetti09}.  Since
\emld\ was not designed to run on a large number of input files, we
grouped the observations in four sets according to the pointing
direction (see Table~\ref{tbl:obsid}), in order to have homogeneous
PSFs across a single set. We summed together the images from the MOS1,
MOS2 and PN cameras, obtaining four images which were used as the
input to \emld. The background maps for \emld\ were calculated with
the method of \citet{cappelluti07}.

We ran \emld\ using the \pwxd\ coordinates as input and without
re-fitting the positions; thus the \emld\ source list is a
subset of the \pwxd\ one. 

\emld\ uses a maximum-likelihood estimator to assign a significance
value to a detection. The \emld\ source list was cut at a likelihood
value of $L=4.6$. This likelihood obeys the law $L=-\mathrm{ln}(p)$,
where $p$ is the probability of a false-positive detection. %
The threshold $L=4.6$ corresponds to a probability of $1.01\e{-2}$.
The number of sources detected at the second stage is 337 and 135 for
the 2-10 and 5-10 keV bands, respectively.

After visual inspection, we identified two cases of source blending,
which correspond to ID210s 228 and 276. These blends were detected in
the 2--10 keV band at $29.1\sigma$ and $22.3\sigma$ by \pwxd, and with
likelihoods $L=565.5$ and 340.1 by \emld, respectively. In both cases,
two sources whose cores can be visually separated were identified by
\pwxd\ as a single one. %
The separation between the two components is $11.5\arcsec$ and
$18.1\arcsec$ for ID210s 228 and 276, respectively (see also
Sect.~\ref{sec:confusion}). Both blends are detected also in the 5--10
keV band, as ID510 1109 and 1134, respectively. In the catalogue, we
do not list the blends; instead we provide counts, rates and fluxes
estimated with aperture photometry for the four individual sources,
which in the 2--10 keV catalogue are numbered 501, 502 (components of
228) and 503, 504 (components of 276).  These sources have
been de-blended also in the 5--10 keV band, and their components are
numbered with ID510s 1501--1504. The main catalogues are defined as
the sources detected at the second stage, minus the two blends, plus
the four blend components, and thus contain 339 and 137 sources in the
2--10 and 5--10 keV bands, respectively.

A comparison of the source counts derived by \pwxd\ and \emld\ showed
a 1-to-1 correlation, as expected, but with a few outliers. Three of
these sources (ID210: 4, 105; ID510: 1001) lie at the south-west rim of the
field and are probably just fluctuations in an area with a strong
exposure gradient. A few others (ID210: 154, 184, 280, 290, 328, 406;
ID510:  1066, 1089, 1100, 1124, 1146)
lie in a crowded area, or very close to brighter sources, and they were
also flagged as ``extended'' by \emld. Their \emld-derived counts are
about one order of magnitude larger than the \pwxd\ estimate, probably
because \emld\ fits them as tails of the PSF of a larger and brighter
source. Aperture photometry showed agreement with the counts derived
by \pwxd. Therefore, in the catalogue these 6 sources are listed with
the \pwxd\ counts; count rates and fluxes were calculated accordingly.

\begin{figure}
  \centering
  \resizebox{\hsize}{!}{\includegraphics[width=\columnwidth,bb=25 160 565 691,clip]{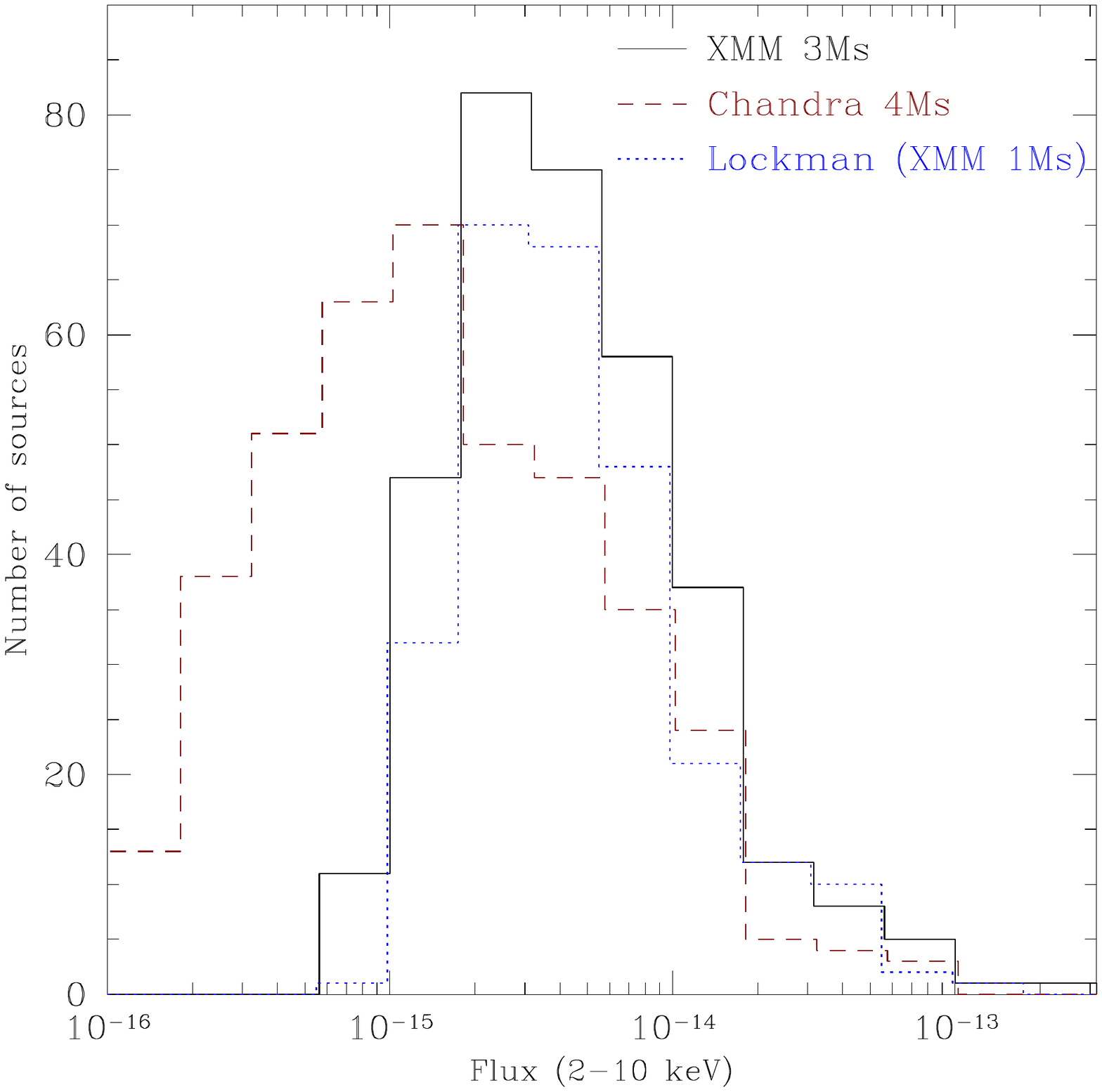}}
  \caption{2-10 keV fluxes of the sources detected in the XMM-CDFS,
    compared with other surveys.  Solid black histogram: main
    XMM-CDFS catalogue; dashed red histogram: \chandra\ catalogue
    (X11); dotted blue histogram: Lockman hole
    \citep{brunner08}. }
  \label{fig:fluxhisto}
\end{figure}

The flux to count rate conversion factors have been calculated for a
power-law spectrum with slope $\Gamma=1.7$ \citep{mainieri07}; their
values are $1.86\e{11}$ and $0.91\e{11}$ counts cm$^{-2}$ erg$^{-1}$
for the 2--10 and 5--10 keV bands, respectively. Had we used a
different slope (for example, 1.4 or 2.0), the conversion factors
would differ by 8--9\%.  
The Galactic column density in the CDFS direction is $8\e{19}$
  cm$^{-2}$, so that the bands 2--10 and 5--10 keV are not
  significantly affected and no correction for Galactic absorption is needed.
A histogram of the
fluxes of the XMM-CDFS 2-10 keV catalogue is shown in
Fig.~\ref{fig:fluxhisto}. The number of detected sources drops
at fluxes $\lesssim 10^{-15}$ \ergscmq, with the faintest source at $6.6\e{-16}$
\ergscmq\ giving the survey flux limit. For comparison, we also plot
the flux distribution from X11 and \citet{brunner08}. While the
\chandra\ 4~Ms survey is deeper as expected, the XMM-CDFS contains a
considerable larger number of sources at medium-faint fluxes
($2\e{-15}$--$2\e{-14}$ \ergscmq), most of which are anyway detected
  in the ECDFS. The Lockman hole survey has a
distribution similar to the XMM-CDFS, though a larger number of
sources (mainly at faint fluxes) is detected in the latter than in the
former.

\subsection{Catalogue significance}
\label{sec:significance}

It is not easy to compute analytically the joint probability of a
false-positive detection in the two stages, though some hints can be obtained
using conditioned probability:
\begin{equation}
\prob(\mathrm{detection})=\prob(P)\ \prob(E | P)
\end{equation}
where $\prob(P)$ is the probability of a \pwxd\ false-positive
detection ($6.3\e{-5}$), and $\prob(E | P)$ is the probability of
an \emld\ false-positive after the \pwxd\ selection.
The value of $\prob(\mathrm{detection})$ can be bounded
considering the two extreme cases of $\prob(E | P)=1$
(i.e., accepting all \pwxd\ sources) and $\prob(E |
P)=\prob(P)\,\prob(E)$ (i.e., considering the two detection
stages as independent experiments; $\prob(E)=1.01\e{-2}$):
\begin{equation}
\prob(P)\,\prob(E) < \prob(\mathrm{detection}) < \prob(P)
\end{equation}
which, in terms of $\sigma$, give a threshold significance
for the catalogues in the interval $4\sigma$--$5\sigma$.

The number of spurious detections can be used to obtain an
approximation for $\prob(E|P)$. This number will be first estimated
in Sect.~\ref{sec:detmask} using simulations, and then refined in
Sect.~\ref{sec:newsources} by comparing with the \chandra\ catalogues
and by inspecting the signal/noise ratios of the sources. Therefore,
considering 12 (4) spurious sources in the 2--10 (5--10) keV band:
\begin{eqnarray}
\prob(\mathrm{detection; 2-10})\sim 6.3\e{-5} \times 12/411 \sim 1.8\e{-6} \\ %
\prob(\mathrm{detection; 5-10})\sim 6.3\e{-5} \times 4/196 \sim 1.3\e{-6}
\end{eqnarray}
which may be quoted as the approximate significance threshold for the catalogues
presented here, and roughly correspond to $4.77\sigma$ and $4.84\sigma$
for the 2--10 and 5--10 keV bands, respectively.

\subsection{Main catalogues}
\label{sec:maincat}

The XMM-CDFS 2-10 keV and 5-10 keV catalogues include 339 and 137
sources, respectively, detected in both stages. The catalogues are
available in electronic form from the Centre de Donne\'es
Astronomiques de Strasbourg (CDS), and from the XMM-CDFS website%
\footnote{http://www.bo.astro.it/xmmcdfs/deeprime/}, where we
  will publish updates to the redshifts whenever they become available. The column
description is for both as follows:
\begin{enumerate}
\item IAU\_IDENTIFIER --- source identifier following International
  Astronomical Union conventions;
\item ID210 (2--10 keV catalogue) or ID510 (5--10 keV catalogue) ---
  \pwxd\ source number;
\item RA --- right ascension (degrees);
\item DEC --- declination (degrees);
\item RADEC\_ERR --- error on position (arcsec; $1\sigma$)
\item COUNTS --- \emld\ sum of the net source counts from MOS1, MOS2
  and PN;
\item COUNTS\_ERR --- error on the COUNTS ($1\sigma$);
\item BKG ---  (\emld) background counts/arcsec$^2$;
\item EXPOSURE --- exposure time, averaged over the three cameras;
\item RATE --- (\emld) net count rate, summed on the three cameras;
\item RATE\_ERR --- error on the RATE ($1\sigma$);
\item FLUX --- (\emld) flux (\ergscmq);
\item FLUX\_ERR --- error on the FLUX ($1\sigma$);
\item SIGNIFICANCE --- (\pwxd) detection significance (no.\ of sigma);
\item DET\_ML --- (\emld) detection likelihood;
\item DET\_SCALE --- (\pwxd) wavelet detection scale;
\item CID --- \chandra\ source number (see Sect.~\ref{sec:optical}). If the letter E appended to the
  number, then the ID is from \citet{lehmer05}; otherwise, it is from X11;
\item REDSHIFT;
\item REDSHIFT\_REF --- reference for choice of redshift;
\item ID510 (5--10 keV catalogue) or ID210 (2--10 keV catalogue) ---
  source number in the other band;
\item NOTES --- mainly about source blends and counterparts.
\end{enumerate}

\subsection{Supplementary catalogues}
\label{sec:supplcat}

The catalogues of the 74 2--10 keV, and 61 5--10 keV sources detected
only in the first stage are also available in electronic form.
These supplementary sources are on
average detected at low significance; many of them are on the borders
of the field of view; and a few are in crowded fields where \emld\ has
trouble separating the different PSF components. Nonetheless, 4 of
these sources are bright enough that a spectrum could be extracted
(Comastri et al., in prep.).

The columns whose values in the main catalogue were derived with
\emld\ have been replaced, where possible, with analogous quantities
derived with \pwxd. The main differences between the output columns of
\pwxd\ and \emld\ are \textit{i)} that \pwxd\ rescales the MOS counts
to the PN response before summing them, while \emld\ does not, and
\textit{ii)} that \pwxd\ sums the exposure times while \emld\ averages
them. For consistency, we renormalized
the source counts and exposure times to the same scale of \emld.

The column description is as follows:
\begin{enumerate}
\item IAU\_IDENTIFIER --- source identifier following International
  Astronomical Union conventions;
\item ID210 (2--10 keV catalogue) or ID510 (5--10 keV catalogue) ---
  \pwxd\ source number;
\item RA --- right ascension (degrees);
\item DEC --- declination (degrees);
\item RADEC\_ERR --- error on position (arcsec; $1\sigma$);
\item COUNTS --- net source counts from MOS1, MOS2 and PN;
\item COUNTS\_ERR --- error on the COUNTS ($1\sigma$);
\item BKG ---  background counts/arcsec$^2$;
\item SUMMED\_EXPOSURE --- exposure time, summed on the three cameras;
\item RATE --- net count rate;
\item RATE\_ERR --- error on the RATE ($1\sigma$);
\item FLUX --- flux (\ergscmq);
\item FLUX\_ERR --- error on the FLUX ($1\sigma$);
\item SIGNIFICANCE --- detection significance (no.\ of sigma);
\item DET\_SCALE --- wavelet detection scale;
\item CID --- \chandra\ source number (see
  Sect.~\ref{sec:optical}). If the letter E appended to the number,
  then the ID is from \citet{lehmer05}; otherwise, it is from X11;
\item REDSHIFT;
\item REDSHIFT\_REF --- reference for choice of redshift;
\item ID510 (5--10 keV catalogue) or ID210 (2--10 keV catalogue) ---
  source number in the other band;
\item NOTES --- mainly about source blends and counterparts.
\end{enumerate}

\section{Coverage and number counts}
\label{sec:completeness}

\begin{figure*}
  \centering
  \includegraphics[width=\columnwidth,bb=13 157 565 693,clip]{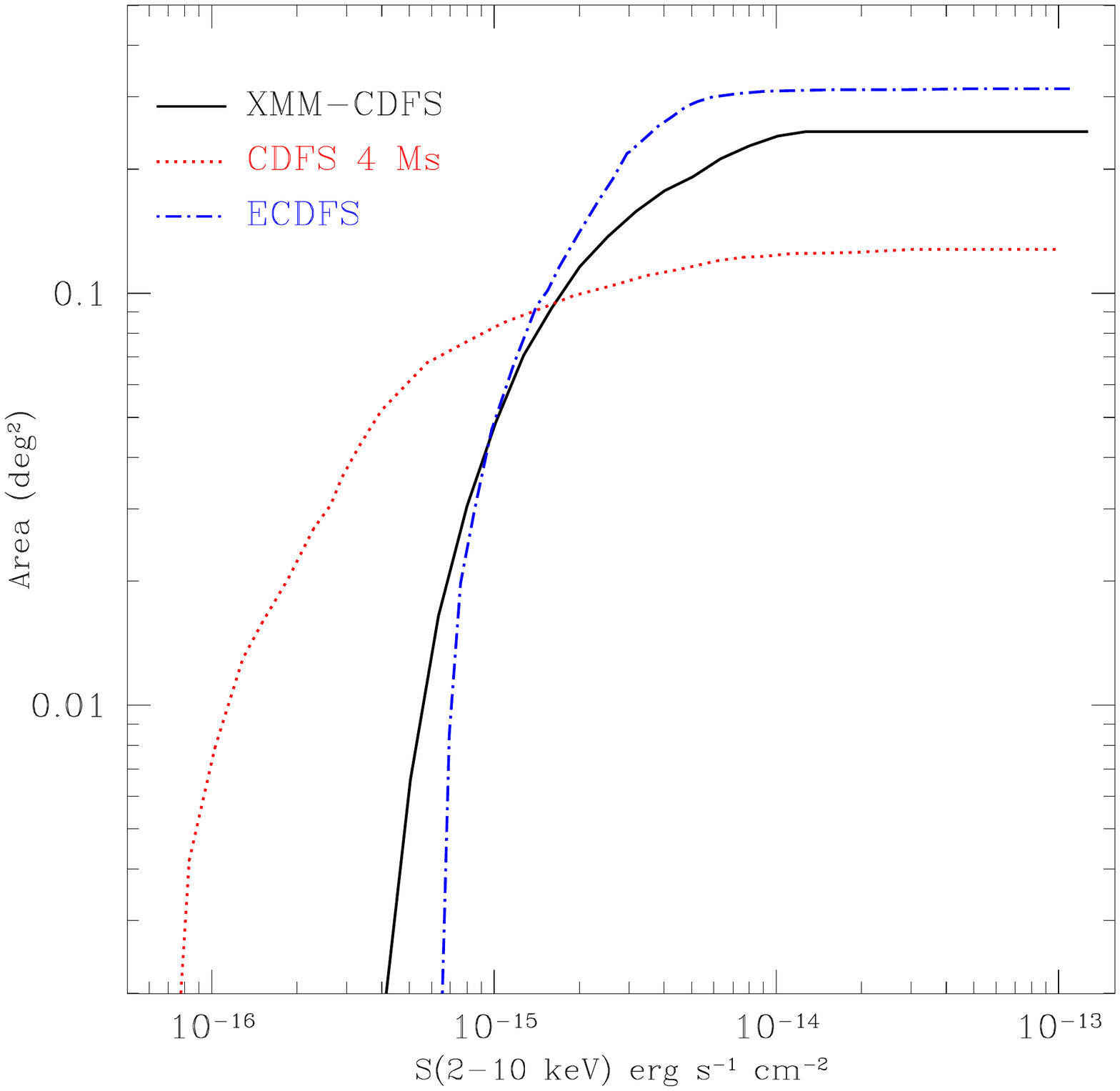}
  \includegraphics[width=\columnwidth,bb=13 157 565 693,clip]{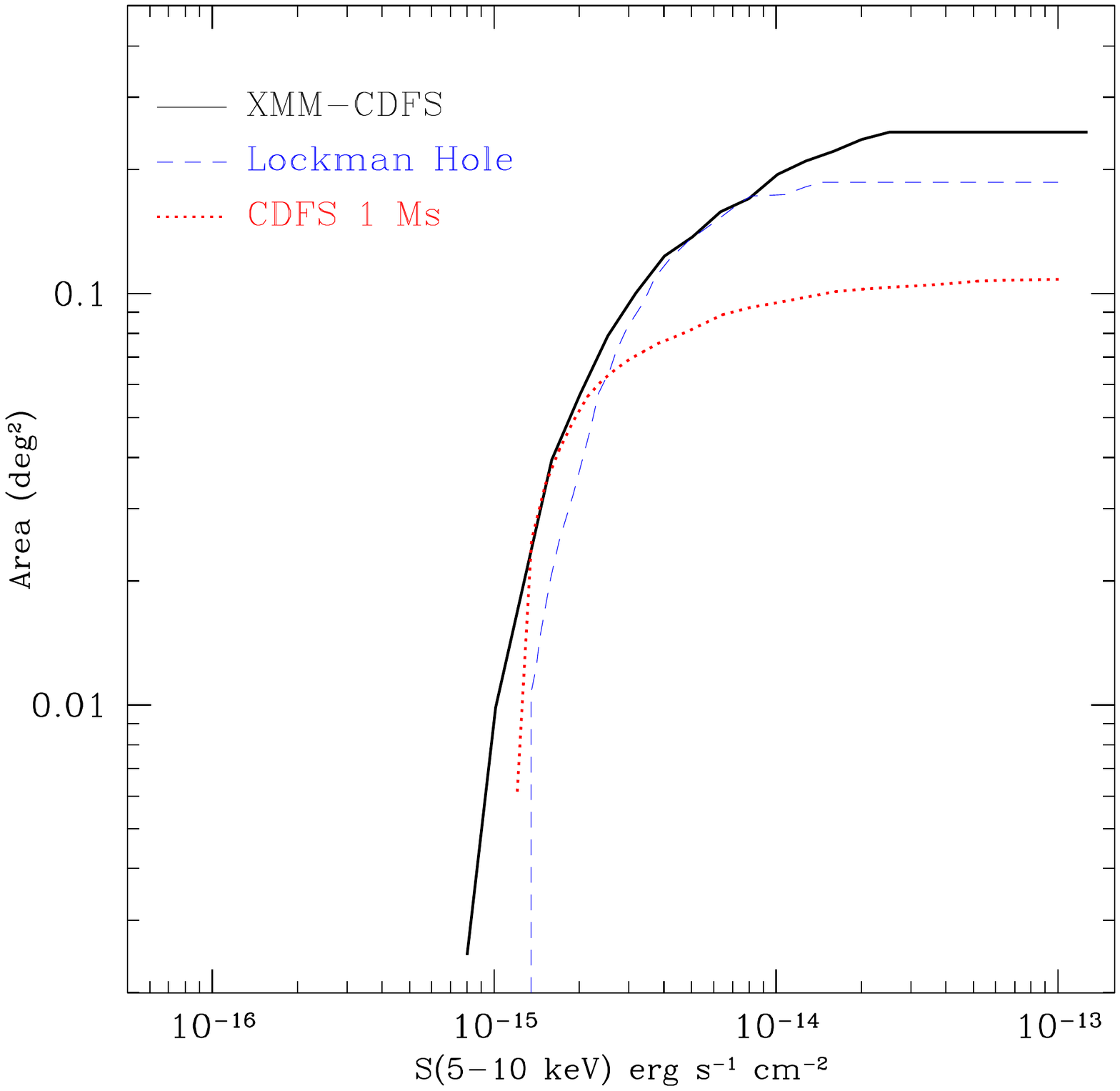}
  \caption{Coverage of the XMM-CDFS survey (solid black lines) in the
    2--10 keV (left panel) and 5--10 keV band (right panel), defined
    as the fraction of simulated sources which were successfully
    detected as a function of their output flux, and normalized to the
    area probed by the catalogues.  The dotted red lines show, for
    comparison, the same quantity for \chandra\ (left panel: 4~Ms
    survey, X11; right panel: 1 Ms survey, \citealt{rosati02}). The
    dashed blue line in the right panel shows the coverage for the
    \xmm\ Lockman Hole \citep{brunner08}. We have not shown in the
    right panel the CDFS 4~Ms \citep{lehmer2012} since formally it is
    in a softer band (4--8 keV).}
  \label{fig:completeness}
\end{figure*}

The survey coverage has been estimated with extensive simulations,
which are described in Sect.~\ref{sec:simulations}.
In Fig.~\ref{fig:completeness} we show the coverage in deg$^2$,
defined as the average fraction of sources per simulation with a
successful match between the output and input catalogues, times the
surveyed area, as function of the flux.  The confidence intervals for
the coverage, estimated according to the \citet{fieller1954}
formula for the ratio of two average quantities, are of the order of
the line width and thus they are not plotted. %
The coverage is
normalized to 0.2466 deg$^2$, i.e.\ the area with exposure$\,\ge
\!\!\mathrm{min}(T)$, where $T$ is the exposure column in the 2--10 keV catalogue.

In Fig.~\ref{fig:completeness} we also show, for comparison, the
coverage for the \chandra\ 4~Ms survey in the 2--10 keV band (X11) and
that for the \xmm\ Lockman Hole 1 Ms survey for the 5--10 keV band
\citep{brunner08}.

\label{sec:lognlogs}

The number counts can be derived from the 2--10 and 5--10 keV
catalogues and the %
coverage curves derive above. The cumulative number counts are the sum
of the inverse areas in which the sources could be detected:
\begin{equation}
N(>S) = \sum_{S_i>S} \frac{1}{A_i}
\end{equation}
where $S_i$ and $A_i$ are the flux and area for a given source
$i$. With this definition, cumulative counts are not binned.

For the differential counts, we bin the sources according to their
fluxes, and define
\begin{equation}
n(S) = \frac{1}{\Delta S} \sum \frac{1}{A_i}
\end{equation}
where the sum is performed on all sources with flux $S_i\in [S-\Delta
S/2,S+\Delta S/2]$. The Poissonian error on the counts in any bin can
be computed with the \citet{gehrels86} approximation.
\smallskip

\begin{figure*}
  \centering

\begin{tabular}{cc}
    \includegraphics[width=.47\textwidth,bb=0 0 545 531,clip]{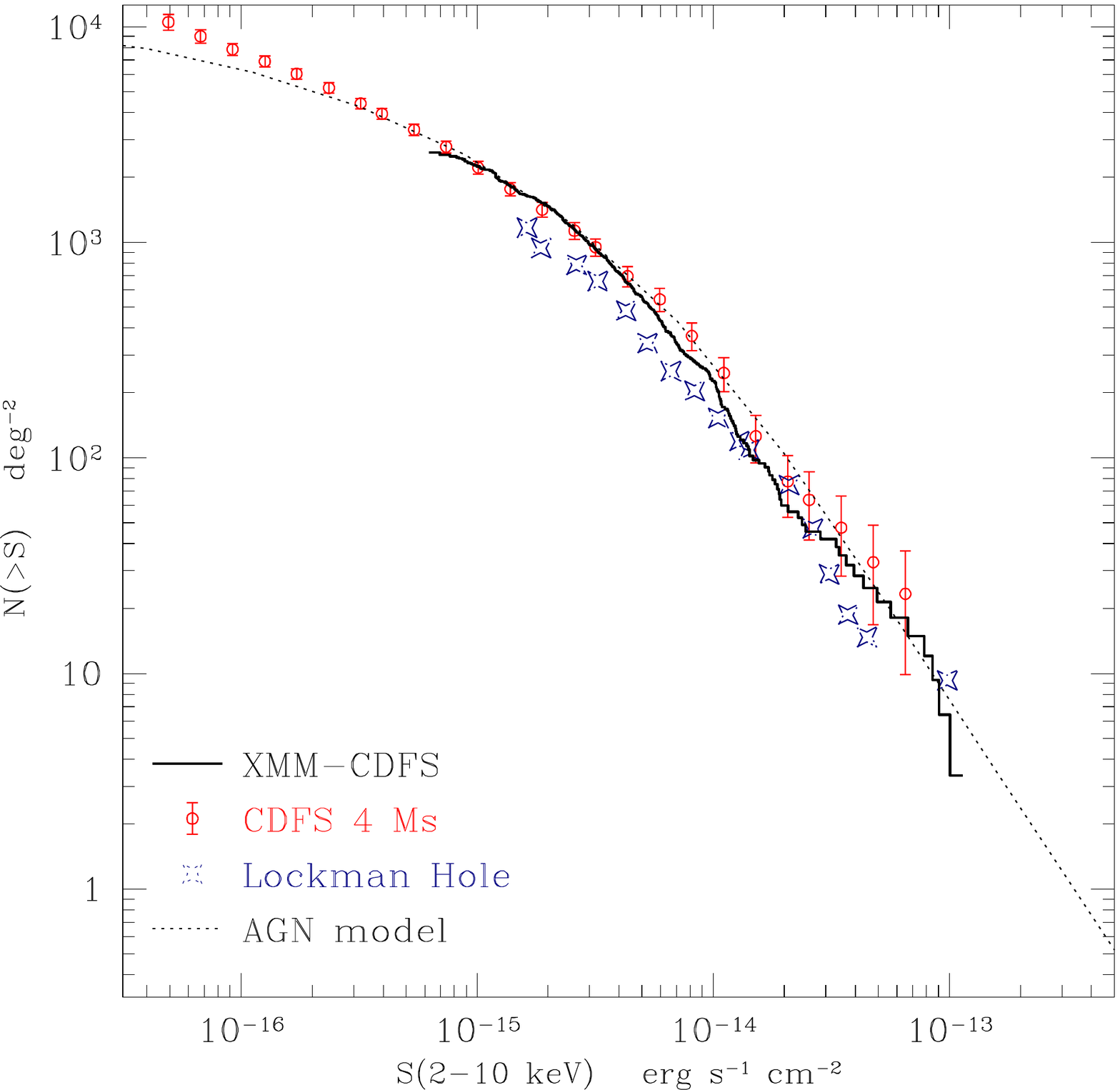}
&   \includegraphics[width=.47\textwidth,bb=0 0 545 531,clip]{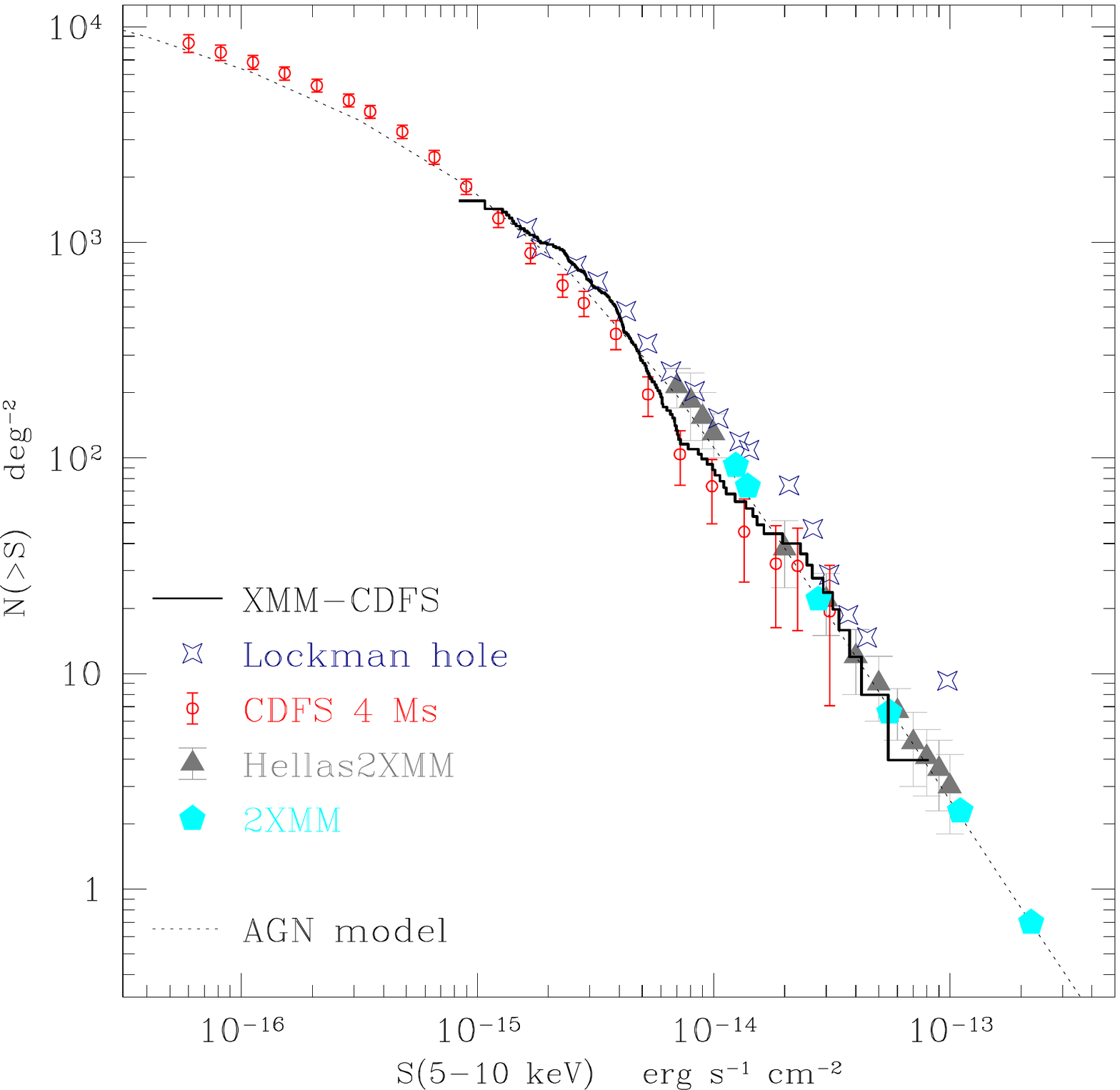}\\
\\
  \includegraphics[width=.47\textwidth,bb=0 348 236 598,angle=90,clip]{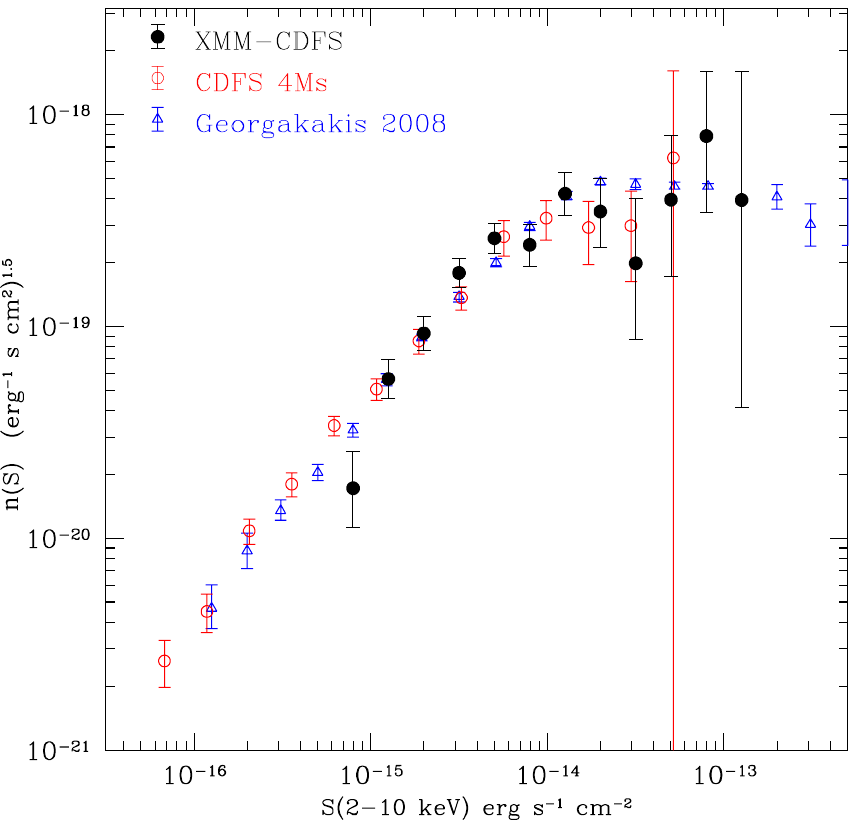}
& \includegraphics[width=.47\textwidth,bb=0 348 236 598,angle=90,clip]{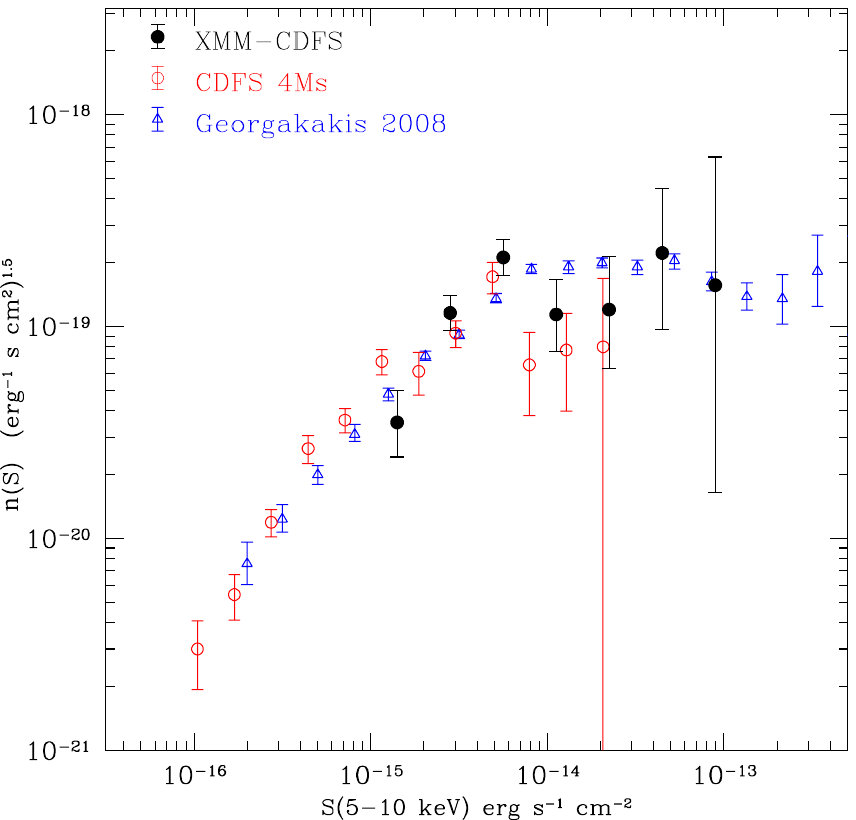}
\end{tabular}
  \caption{Upper panels: cumulative number counts in the 2--10 keV
    (left) and 5--10 keV (right) bands.
    Lower panels: differential number counts (left: 2--10; right:
    5--10 keV band) in Euclidean scale (i.e., $S^{2.5} \de
      N/\de S$).
    For comparison, in all panels
    we plot the \chandra\ number counts \citep[red circles]{lehmer2012}.
    In the upper panels, we also show the cumulative number counts
    from the \xmm\ survey in the  Lockman Hole \citep[blue
    stars]{brunner08}, and the model cumulative counts for AGN
    \citep[black dotted line]{gilli07}. In the upper right panel, we also plot the number counts from
    the Hellas2XMM \citep{baldi02} and 2XMM
    \citep{mateos08}.
    In the lower panels, we also show the differential number counts from \citet{georgakakis2008}.
}
  \label{fig:lognlogs}
\end{figure*}

The number counts are shown in
Fig.~\ref{fig:lognlogs} (upper panels: cumulative number counts; lower
panels: differential). 
The cumulative and differential number counts observed by \chandra\
\citep{lehmer2012} and the cumulative model for AGN \citep{gilli07}
are also shown. The
\chandra\ number counts have been converted from the 2--8 and 4--8 keV
bands to the 2--10 and 5--10 keV bands, respectively, assuming the
same spectrum used for the XMM-CDFS catalogue.

In both bands there is, as expected, a near-perfect agreement between
the \chandra\ and \xmm\ estimates.  Some low-significance deviations
at fluxes around $6\e{-15}$--$10^{-14}$ (2--10 keV) and 2--$6\e{-15}$
\ergscmq\ (5--10 keV) might be due to the larger field of view of the
XMM-CDFS with respect to \chandra. 

A cross-correlation analysis between the \chandra\ ACIS-S3 detector
chip and the three \xmm\ cameras in the 2--8 keV band
\citep{tsujimoto2011} showed that \chandra\ finds fluxes which are
larger than the \xmm\ ones by 10--$20\%$. The 4~Ms survey however was
performed with the ACIS-I detector, which was not analyzed by
\citet{tsujimoto2011}. If the ACIS-S3 results also hold for the ACIS-I
detector, than the larger fluxes found by \chandra\ would be
consistent with the minor deviations found in the \lognlogs.  As to
the 5--10 keV band, Fig.~\ref{fig:lognlogs} (upper right panel) would
suggest that any deviation would go in the opposite direction
(\chandra\ fluxes being lower than the \xmm\ ones); however the very
hard band was not considered by \citet{tsujimoto2011}.

In both bands the number counts are also in agreement with the model.

The Lockman Hole was also observed with \xmm; \citet{brunner08}
presented a 5--10 keV \lognlogs\ which is in agreement, within
poissonian errors, with the XMM-CDFS. At brighter fluxes, we compare
with the counts from the Hellas2XMM \citep{baldi02} and 2XMM
\citep{mateos08} surveys; all of them
are in agreement, within errors, with the XMM-CDFS.
Poissonian errors are plotted only for the CDFS and Hellas2XMM,
to avoid cluttering the figure.

One interesting thing to note is that in the interval
$7\e{-15}$--$1.2\e{-14}$, the CDFS and the XMM-CDFS 5--10 keV number
counts stay below all other determinations. This may be a
low-significance feature of the CDFS field (though formally all number
counts are consistent with each other within Poissonian errors).

Finally, we show the differential number counts derived by
\citet{georgakakis2008} from a combination of deep (CDFS and \chandra\
Deep Field North) and shallow surveys. \citet{georgakakis2008} report
as their best-fit model a broken power-law of the form:
\begin{equation}
  \label{eq:brokenpo}
 \frac{\de N}{\de S} \propto \left\{
   \begin{array}{ll}
      S^{\Gamma_1}, &S\le S_\mathrm{break}\\
      S^{\Gamma_2}, &S\ge S_\mathrm{break}
   \end{array}
  \right.
\end{equation}
with slopes $\Gamma_1=-1.56\pm0.04$ and
$\Gamma_2=-2.52_{-0.09}^{+0.07}$, and with a break flux
$\mathrm{Log}(S/$\ergscmq$)=-13.91_{-0.05}^{+0.08}$ for the 2--10 keV
band ($-1.70_{-0.06}^{+0.08}$, $-2.57_{-0.09}^{+0.07}$,
$-14.09_{-0.05}^{+0.08}$ respectively for the 5--10 keV band). The
XMM-CDFS data points are consistent within 1 or 2$\sigma$ with the
broken power-law model.  
A binned $\chi^2$ fit yields for the XMM-CDFS (quoted errors
  are $1\sigma$):
$\Gamma_1=-0.85\pm1.2$,
$\Gamma_2=-2.29\pm0.36$
and break flux $\mathrm{Log}(S/$\ergscmq$)=-14.46\pm0.30$ for
the 2--10 keV band, and $-0.77\pm2.6$, $-2.49\pm0.48$ and $-14.47\pm0.41$ for the 5--10
keV band. The best-fit parameter are in agreement,
within errors, with the \citet{georgakakis2008} ones, with the
possible exception for the 2--10 keV break
flux which differs by $2\sigma$.
This may be the result of the smaller number of data points at fluxes
fainter than the break in the XMM-CDFS than in
\citet{georgakakis2008}, and of the larger error on the faintest data
point.

\section{Simulations}
\label{sec:simulations}

We have performed extensive simulations to address the XMM-CDFS
properties in terms of survey coverage and source confusion. For each
simulation we built a mock survey, and ran the two-stage detection
process.

We have used a simulator developed with the aim to reproduce the
background and the PSFs of any given observation as closely as
possible. Though currently tuned for \xmm, the simulator can be
adapted to other missions. While more details are given in the
Appendix, the main features are:
\begin{itemize}
\item the simulator attempts to reproduce the source and background
  distribution for each of the 99 exposures
  (33 \obsid s times 3 cameras) of the XMM-CDFS independently;
\item the background level is decomposed into its constituents (cosmic,
  particle and residual soft protons), whose spatial distributions are
  reproduced separately;
\item the cosmic X-ray background is modeled after \citet{gilli07};
  the particle and proton backgrounds are resampled from maps
  distributed as part of the \xmm\ calibration;
\item the distribution of the XMM-CDFS background counts is reproduced
  by the simulator with an error of $\sim 10\%$;
\item the two-stage source detection is run on each simulation;
\item a sizable number of simulations is run so that average
  quantities (number of sources, coverage, number of confused sources)
  can be derived.
\end{itemize}

Given the long exposure time ($\sim 3$ Ms) of the XMM-CDFS
survey, the average number of counts per pixel is much larger
than that for other \xmm\ survey projects. Because of the
difference in the pointing positions among the 33 \obsid s, the
background counts have a complex spatial pattern, mainly given by the
chip gaps and borders occurring in slightly shifted positions.
In any \xmm\ observation, the following background components are
present in addition to the detectable sources:
\begin{itemize}
\item \textit{cosmic X-ray background} (CXB), produced by the many sources which
are below the survey flux limit;
\item \textit{particle background} (PB; sometimes also called ``non X-ray
  background''), which basically includes the electronic noise and a
  few fluorescent lines produced by the telescope assembly;
\item residual \textit{soft protons} (SP): a quiescent background due to
  soft proton clouds;
\item \textit{solar wind charge-exchange}, negligible at energies
  $\gtrsim$ 2 keV and therefore not considered here.
\end{itemize}
The modelling of each simulated component will be discussed in detail
in the next Sections. Our treatment of the different background
components builds directly on the work by \citet{deluca04},
 \citet{snowden04} and \citet{snowden08}; see also \citet{esas_cookbook}.

\subsection{X-ray sources and cosmic background}
\label{sec:cosmic}

For each simulation, an input catalogue of cosmic sources is generated
from a \lognlogs\ obtained as the sum of: i) the 2-10 keV
\lognlogs\ of AGN, calculated with the \citet{gilli07} model over its
full range of parameters, and ii) the 2-10 keV galaxies number counts
computed from the \citet{rcs05} model. This \lognlogs\ extends from
$2.8\e{-20}$ to $1.0\e{-11}$ \ergscmq, thus allowing simulation of both
the detectable sources, and the cosmic X-ray background. It is sampled
on a square area slightly larger than the CDFS, assuming uniform
distributions in RA and DEC, thus giving a number of $\sim 50000$
sources per simulation. 

Most of these sources are too faint to contribute, on average, even a
single photon to an observation; however when they are taken together,
the sources below the survey flux contribute to the total
(cosmic+particle+proton) observed background.

The flux to count rate conversion factors have been calculated for a
power-law spectrum with slope $\Gamma=1.7$, as done for the real
catalogue (considering $\Gamma=1.4$ or 2.0 would change the fluxes by
$\sim 8\%$); in doing this, we have assumed a single average spectral
slope for all sources, which is acceptable since we are interested only
in broad-band detections.

\subsection{Particle background}
\label{sec:particle}

The PB is mainly due to the instrument
electronics and to reflection lines from the telescope and its
mounting.  The spatial distribution of the PB has been taken from the
Filter Wheel Closed (FWC) observations provided by the ESAS
CALDB\footnote{The Extended Source Analysis Software (ESAS), developed
  by S. Snowden and K. Kuntz, is a part of the standard \xmm\ SAS
  software and includes several calibration files.}.  The FWC were made
with the filter wheel in the closed position, and without the
calibration source on\footnote{Conversely, the more common CALCLOSED
  observations do have the calibration source visible, in addition to the PB.}.

The actual PB level in any observation may be determined by considering
the \textit{corner areas}, i.e., the parts of the MOS and PN detectors
which lie outside of the FOV and are neither exposed to the CXB nor to
the protons. The data for the corner areas are normally filtered out
by the \xmm\ pipeline, but can be obtained by reprocessing the ``raw
data'' (i.e., the Observation Data Files, ODF). The surface brightness
of the corner areas needs to be appropriately rescaled to obtain the PB
brightness inside of the FOV.

\begin{figure*}
  \centering
  \includegraphics[width=.98\textwidth]{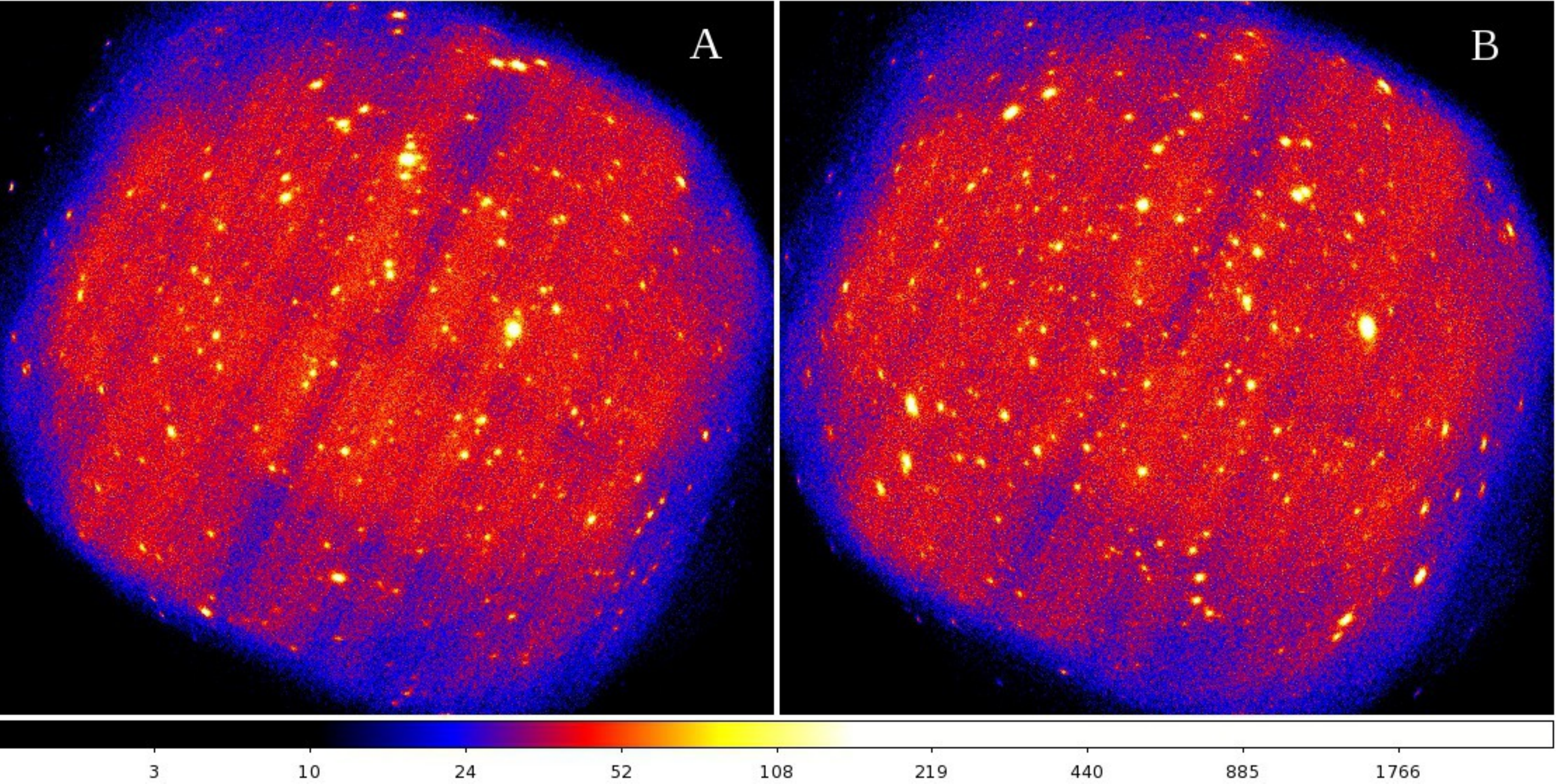}
  \caption{Comparison of the real XMM-CDFS (A) with one simulation
    (B). The colour scale is the same for both panels and maps the
    2--10 keV intensity.}
  \label{fig:sim}
\end{figure*}

Let us define the total average surface brightnesses which are
observed inside and outside of the FOV as $\Sigma_\mathrm{in}$ and
$\Sigma_\mathrm{out}$, respectively.  The value of both $\Sigma$s is
different for any obsid and camera, and can also be measured for the
FWC.  To minimize uncertainties due to small number statistics for
bright sources, the positions corresponding to the 30 brightest
sources have been masked out during the determination of the above
surface brightnesses. This corresponds to removing sources brighter
than the $2\e{-14}$ \ergscmq\ threshold; at this flux level, 43\% of
the background is resolved (using the HEAO-1 value,
\citealt{gruber1999cxb}).

On the one hand, the FWC only contain PB, therefore:
\begin{eqnarray}
\label{eq:sigFWC}
\left\{
  \begin{array}{lllll}
  \Sigma_\mathrm{in,FWC}  &= &\Sigma_{\mathrm{PB,FWC}}\\
  \Sigma_\mathrm{out,FWC} &= &k\ \Sigma_{\mathrm{in,FWC}} &= &k\ \Sigma_{\mathrm{PB,FWC}}\\
\end{array}
\right.
\end{eqnarray}
where $k$ accounts for spatial variations of the PB surface brightness.

On the other hand, for a generic observation, the FOV will contain PB,
proton and cosmic background components, while the corners will only have PB:
\begin{eqnarray}
\label{eq:sigOBS}
\left\{
\begin{array}{lll}
  \Sigma_\mathrm{in}  &= &\Sigma_\mathrm{PB} + \Sigma_\mathrm{SP} + \Sigma_\mathrm{CXB}\\
  \Sigma_\mathrm{out} &= &k\ \Sigma_\mathrm{PB}\\
\end{array}
\right.
\end{eqnarray}
where $\Sigma_\mathrm{PB}$, $\Sigma_\mathrm{SP}$, and $\Sigma_\mathrm{CXB}$ are surface brightnesses.

The important assumption here is that the $k$ is the same for the FWC
and for all the observations, which is equivalent to assuming that the
spatial distribution of the PB is the same in all observations.
Taking $k$ from Eq.~(\ref{eq:sigFWC}), from Eq.~(\ref{eq:sigOBS}) we get
\begin{equation}
\Sigma_\mathrm{PB}=\Sigma_\mathrm{out} \frac{\Sigma_{\mathrm{in,FWC}}}{\Sigma_\mathrm{out,FWC}}.
\end{equation}
The average fractions of PB with respect to the total of
CXB+PB+protons are reported in Table~\ref{tbl:bkglevels}.  The number
of PB photons to be simulated is $\Sigma_\mathrm{PB}\times
A\times E$ where $A$ is the FOV area and $E$ the exposure time.

\subsection{Soft protons}
\label{sec:protons}

Clouds of soft protons are often encountered by \xmm, especially when
the satellite is close to perigee, and cause strong background
flares.  When such a flare is detected, the time intervals in which it
occurs are usually excluded from further analysis. However, there is
also a quiescent component of soft proton background, with much
smaller brightness, that may occur for the entire length of an
observation, and for which the standard recipe of identifying and
excluding high-background periods is not applicable --- or which may
remain even after the flares are excluded. The quiescent soft protons
are subject to the telescope vignetting, though to a lesser extent
than X-ray photons (see Fig.~17 in \citealt{kuntz08}).  Both
\citet{deluca04} and \citet{snowden04} have considered this problem
and we refer to these papers for further details.

Here we estimate the residual soft proton level from Eq.~(\ref{eq:sigOBS}):
\begin{equation}
\label{eq:sigSP}
\Sigma_\mathrm{SP}=\Sigma_\mathrm{in} - \frac{\Sigma_\mathrm{out}}{k}
     - \Sigma_\mathrm{CXB}
\end{equation}
which is also shown in Fig.~\ref{fig:secularbkg} and
Table~\ref{tbl:bkglevels}.

We assume the value $\Sigma_\mathrm{CXB} =7.12\e{-12}$ \ergscmqdegq\
for the CXB brightness, which results
from integrating the AGN and galaxy number counts from zero to the
$2\e{-14}$ \ergscmq\ threshold (Sect.~\ref{sec:particle}). 
The number counts are normalized to the HEAO-1 value for the X-ray
background \citep{gruber1999cxb}; had we normalized to any of the higher
values derived from \xmm\ data (e.g., \citealt{lumb02} or
\citealt{deluca04}), this number would have been proportionally
larger. However, the increase would have been compensated by a
decrease in the soft proton component, with the net effect of a
slight change in the amount of background vignetting. This effect is
probably within the uncertainties with which the background
brightness is measured.  

We now can determine the number of photons contributed by the soft
protons as $\Sigma_\mathrm{SP}\times A\times E$, where $A$ and $E$ are
again the FOV area and the exposure time.

A simulation produced with this recipe is shown in Fig.~\ref{fig:sim},
where an image of the real XMM-CDFS is shown with the same colour scale
for comparison.

\begin{table}[h]
  \caption{\label{tbl:bkglevels}Average percentage contribution of the different
  background components to the overall level.}
\centering
\begin{tabular}{lclll}
\hline\hline
Camera &Band (keV) &CXB &Particle &Proton \\
\hline
MOS1   &2-10 &$1.1\pm0.4$ &$81\pm 10$ &$19\pm 10$ \\  
MOS2   &2-10 &$1.3\pm0.4$ &$80\pm 9$  &$18\pm 9$  \\
PN     &2-10 &$2.2\pm0.8$ &$90\pm 6$  &$7.4\pm6.0$\smallskip\\
MOS1   &5-10 &$0.5\pm0.2$ &$88\pm 8$  &$12\pm 8$ \\
MOS2   &5-10 &$0.6\pm0.2$ &$87\pm 8$  &$12\pm 8$ \\
PN     &5-10 &$1.5\pm0.5$ &$95\pm 4$  &$3.9_{-3.9}^{+4.4}$\\
\hline
\end{tabular}
\tablefoot{
  All values are per cent.
  When calculating the brightness levels, the areas around
  the 30 brightest sources have been excluded. Thus, the cosmic
  background refers to sources with fluxes $\le 2\e{-14}$ \ergscmq.
}
\end{table}

\subsection{Simulated fields and catalogues}
\label{sec:simresults}

\begin{figure}
  \centering
  \includegraphics[width=\columnwidth,bb=13 160 568 699,clip]{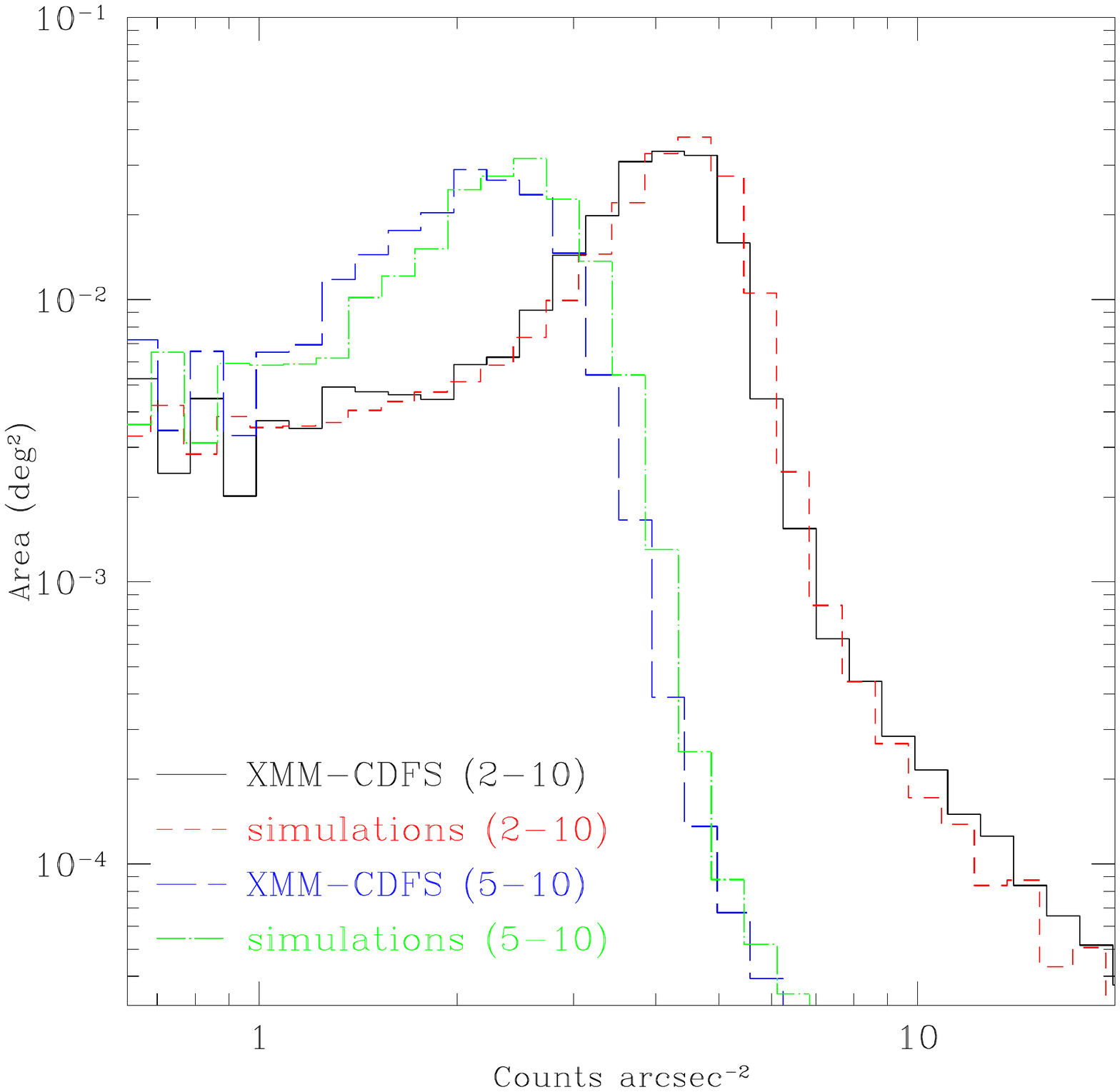}
  \caption{Distribution of the counts per arcsec$^2$ in the
    simulations (average over 5 simulations), compared with the
    XMM-CDFS, for the 2--10 and 5--10 keV bands. The vertical axis
    shows the area covered by each bin. }
  \label{fig:bkgreproduction}
\end{figure}

A comparison between the distribution of counts reproduced by the
simulator and that of the XMM-CDFS is shown in
Fig.~\ref{fig:bkgreproduction}, where the histograms of the counts per
arcsec$^2$ are plotted in the 2--10 and 5--10 keV bands. The counts
include both the background (mainly the peak of the distribution) and
the sources (the right tail).  In both bands, the agreement seems
good, with a horizontal shift between simulations and the XMM-CDFS of
$\sim 10\%$ around the peaks.

We ran about 150 simulations in the 2--10 and 5--10 keV bands,
respectively, to assess the survey properties.  For each simulation,
the two-stage detection process was performed in the same way as for the (real)
XMM-CDFS main catalogue. %
The output catalogues were cross-correlated with the (input) list of
simulated sources. For each detected source, its input counterpart is
defined as the one which minimizes the score $S$ \citep{xmm-cosmos}:
\begin{equation}
\label{eq:crosscorrchi}
S = \left( \frac{\Delta \mathrm{RA}}{\sigma_{\mathrm{RA}}} \right) ^2 +
    \,\left( \frac{\Delta \mathrm{DEC}}{\sigma_{\mathrm{DEC}}} \right) ^2 +
    \,\left( \frac{\Delta \mathrm{RATE}}{\sigma_{\mathrm{RATE}}} \right) ^2
\end{equation}
where $\Delta$(RA, DEC, RATE) are the differences between the output
and input right ascension, declination, and count rate, respectively,
and the $\sigma$ are the corresponding errors.

\begin{figure}
  \centering
  \resizebox{\hsize}{!}{\includegraphics[width=\columnwidth,bb=29 155 564 690]{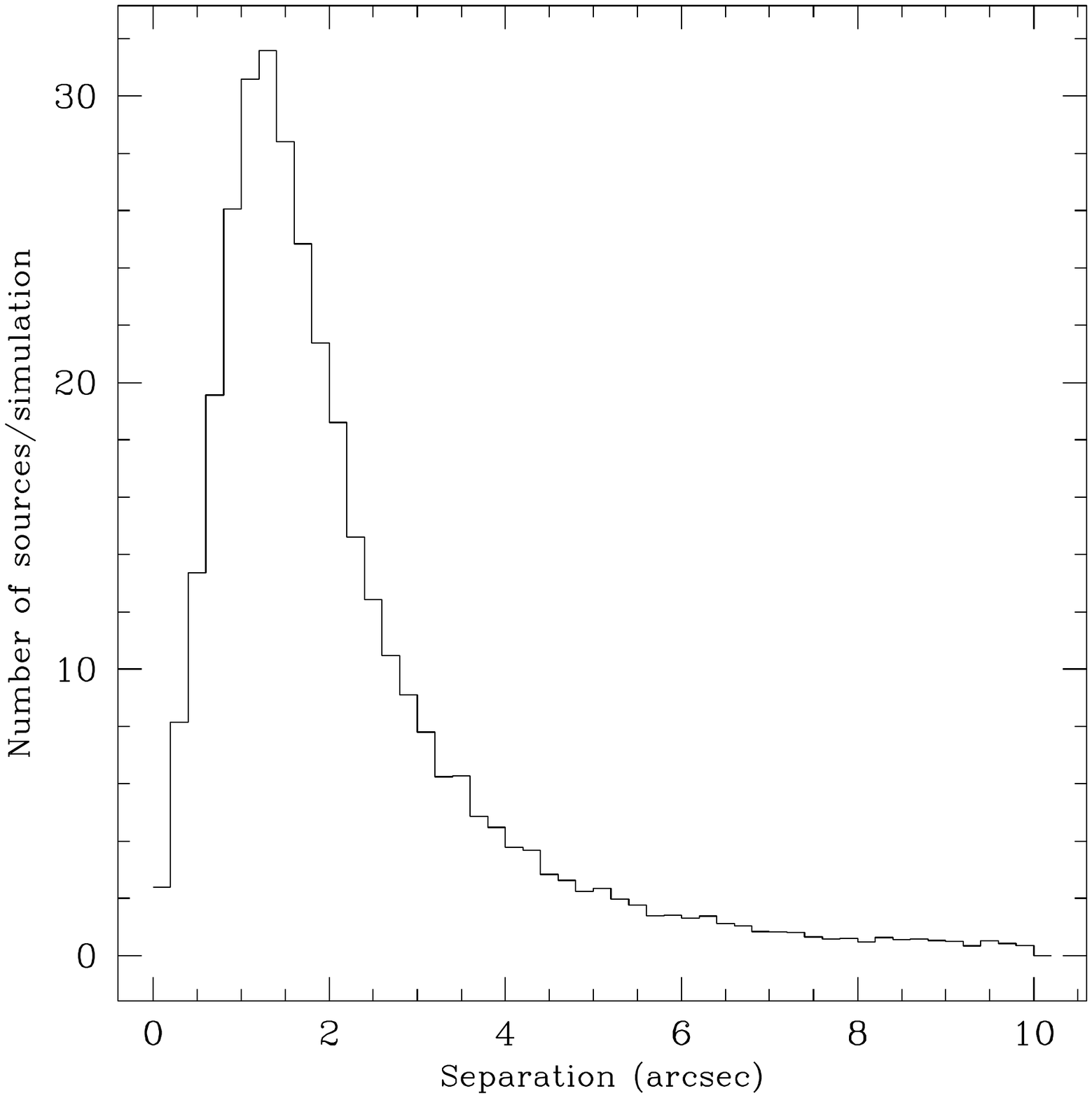}}
  \caption{Spatial separation between the input and output positions
    of simulated sources.
  }
  \label{fig:separation}
\end{figure}

We show in Fig.~\ref{fig:separation} the histogram of the spatial
separation between the output and input positions. Most sources show a
displacement of $\sim 1.25\arcsec$, which was already highlighted by
\citet{xmm-cosmos} and \citet{pineau2011}, and therefore it seems to
be a characteristic of \xmm. The displacement does not occur in any
particular direction and occurs even if the source detection is run on
just one single camera, or on just one obsid. The number of matches
drops sharply for displacements $\gtrsim 2\arcsec$.

\begin{figure}
  \centering
  \includegraphics[width=\columnwidth,bb=0 0 547 534,clip]{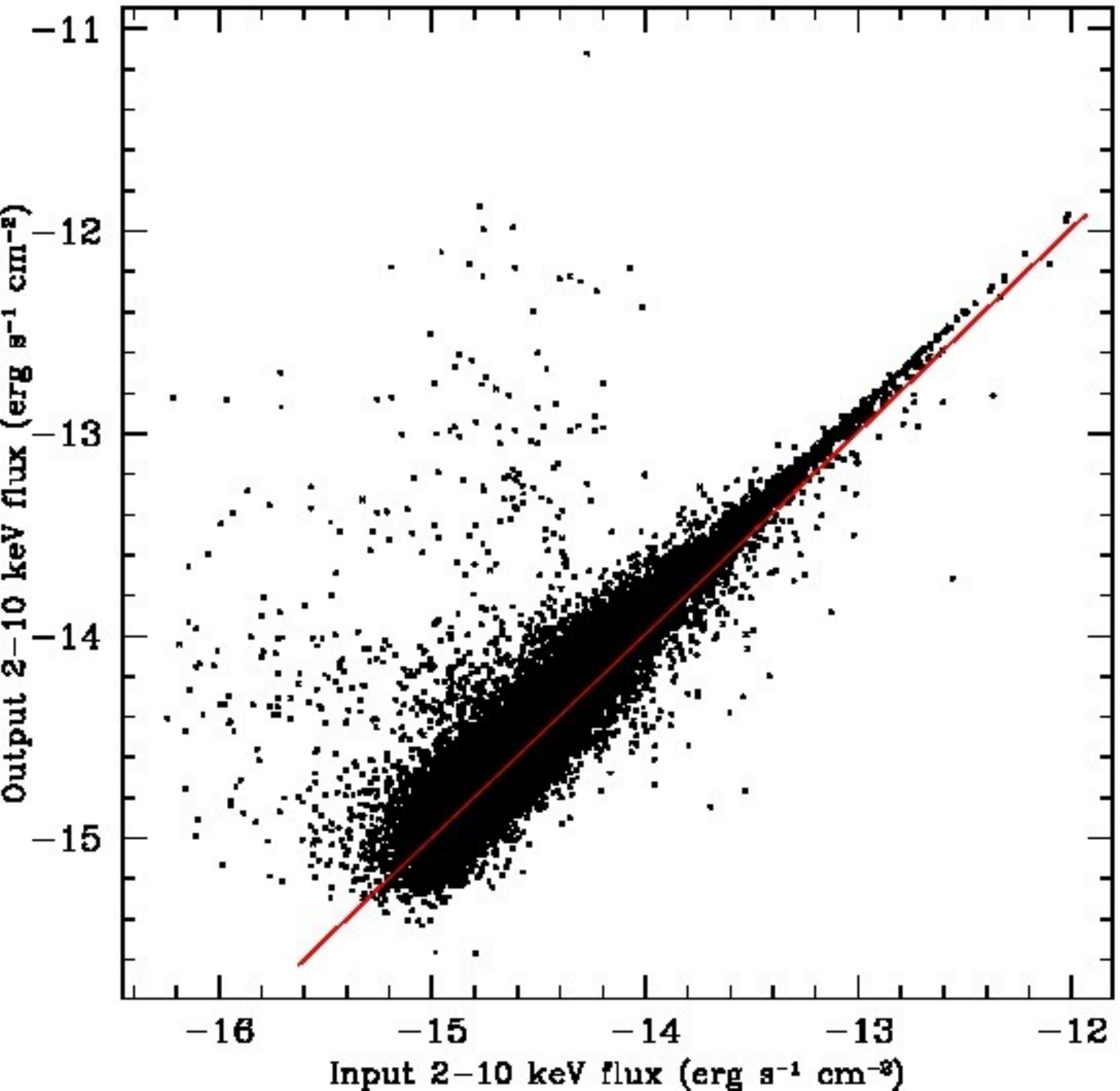}
  \caption{Input vs.\ output fluxes of the simulated sources, from 100
    2--10 keV simulations. Only sources with a match within $6\arcsec$
    are considered. Some outliers are present, due to \emld\
    uncorrectly associating the source with extended emission
    especially in crowded regions, and for which we rather
    use the \pwxd\ fluxes; the number of outliers is a few per
    simulation. The red line shows the identity relation. At
      bright fluxes ($\gtrsim 10^{-14}$ \ergscmq), the output values are about 20\% larger than the
      input ones.}
  \label{fig:inoutrate}
\end{figure}

The input and output fluxes of the simulated sources are shown in
Fig.~\ref{fig:inoutrate}; only sources with displacements $\le
6\arcsec$ are considered here. A few outlier per simulations are
present, which is compatible with the amount found in the XMM-CDFS
(Sect.~\ref{sec:catalogue}). These outliers are mainly due to
\emld\ identifying some sources, especially in crowded areas, as
extended sources. In the compute of the coverage, we identify the
outliers has having an \emld\ flux larger than 3 times the
\pwxd\ flux; the latter is then used for the calculation.  At bright
fluxes ($\gtrsim 10^{-14}$ \ergscmq), the output values are
systematically larger by about 20\% than the input ones. This
uncertainty does not affect the flux range where most of the sources
(82\%) are detected.

\section{Number of spurious sources} %
\label{sec:detmask}

\begin{table}[t]
  \caption{\label{tbl:spurious}Number of spurious sources in the 2--10
    keV band simulations.}
\centering
\begin{tabular}{lrrrrr}
\hline\hline
Exp.    &$\!\!\!\!\!\!$CDFS          &Total sim.  &W.count.     &Spurious   &$\!\!\!\!\!\!$Fract.\\
\hline
any         &335           &$358\pm 16$ &$328\pm 17$  &$30\pm 5$  &8.4\%\\ %
$\ge 300$   &302           &$306\pm 15$ &$296\pm 15$  &$10\pm 3$  &3.3\%\\ %
\hline
\end{tabular}
\tablefoot{
  The column ``CDFS'' gives the number of sources found in the XMM-CDFS
  which respect the same selection of the simulations (i.e.\
  \texttt{extent}$<0.5$ and no de-blending; see text). The columns ``Total
  sim.'', ``W.count.'', ``Spurious'' give the average and standard
  deviation of the number of sources per simulation, the number of
  sources with a counterpart in the input catalogues, and the
  number of spurious, respectively.  The column ``Fract.'' gives the
  fraction of spurious sources over the total number. Exposure
  thresholds are expressed in ks.
}
\end{table}

We estimate the number of spurious sources in the 2--10 keV %
band by running the detection process on the simulated
observations. 
The total number of sources detected in the simulations is reported in
Table~\ref{tbl:spurious}. In the 2--10 keV band, the simulations
produce catalogues which contain on average $358\pm 16$ \srcsim, i.e.\
a larger number of sources than the XMM-CDFS, though the difference is
less than twice the simulations' standard deviation; in around 10\%
of the simulations a number of sources lower than or equal to the
XMM-CDFS is detected. %
The number of detected
sources is in general dependent on the background level; residual
variations between the simulations and XMM-CDFS background (see
Sect.~\ref{sec:simresults}) may help to explain the difference observed in
the 2--10 keV band.

In the \chandra-COSMOS survey, which used the same detection software
as this paper, \cite{puccetti09} found that a large number of spurious
sources, only present in simulations, were characterized by source
sizes smaller than the PSF sizes ---a clear indication of fluctuations
detected as sources--- and removed them. \pwxd\ uses the
\texttt{extent} column to express the source size in ``PSF units''
(i.e., adimensional); in particular, all sources with
\texttt{extent}$<0.5$ are likely to be spurious. Thus, in all
simulations we filtered the catalogue
 to remove source candidates
satisfying this criterion. %

Spurious sources may arise mainly from background fluctuations, or
from an incorrect reproduction of the spatially-dependent background
level.  For instance, the simulations made use of spatial maps of the
soft proton background, which were derived by time filtering of highly
flared observations, and may still contain small local enhancements
originally due to astronomical sources. Local enhancements are also
present in the particle background maps, and further investigation
would be needed to determine wether these are representative of any
given observation.  In general, any background enhancement in the
proton and particle maps is likely to appear in many simulations,
since these maps were used as probability distributions without
smoothing on significantly large scales (e.g., larger than a few PSF
cores). The number of this kind of spurious sources is non-trivial to
estimate; for this reason, the numbers of spurious sources quoted in
Table~\ref{tbl:spurious} should be regarded as upper limits if applied
to the XMM-CDFS; and in Sect.~\ref{sec:conclusion} the comparison of
the \chandra\ and \xmm\ catalogues will be used to further refine the
estimate of the spurious fraction.

A check on the position of the candidate spurious sources shows that
about half of them (i.e.\ $20\pm 4$ in the 2--10 keV band) 
lie in the outer parts of the field, where the exposure is $\le 300$
ks. Under the latter threshold, the number of sources detected in the
simulations with a counterpart in the input catalogues, and the number
of sources in the XMM-CDFS, match more closely.

The main reason for the larger fraction of spurious sources at low
exposure ($\le 300$ ks) is that fluctuations are more important at the
borders of the FOV than in the inner area --- or, in other words, that
fluctuations are ``averaged out'' when the exposure is large. The
spatial distribution of spurious sources has well-defined peaks,
falling in the low-exposure areas, which can be traced back to
localized high background features.

For the rest of this Sect., we define spurious sources as those
detected in the simulations, but without any counterpart in the input
catalogues within $6\arcsec$ from their position; the numbers are
reported in Table~\ref{tbl:spurious}. The threshold of $6\arcsec$ was
chosen after the check for \chandra\ counterparts
(Sect.~\ref{xmmchandra_xcorr}), in which a tail of reliable
associations is present up to $6\arcsec$ especially for sources at
large off-axis angles.
For the 2--10 keV, the average
number of spurious sources is $ 30\pm 5$ \srcsim.  The number of XMM-CDFS
sources to which the comparison can be done is 335 (337 sources
detected at the second stage, without de-blending sources no.\ 501--504,
and with 2 sources not considered because they have
\texttt{extent}$<0.5$).  %

Confused sources may also give rise to spurious sources, in the case
that a pair of input sources is detected as a single one, and that the
detected position is distant more than $6\arcsec$ from both the
components of the input pair (see Sect.~\ref{sec:confusion},
``double-detection pairs''). The number of this kind of spurious
sources is $\sim 3$ per 
simulation, and does not depend on the position in the field.

A check on the number of spurious sources, based on the probability of
association with \chandra\ counterparts and on the signal/noise ratio
in the \xmm\ images, is discussed in Sect.~\ref{xmmchandra_xcorr}.

The fraction of spurious sources is higher in the 5--10 keV band than
in the 2--10 keV, probably because localized background features
contribute with a larger proportion to the total counts. Improving
this would require a deeper study of the \xmm\ background and an
update of the simulator, which we defer to a future paper.

\section{Source confusion}
\label{sec:confusion}

Because of the size of the XMM-CDFS average angular resolution
($8.5\arcsec$ FWHM, which is the median across the FOV; see
Sect.~\ref{sec:astrometry}) and the faint flux limit of the
XMM-CDFS, it is important to understand how the detection process
treats close sources. Depending on the distance of its components, a
pair of sources may be mistakenly detected as a single source
(``source confusion''); the off-axis angle dependence of the PSF and
local small-scale variations of the background also play a
role. This may also impact the estimated number of spurious
sources, if the detected position is distant enough from that of the
pair components. Here, we use the simulations to estimate the amount
of source confusion in the XMM-CDFS survey, and compare at end
with \chandra\ data. In this Section, we only consider the 2--10 keV
band; the 5--10 keV range is expected to be less affected by source
confusion.

\begin{figure*}
  \centering
  \includegraphics[width=\columnwidth,bb=15 157 570 692,clip]{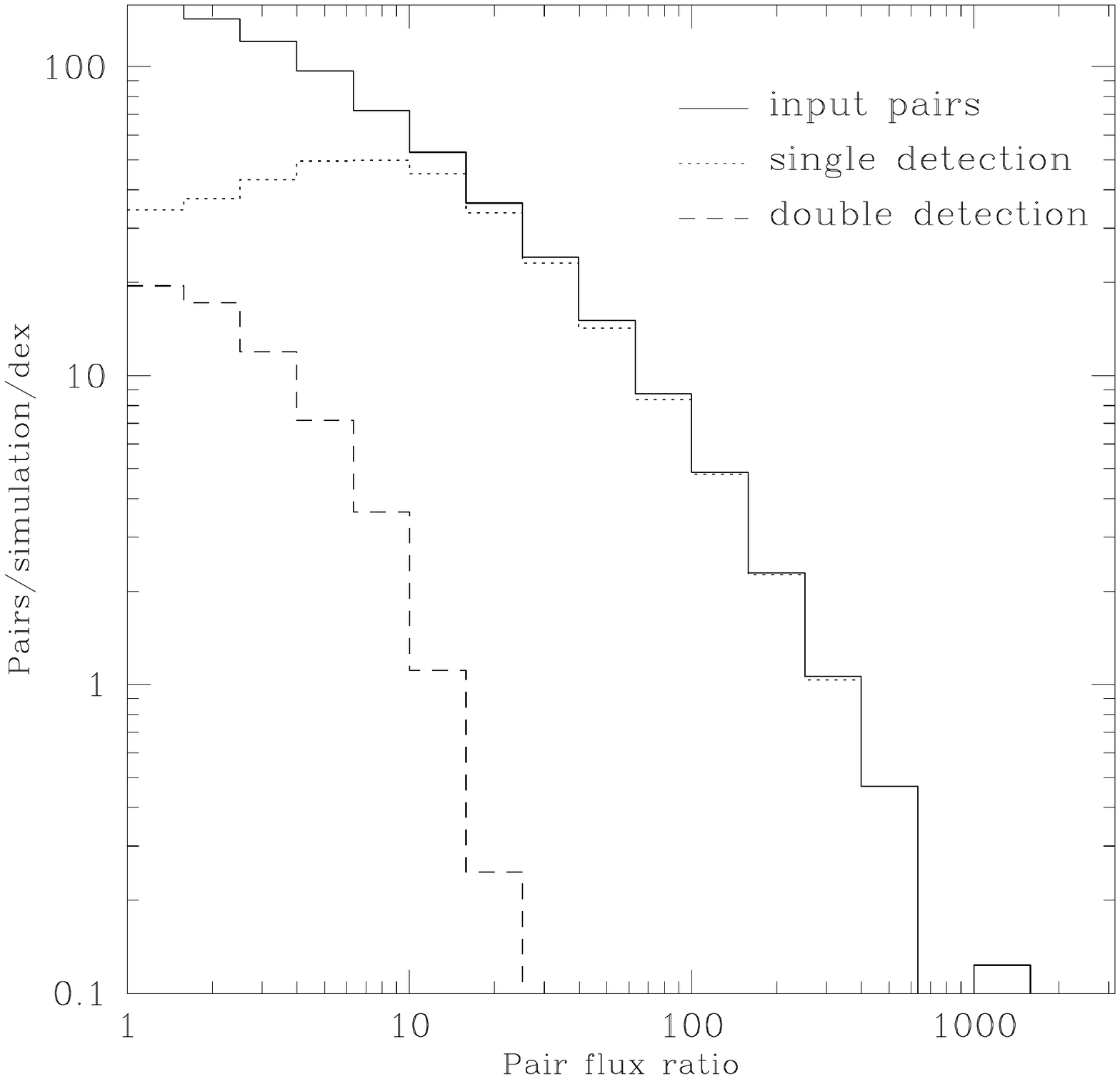}
  \includegraphics[width=\columnwidth,bb=15 157 570 692,clip]{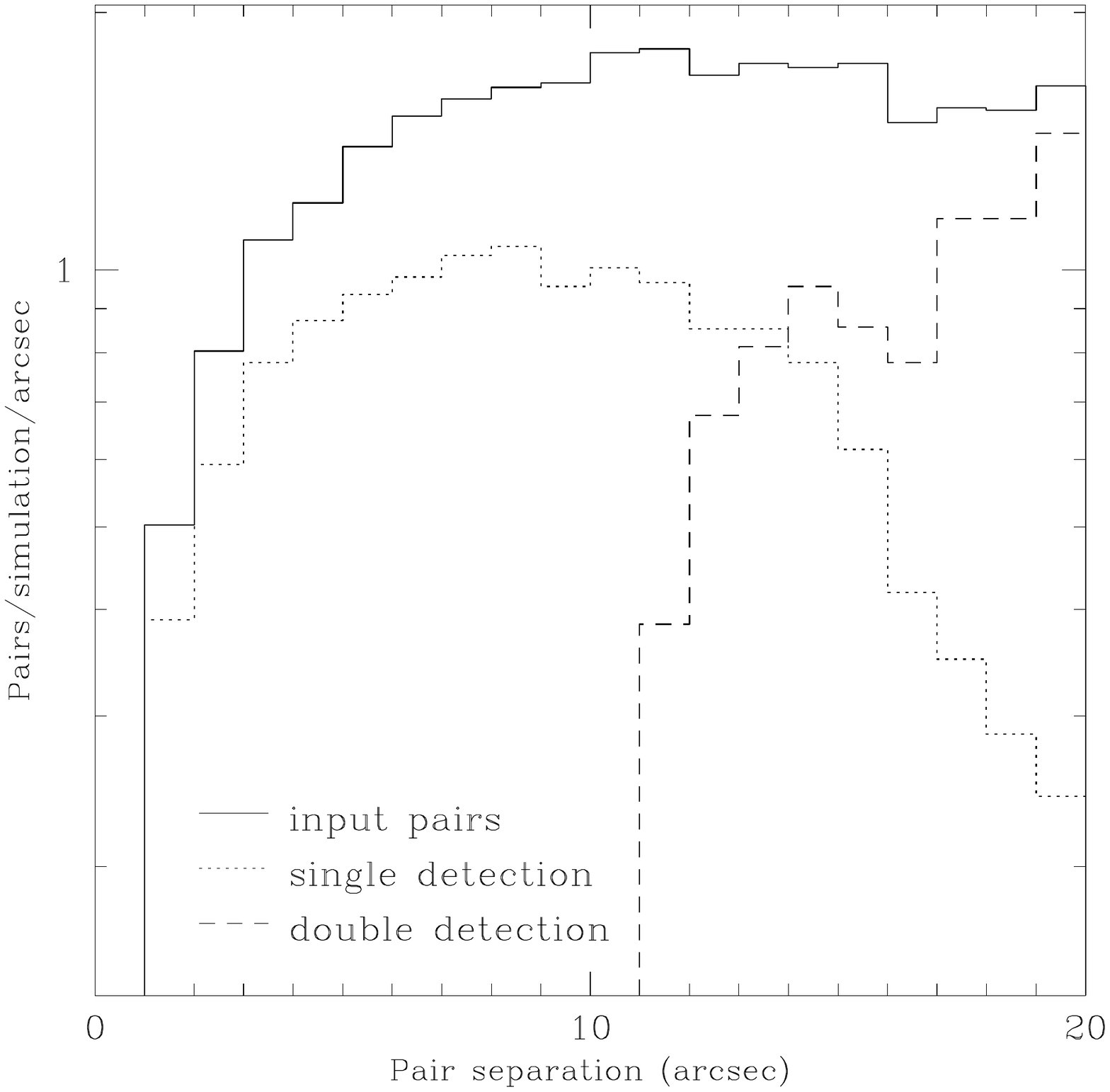}
  \caption{
    \textit{Left panel:} Number of pairs of sources, as function of the
    flux ratio between pairs of sources.
    Sources with fluxes 10 times fainter than the local (i.e.\
    position-dependent) flux limits are included.
    Solid histogram: pairs of input sources. Dotted
    histogram: pairs detected as a single output source.  Dashed
    histogram: pairs where both components have a detection.
    \newline
    \textit{Right panel:} Number of pairs of
    sources, as function of the spatial separation between the two
    pair components.  Line styles as in the left panel.
    Only pairs with flux ratios lower than 3,
    and whose components are brighter than $6.6\e{-16}$ \ergscmq\ 
    are considered.
    The larger number of input pairs with respect to output ones in the
    3$\arcsec$--10$\arcsec$ range is due to pairs where the distance
    between the centroid and the position of the detection is larger
    than $6\arcsec$; this happens for $\sim 3$ sources per simulations.
    }
  \label{fig:pairsep}
\end{figure*}

We start our analysis by considering input sources with fluxes greater than
10\% that of the faintest source detected in the real XMM-CDFS, and
scaled for lower sensitivity areas through an exposure map (i.e., $0.1
\times 6.6\e{-16} / \mathrm{Exp}(\mathrm{RA},\mathrm{DEC})$ \ergscmq,
where $\mathrm{Exp}(\mathrm{RA},\mathrm{DEC})$ is normalized so that
$\mathrm{max}(\mathrm{Exp}(\mathrm{RA},\mathrm{DEC}))=1$).  This
avoids cluttering the results with sources too faint to have a
significant impact on the detection process. For simplicity, we also
do not consider triplets of sources, nor groups with a larger number
of members. With
these constraints, we identify an average of 61 (108) pairs per simulation
with separation $\le 10\arcsec$ ($\le 15\arcsec$) in the input catalogues. 

The number of pairs per simulation, as function of the flux ratio
between the two components of the pair, is shown in
Fig.~\ref{fig:pairsep} (left panel) as the solid histogram (``input
pairs'').  Most pairs have a flux ratio $\lesssim 10$ between the
two components. Pairs with a larger difference between the fluxes of
the components are considerably more rare.

The right panel of Fig.~\ref{fig:pairsep} shows the number of pairs as
function of the separation between the pair components. To be more
representative of what is observable in practice in the XMM-CDFS, we
restrict this panel to pairs with a flux ratio $<3$, and we also
drop any pair with a component fainter than the XMM-CDFS flux limit
($6.6\e{-16}$ \ergscmq). In the following we will refer to this
sub-selection as the ``detectable pairs''.
The number of pairs with very small separation ($\lesssim
2\arcsec$) is quite small ($\sim 0.4$ pairs/sim). It grows rapidly at
larger separations, stabilizing at $\gtrsim 8\arcsec$ with a value of
1.7 pairs/sim/arcsec (i.e., \textit{every arcsec of separation adds
  1.7 pairs}).

We consider the following possibilities for the
detection of a pair of input sources:
\begin{enumerate}
\item detected as two distinct sources;
\item detected as one single source.
\end{enumerate}

\subsection{Pairs detected as two distinct sources}

We have searched for source pairs where both components are
individually detected in the output catalogues of simulated sources
(``double-detections'').  

The number of double-detection pairs, when seen as a function of the
pair flux ratio (Fig.~\ref{fig:pairsep}, left panel, dashed
histogram), roughly follows the number of input pairs, though it stays
lower by $\sim 1.5$ orders of magnitude and has a sharper drop (a
factor 10 vs.\ a factor 2 for the double-detections and input pairs,
respectively, at flux ratios $<10$). Thus, double detections are
more probable when the flux ratio is small.

While the number of detectable input pairs rises with the separation from 0 to
$8\arcsec$ and remains constant afterwards, the number of
double-detections is consistent with 0 up to $11\arcsec$ and only
starts to rise afterwards (Fig.~\ref{fig:pairsep}, right panel). This
means that input pairs are not recognised as such by the detection
process if the separation is $\lesssim 11\arcsec$.

\subsection{Pairs detected as a single source}

Input pairs detected as a single output source are what is usually
referred to as confused sources. We have searched for them by
requiring: i) one output source within $6\arcsec$ from the pair
centroid; ii) no other output sources within $6\arcsec$ from the
positions of the two input sources. The pair centroid is defined as
the weighted mean of the pair coordinates, using the count rates as
the weights. The number of single-detection pairs is also shown in
Fig.~\ref{fig:pairsep} (both panels, dotted histograms). By
considering only pairs with separation $\le 15$ arcsec,
their average number is 13 pairs/simulation. When comparing with
the XMM-CDFS (see Sect.~\ref{xmmchandra_xcorr}), this number should be
taken as a lower limit, since the simulations do not include
clustering; in Sect.~\ref{xmmchandra_xcorr}, the number of \xmm\
sources with two or more \chandra\ counterparts is estimated to be 21.

Almost all pairs with a large flux ratio ($\gtrsim 10$) are detected
as a single source (Fig.~\ref{fig:pairsep}, left panel), probably
because the fainter source cannot be distinguished from a fluctuation
in the wings of the PSF of the brighter source. Conversely, at
small flux ratios the fraction of single-detections drops, and
$\lesssim 1/4$ of the input pairs with flux ratio $\le 2.5$ are
single-detected. The reason is probably that it is easier to have a
double-detection at small flux ratios, where the cores of the two PSFs
can be better recognised. Also, part of the explanation might involve
the inclusion of very faint sources in the input list.

Considering the detectable pairs, the number of single-detections as
function of the pair separation (Fig.~\ref{fig:pairsep}, right panel)
peaks at $\sim 9$ arcsec, and it always stays lower than the number of
input pairs, probably because our selection of ``detectable pairs'' is
conservative and allows for pairs where one of the components is
just below the local flux limit. At separations $\gtrsim 9\arcsec$ the
number of single-detections starts to decline, and matches the number
of double-detections at $14\arcsec$. The total number of
single-detections with separation $\le 20\arcsec$ is 14, which may
taken as an estimate of the expected number of confused sources in the
XMM-CDFS.

\smallskip

Summarizing, sources closer than $11\arcsec$ cannot be individually
detected. Sources separated by larger angles \textit{may} be
individually detected, the odds for it increasing with the separation,
though the PSFs may still be blended. Confusion is no longer an issue
at angles larger than $\sim 16\arcsec$.

\section{\chandra\ and optical identification and redshifts}
\label{sec:optical}

In the catalogue tables we include the most probable association
between our catalogue and the \chandra\ 4~Ms and ECDFS ones,
determined with the procedure explained in Sect.~\ref{sec:pineau}
(blends and other possible associations are reported in the NOTES column).

For the optical identification and for the redshift determination, we
rely mainly on the work done by X11, and report the redshift of the
optical counterparts from this catalogue. To search for more redshifts
and for updates, we checked spectroscopic redshifts in
\citet{treister09zcdfs}, \citet{silverman2010}, \citet{cooper2011},
and \citet{kurk2012}; and photometric redshifts in
\citet{cardamone2010} and \citet{rafferty2011}.

\subsection{Theory} \label{sec:pineau}

We have cross-correlated the XMM-CDFS 2--10 keV catalogue with the
\chandra\ 4~Ms (X11) and ECDFS \citep{lehmer05} catalogues. We
have used the likelihood-ratio estimator with a simple prior as
described in \citet{pineau2011}\footnote{This method has been
  implemented in a plugin to Aladin, and is available from {\texttt
    http://saada.u-strasbg.fr/docs/fxp/plugin/ .}}. Briefly, this method
works in two steps. First, for each XMM-\chandra\ source association we define
the likelihood-ratio between the probability of the association being
real vs.\ it happening by chance:
\begin{equation}
LR(r)={e^{-r^2/2}\over 2\lambda}
\end{equation}
(Eq. 9 in \citealt{pineau2011}) where both $r$ and $\lambda$ are
adimensional. $r$ is the angular distance between both sources divided
by the combined (in quadrature) positional error (i.e.
$r=d/\sqrt{\sigma_X^2+\sigma_C^2}$ where $d$ is the angular distance
and the $\sigma$ are the positional errors), and $\lambda$ is the
angular density of \chandra\ sources with flux higher or equal to that
of the \chandra\ member of the association (using the number counts in
\citealt{luo08}\footnote{The 2--8~keV number counts were used if the
  \chandra\ source was detected in that band, otherwise the 0.5--2~keV
  number counts were used. If the source was only an upper limit in
  both bands, the upper limit in the 2--8~keV band was assumed to be
  the flux of the source and the number counts in that band were
  used.}) multiplied by the combined square positional error (i.e.,
$\lambda=(\sigma_X^2+\sigma_C^2)\times N(>S)$).  Second, to obtain the
probability of real association we use Eq. 11 in \citet{pineau2011}.
\begin{equation}\label{PHcp}
P(H_\mathrm{cp}\mid r)= \left( 1+ \left(  {{P(H_\mathrm{cp})\over
          1-P(H_\mathrm{cp})}LR(r)} \right) ^{-1}  \right) ^{-1}
\end{equation}
where $P(H_\mathrm{cp})$ is the (unknown) prior probability
of a \chandra\ source to be the counterpart of an \xmm\ source. Given
the different characteristics and depths of the X11 and ECDFS
catalogues, we assume $P(H_\mathrm{cp})$ with two constant values: one
if the source comes from X11, another if it comes from the ECDFS.

The (unknown) number of real \chandra\ counterparts is the sum of the
probabilities of association, over all associations:
\begin{equation}\label{Nreal}
N_\mathrm{real}=\sum P(H_\mathrm{cp}\mid r)
\end{equation}
which can be approximated by using the total number of candidate
\chandra\ counterparts $N$:
\begin{equation}\label{Pcpspur}
  P(H_\mathrm{cp})\sim {N_\mathrm{real} \over N}
\end{equation}

We finally arrive at
\begin{equation}
N_\mathrm{real}=\sum \left( {1+ \left( {N_\mathrm{real}\over N-N_\mathrm{real}}LR(r)
    \right) ^{-1} } \right) ^{-1}
\end{equation}
\noindent combining Eqs. \ref{PHcp}, \ref{Pcpspur} and \ref{Nreal}
above. This equation can be resolved iteratively for $N_\mathrm{real}$
(starting with $N_\mathrm{real}=N/2$ for instance) to get $P(H_\mathrm{cp})$ and
hence $P(H_\mathrm{cp}\mid r)$ for each association. %

\subsection{Cross-correlation of \xmm\ and \chandra\ sources}
\label{xmmchandra_xcorr}

We have used the method outlined above to cross-correlate
independently our \xmm\ 2--10 keV and 5--10 keV catalogue with the
\chandra\ 4~Ms (using Tables 3 and 6 in X11) and the ECDFS (using
Tables 2 and 6 in \citealt{lehmer05}) catalogues.
                     
Then the cross-correlations were merged; when a source was present in
both X11 and the ECDFS, we only considered the X11 position.  The
cross-correlation was performed considering all XMM-\chandra\
associations with $r\leq5$.  For each \xmm\ source with more than one
possible \chandra\ counterpart, we dropped the association whose
probability was $\le 3\%$ of that of the most probable one. We flagged
initially as ``Good'' the associations with probabilities of
association $P(H_\mathrm{cp}\mid r)\geq 90$~per cent.

The restrictions in the cross-correlation catalogue above were chosen
to provide a manageable size, while likely keeping all possible
associations of interest.  This restricted catalogue was extensively
inspected by eye (taking also into account the \xmm\  and
\chandra\ fluxes) to get our definitive cross-correlation catalogue.
The visual inspection was carried out by superimposing contours of
signal/noise ratio (SNR) on top of the \xmm\ and \chandra\ images, to help
locate the peaks of emission.  The SNR contours were also the basis for the
annotations on the source blending present in the NOTES column of the
catalogue.

As a result of this inspection, 10 \chandra\ counterparts to 2--10 keV
\xmm\ sources had their ``Good'' flag revoked (2 counterparts to 5--10
keV sources), because either the corresponding \xmm\ source was
spurious, the \chandra\ member of the association had only an upper
limit to its 2--8~keV flux or the counterpart was from the ECDFS and
there was a better X11 counterpart. Conversely, 13 (5)
counterparts were considered ``Good'' despite having probabilities of
association below 90~per cent: this was changed because the position
of the \xmm\ source from our source detection did not coincide with
the peak emission on the corresponding \xmm\ image, or blending with
nearby \xmm\ or \chandra\ sources. In all cases we have kept the
positional information from {\tt PWXDdetect}.

Finally, we obtained 336 ``Good'' XMM-\chandra\ associations for \xmm\
sources in the main 2--10 keV catalogue (137 in the main 5--10 keV
catalogue), corresponding to 339 (137) unique \xmm\ sources. Out of
the 315 (130) \xmm\ sources with one or more \chandra\ counterparts,
295 (124) had a single counterpart, 19 (5) had two and 1 (1) had three.
24 (7) \xmm\ sources did not have any \chandra\ counterpart.

\subsection{New sources}
\label{sec:newsources}

To identify candidate real XMM-CDFS sources not previously detected by
X11 or \citet{lehmer05} with \chandra\ (as opposed to candidate
spurious \xmm\ detections) we adopt the following baseline criterion:
a source is considered real if there is at least a signal/noise ratio
(SNR) contour where SNR$>3$ in the band where it was detected.

The SNR was computed, in the Bayesian framework, by treating a source's
count rate $c$ as a random variable constrained to be $c\geq0$. Since
the background images have more than 10 counts over the area where the
exposure map ${t_{kl}>10^6}$~s, we use a Gaussian approximation to the
Poisson statistics, so that for a pixel $(i,j)$ the probability that a
source is in that pixel with count rate $c$ is
\begin{equation}
P_{ij}(c)\propto\prod_{kl}\frac{\exp{-(ct_{kl}w_{kl}-T_{kl}+B_{kl})^2 \over
2(T_{kl}+B_{kl})}}{\sqrt{2\pi(T_{kl}+B_{kl})}}
\end{equation}
where the $(k,l)$ indices run over a circular aperture of
radius 16$\arcsec$ around pixel $(i,j)$, $w_{kl}$ is the value of a PSF
centered in $(i,j)$ at pixel $(k,l)$ (approximated here by a gaussian
of FWHM$=8.5\arcsec$), and $T_{kl}$ and $B_{kl}$ are the total and
background image values. This can be rewritten as
\begin{equation}
P_{ij}(c)={ \exp{-(c-c_{0,ij})^2 \over 2\sigma{ij}^2}\over
\sigma_{ij}\sqrt{\pi \over 2}\ \mathrm{erfc}
\left({c_{0,ij}\over\sigma_{ij}\sqrt(2)}\right)}\,\,\,c\geq0
\end{equation}
where
$\mathrm{erfc}(x)=2\pi^{-1}\int_x^{+\infty}e^{-t^2}\de t$ is the complementary error function.
$P_{ij}(c)$ is normalized taking into account that $c\geq0$ and
$c_{0,ij}$ and $\sigma_{ij}$ are functions of $t_{kl}$, $w_{kl}$,
$T_{kl}$ and $B_{kl}$ around pixel $(i,j)$. Since $P_{ij}(c)$ is
symmetric around $c_{0,ij}$ for $c\geq0$, by analogy with the "pure"
Gaussian case, the probability of having detected a source at pixel
$(i,j)$ is then
\begin{equation}
\mathrm{SNR}_{ij}=\int\limits_0^{2c_{0,ij}} P_{ij}(c)\, \de c=
\frac{
  2-2\,\mathrm{erfc}(c_{0,ij}/\sigma_{ij}\sqrt(2))
}{
  2-\mathrm{erfc}(c_{0,ij}/\sigma_{ij}\sqrt(2))
}
\end{equation}
\noindent We have drawn the contours in terms of "Gaussian sigmas"
taking into account that, for example, SNR=3 corresponds to
$\mathrm{SNR}_{ij}=0.9973$.

Of the 24 (7) sources in the 2--10 (5--10) keV main catalogue without
a \chandra\ counterpart, a
$\mathrm{SNR}>3$ contour could be drawn on the 2--10 (5--10) keV images for 11
(2) of them, namely ID210 5, 85, 176, 189, 207, 224, 280, 348, 381,
402, 407, and ID510 1098 and 1150. Of this list, ID210 207 and ID510
1098 correspond to the same source. Two sources from the supplementary
2--10 keV catalogue (ID210 186, 410) also satisfy this criterion and
are thus considered real. The SNR$>3$ contour of source ID510 1150
includes a blend of ID210 304 (from the supplementary sample) and 306
(main), but it peaks around ID210 304; ID210 304 has no \chandra\
counterpart, while 306 does (source 443 in X11, whose position
coincides with a peak of 0.5--2 keV emission in \xmm). Thus we
consider ID210 304 as the counterpart to ID510 1150, and include 304
among the candidate real sources.

We also considered a source as real, if it has at least a $\mathrm{SNR}>2$
contour in the same band of its detection, plus a $\mathrm{SNR}>3$ contour in
the other band. This happens for one source: ID510 1149 (not detected
in the 2--10 keV band).

Thus the total number of candidate ``new'', \xmm-only sources, is 15,
divided as follows: 11 main 2--10 keV, 3 supplementary 2--10 keV, 3
main 5--10 keV (two sources were detected in both bands).  Conversely,
the number of candidate spurious sources is 13 in the main 2--10 keV
sample and 4 in the 5--10 keV. Thus, the spurious
fraction in the main 2--10 keV catalogue is $(24-11)/339\sim 3.8\%$.

\begin{figure*}[p]
  \centering
  $\begin{array}{ccccc}
  \includegraphics[height=.19\textwidth,bb=0 0 300 315,clip]{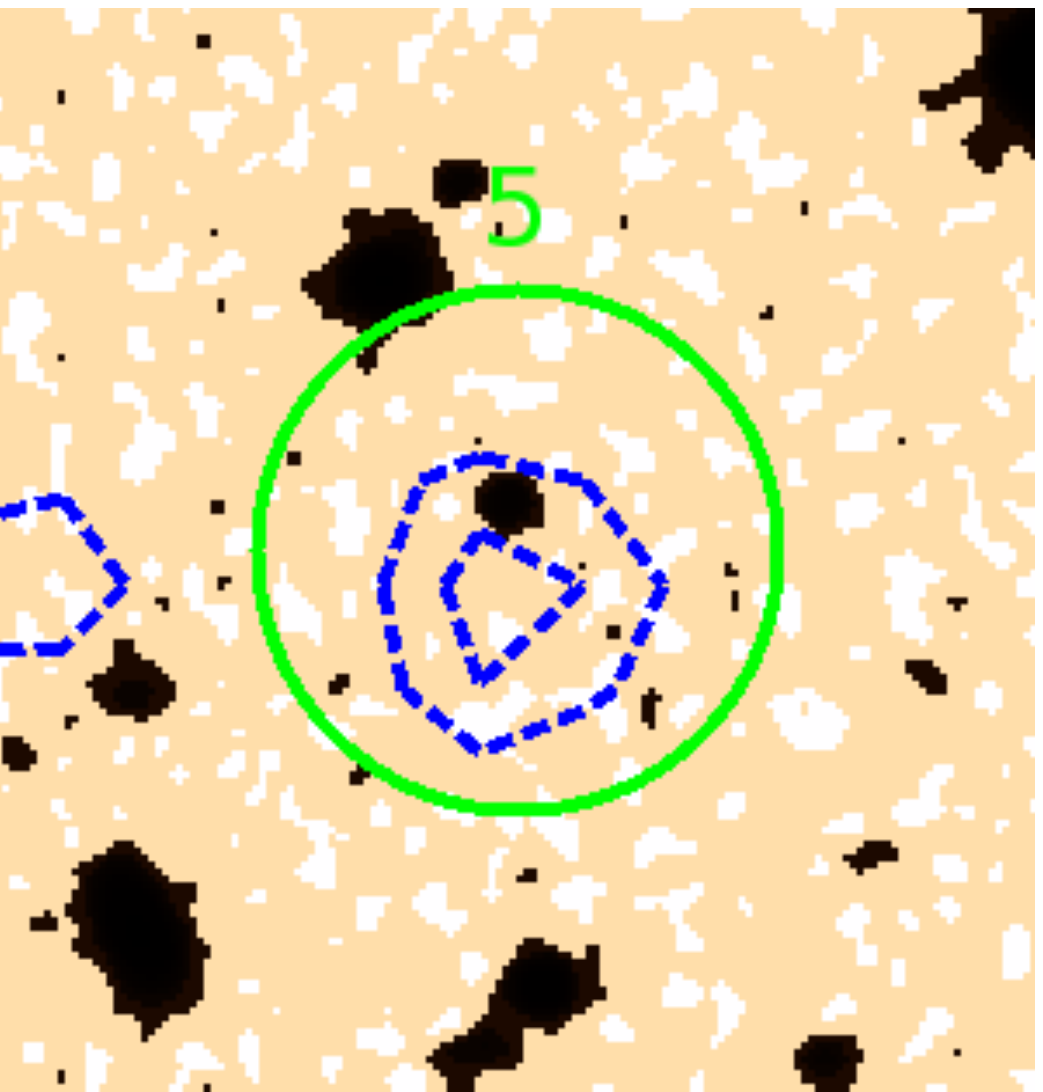}
  &\includegraphics[height=.19\textwidth,bb=0 0 300 315,clip]{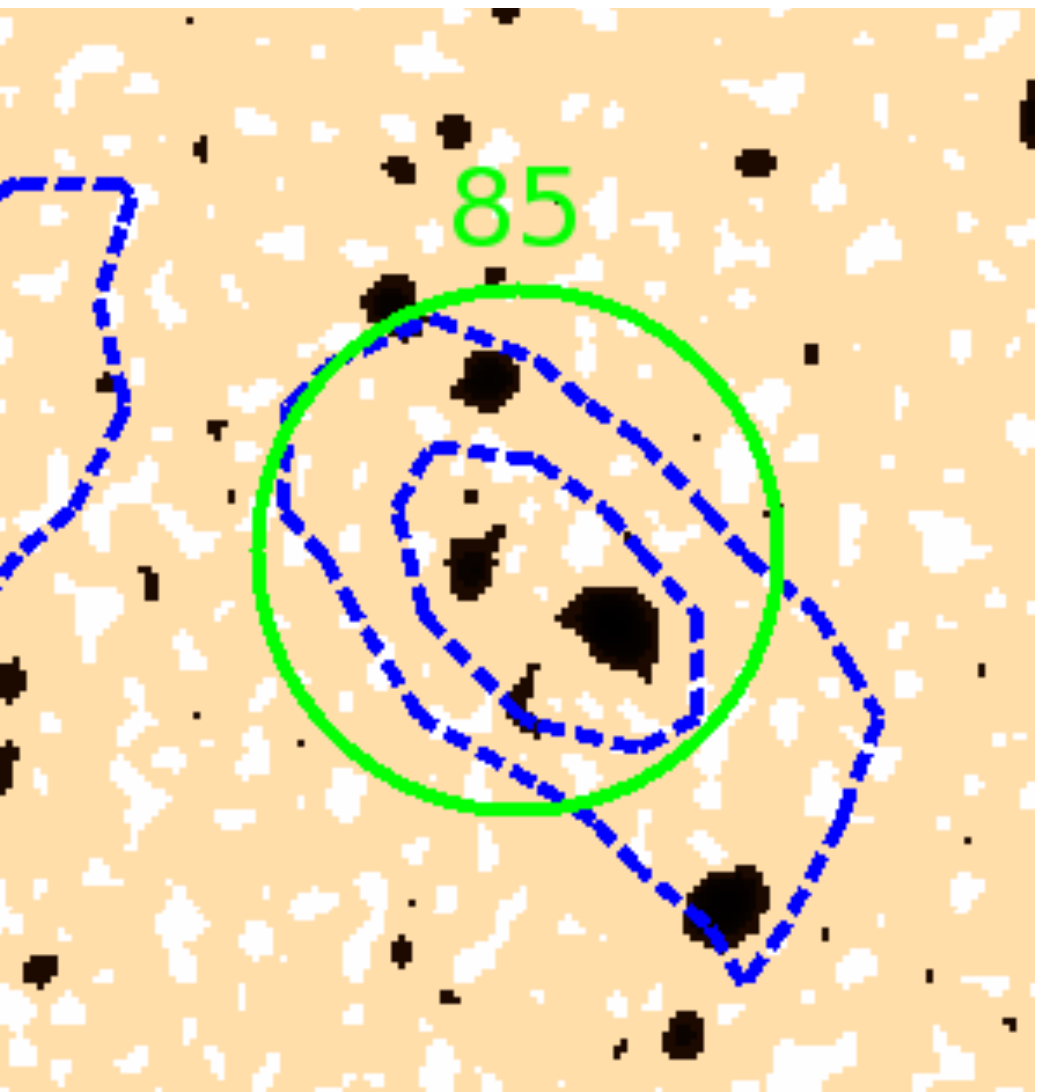}
  &\includegraphics[height=.19\textwidth,bb=0 0 300 315,clip]{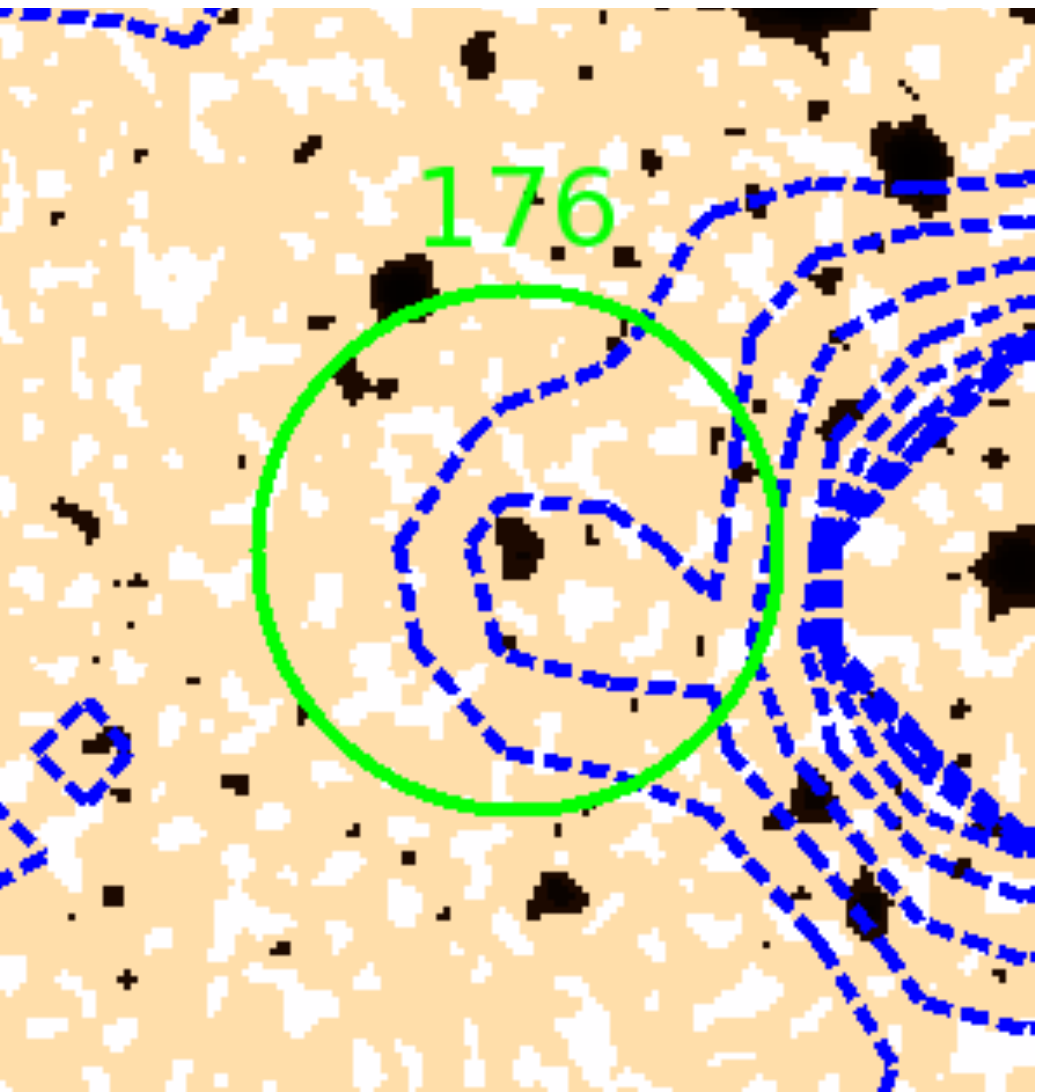}
  &\includegraphics[height=.19\textwidth,bb=0 0 300 315,clip]{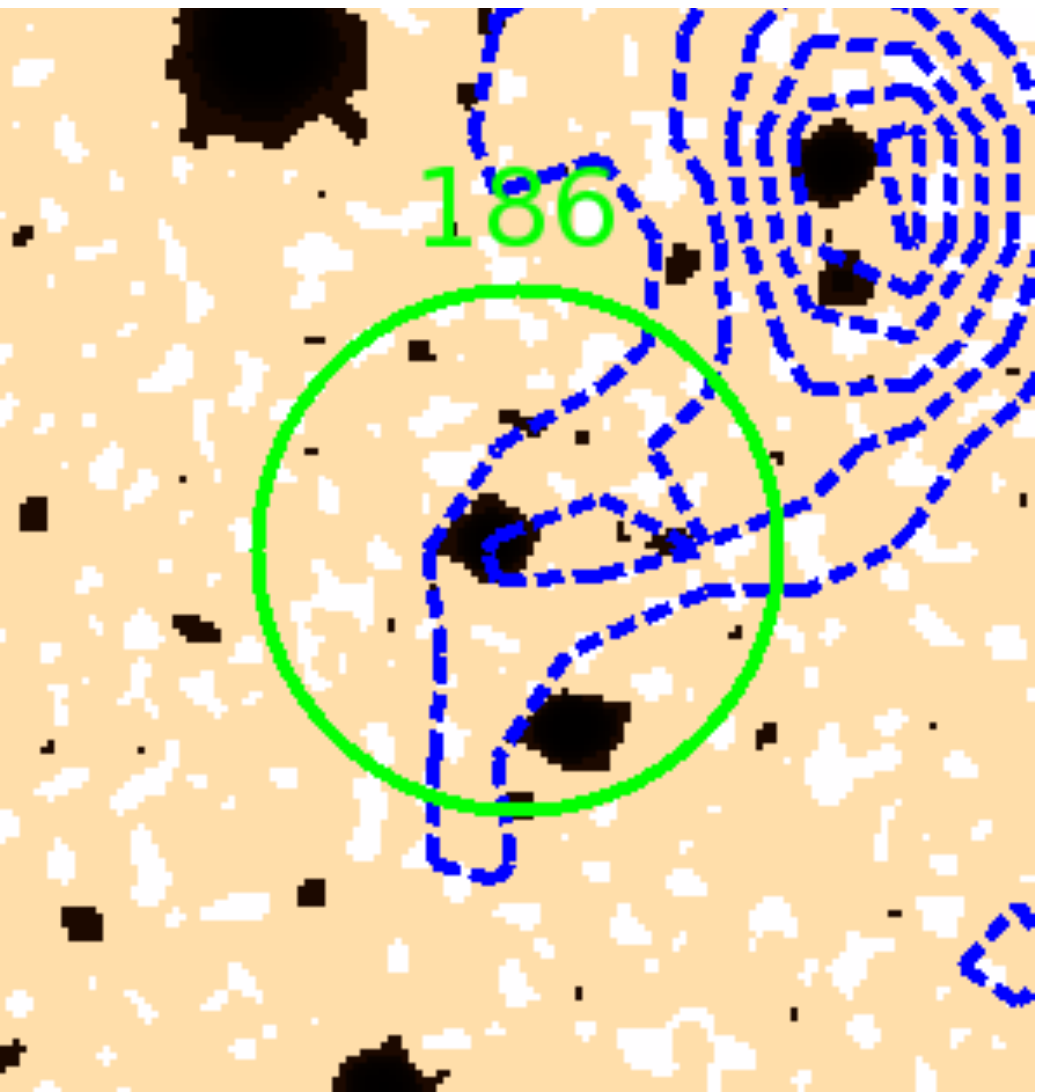}
  &\includegraphics[height=.19\textwidth,bb=0 0 300 315,clip]{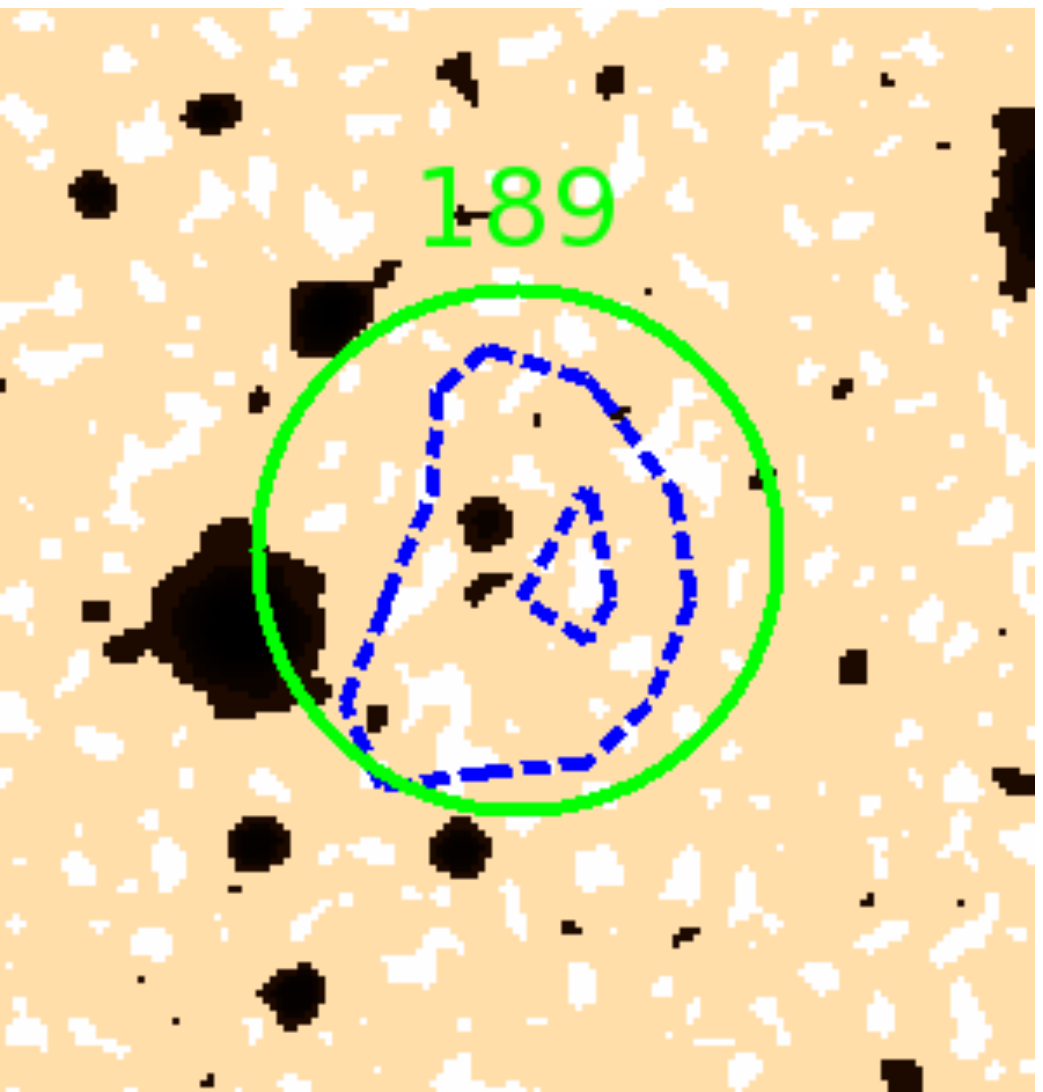} \\
  \includegraphics[height=.19\textwidth,bb=0 0 300 315,clip]{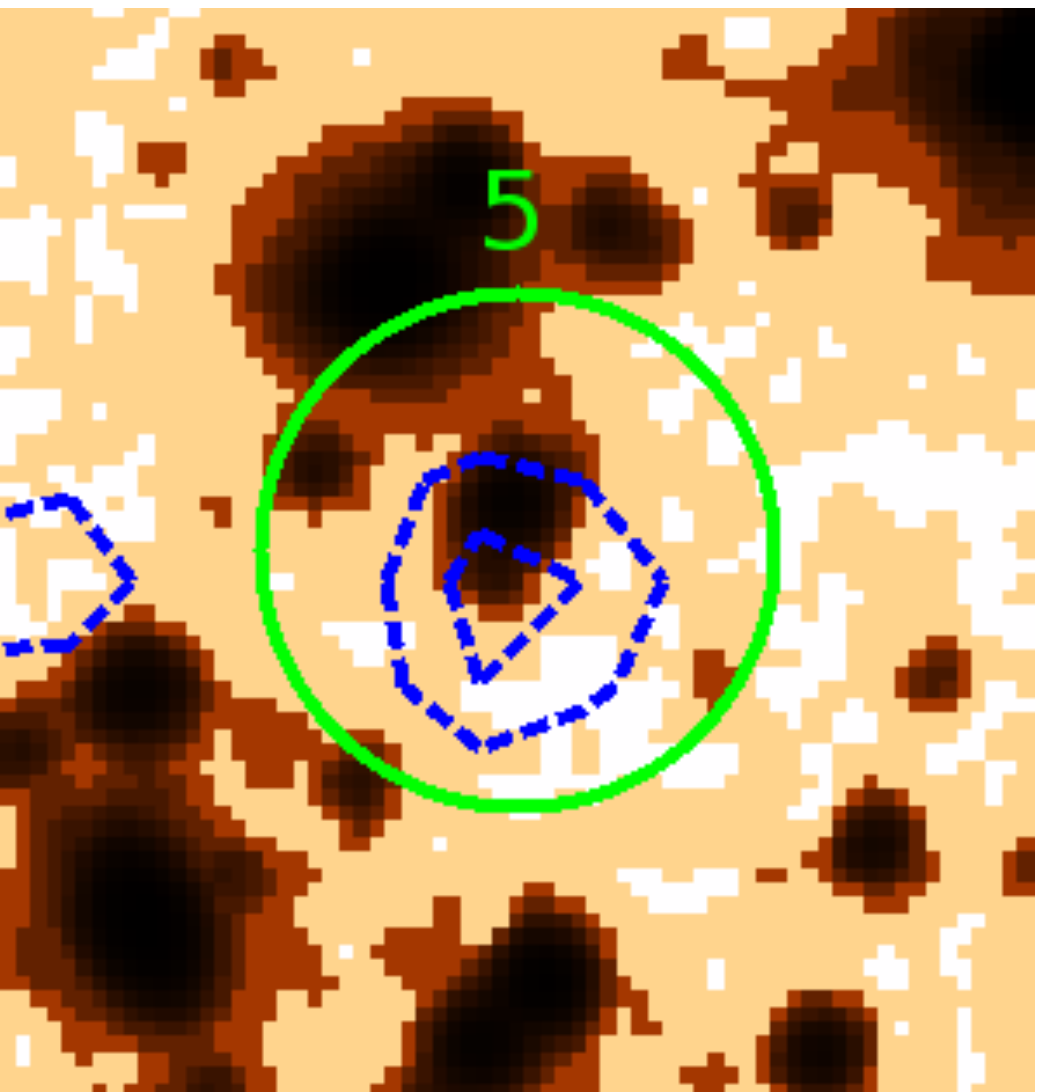}
  &\includegraphics[height=.19\textwidth,bb=0 0 300 315,clip]{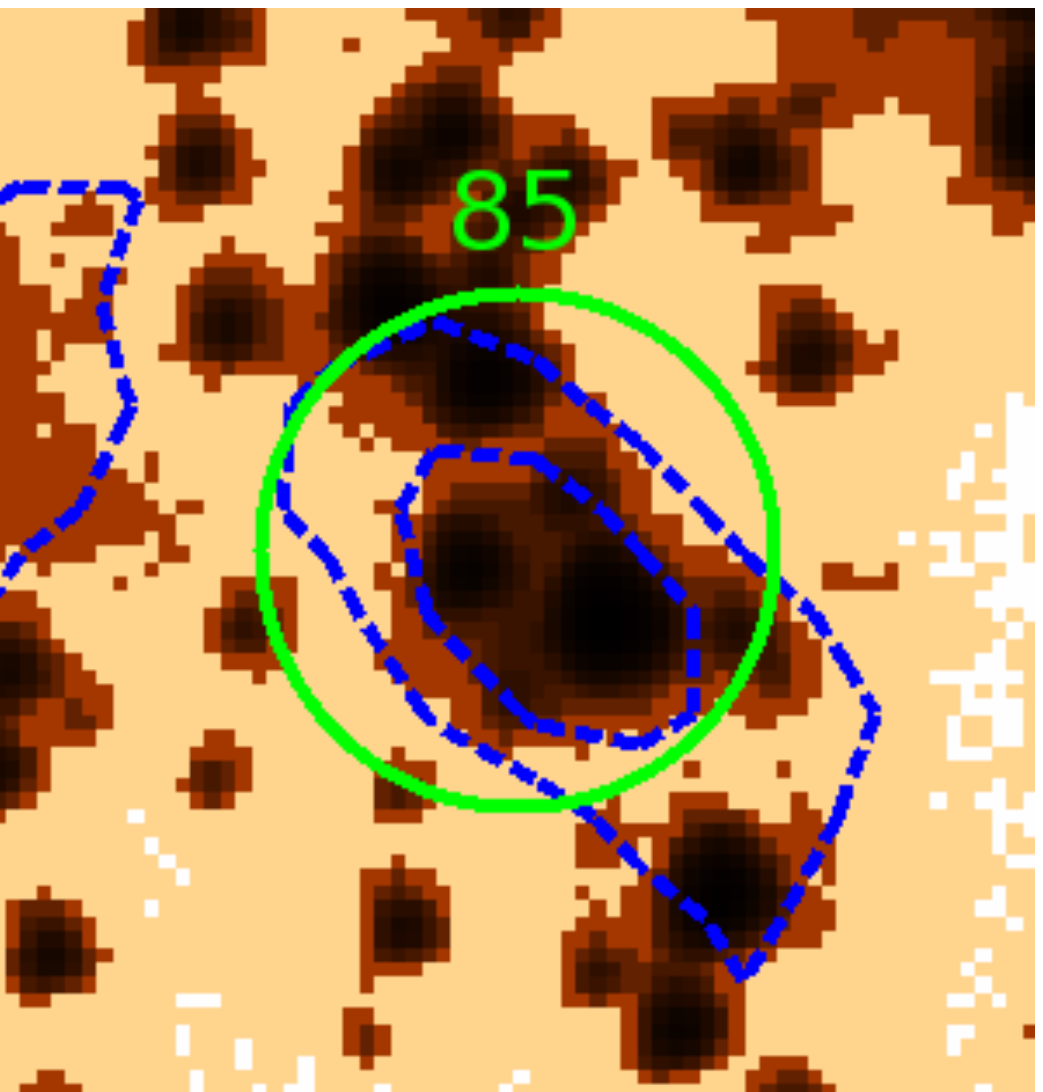}
  &\includegraphics[height=.19\textwidth,bb=0 0 300 315,clip]{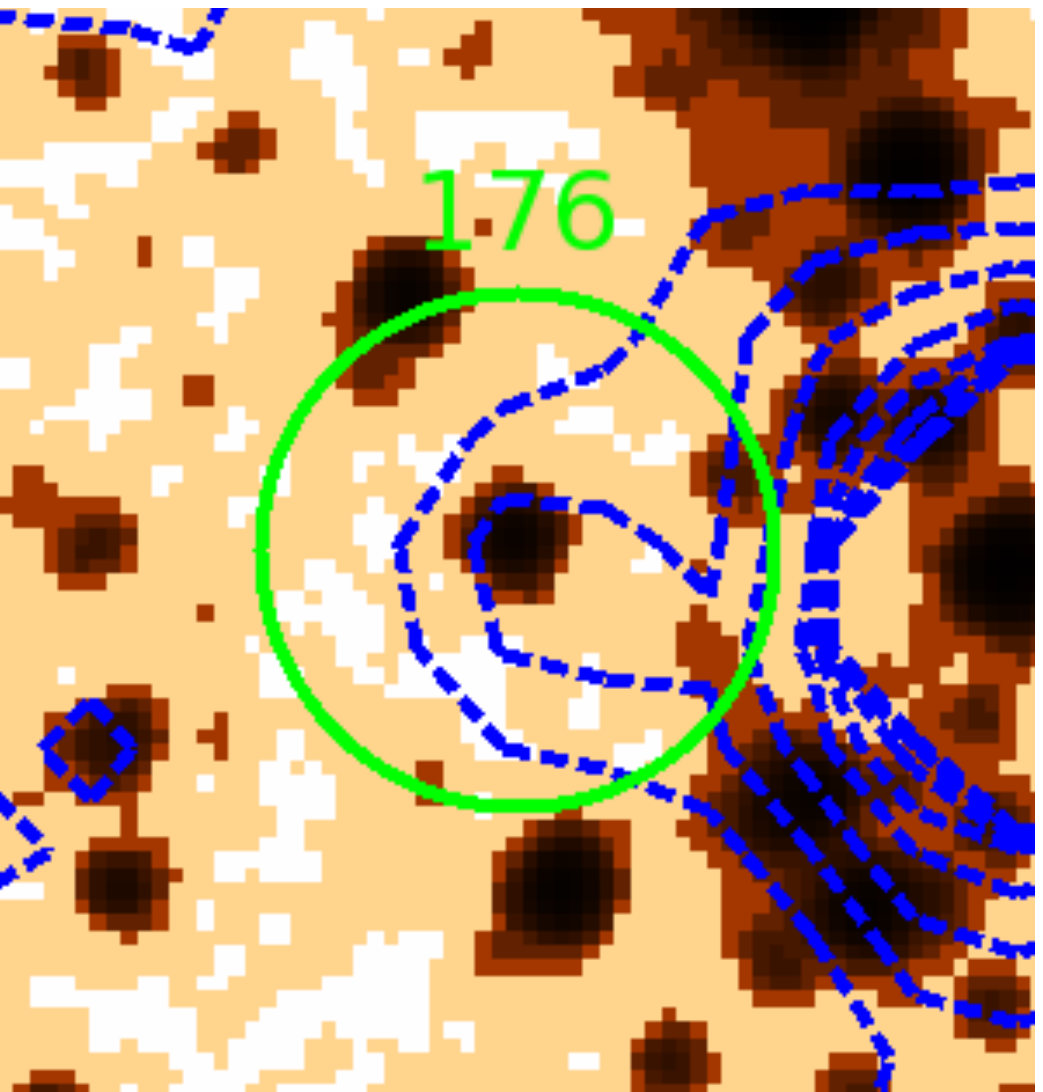}
  &\includegraphics[height=.19\textwidth,bb=0 0 300 315,clip]{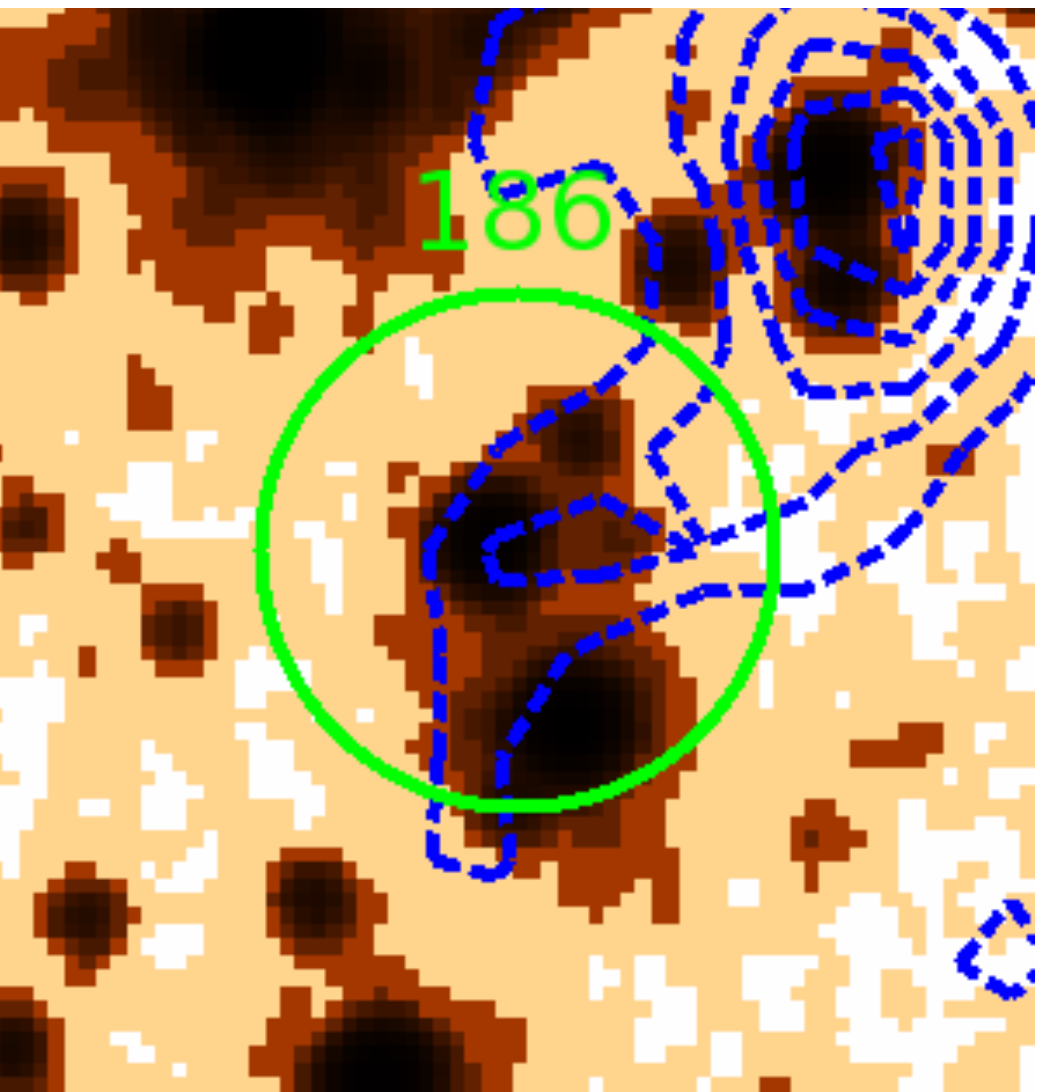}
  &\includegraphics[height=.19\textwidth,bb=0 0 300 315,clip]{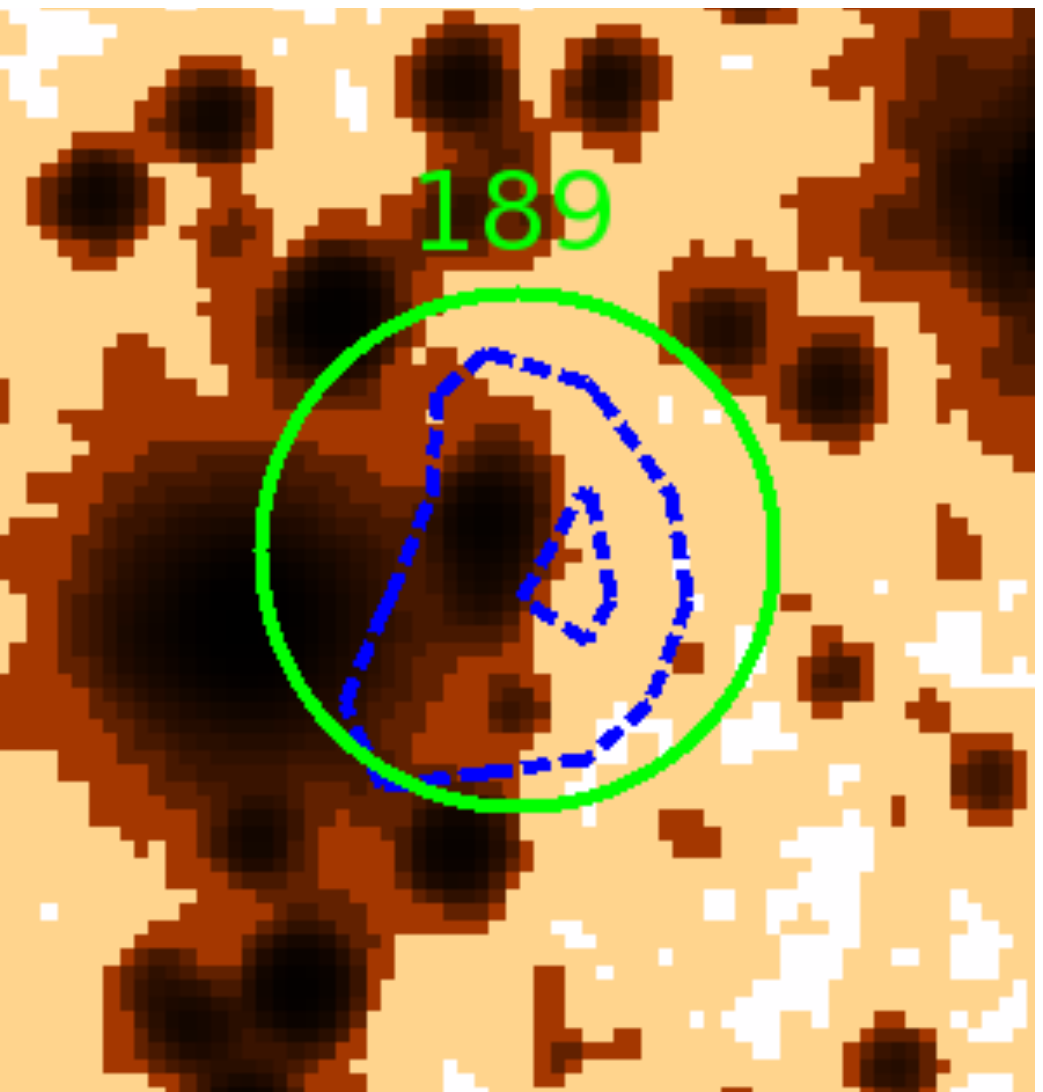} \\
  \includegraphics[height=.19\textwidth,bb=0 0 300 315,clip]{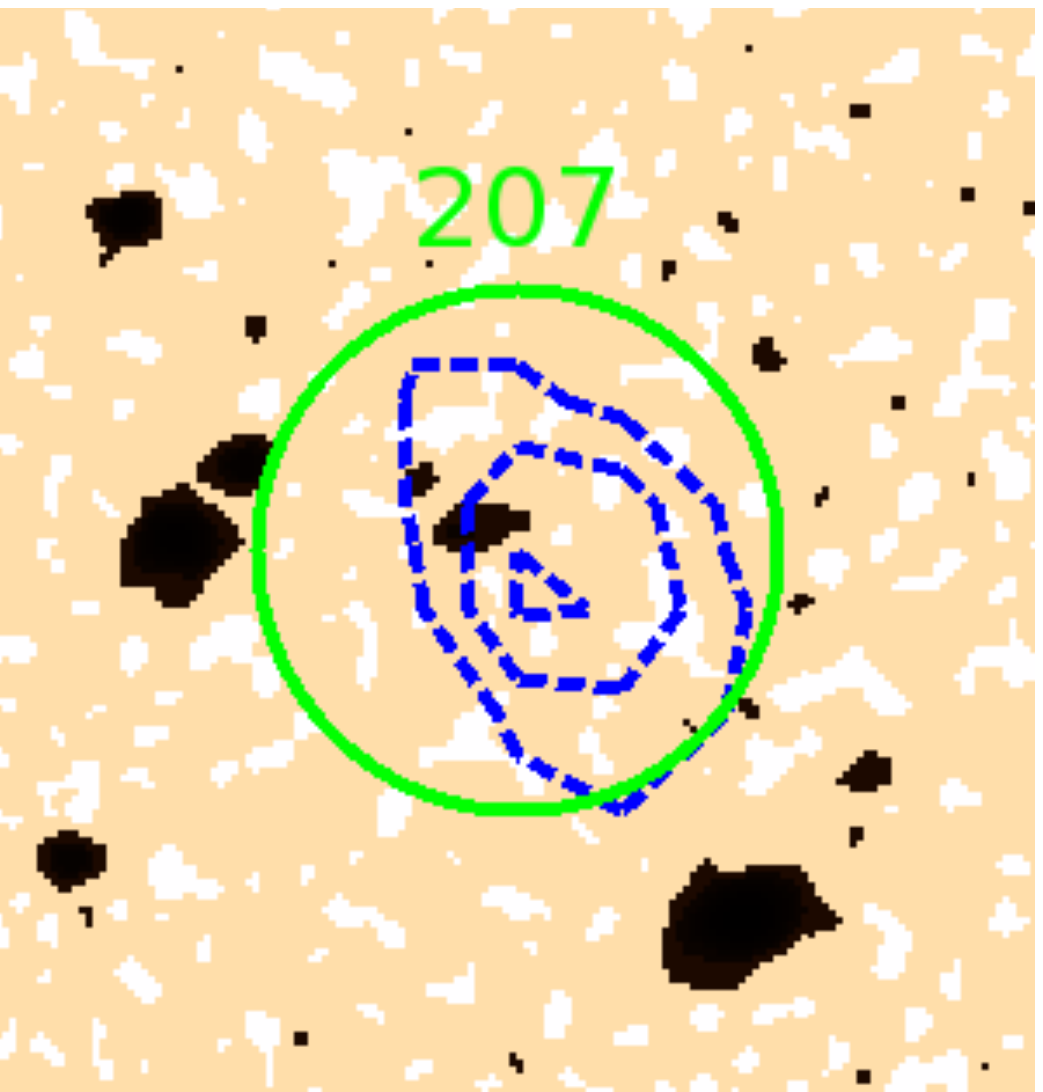}
  &\includegraphics[height=.19\textwidth,bb=0 0 300 315,clip]{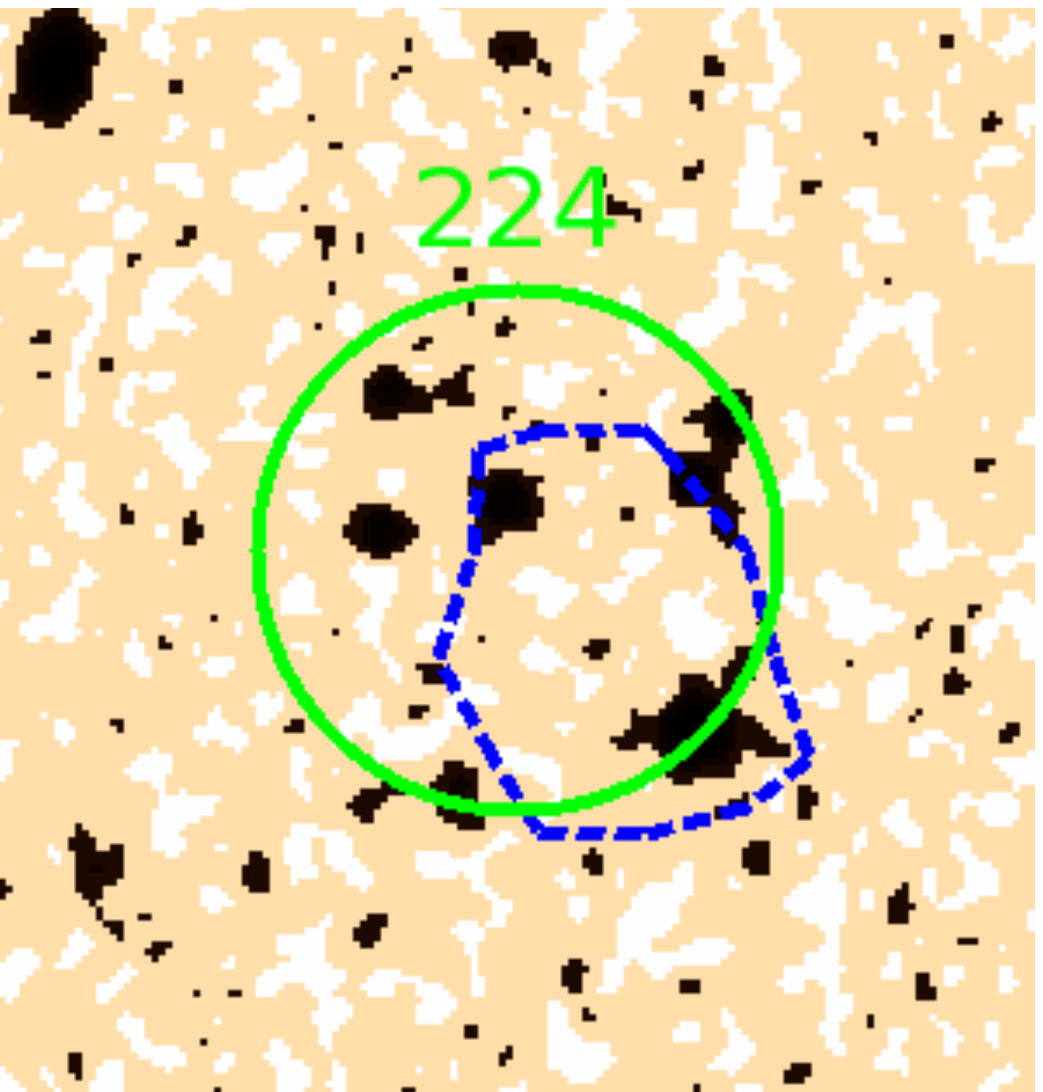}
  &\includegraphics[height=.19\textwidth,bb=0 0 300 315,clip]{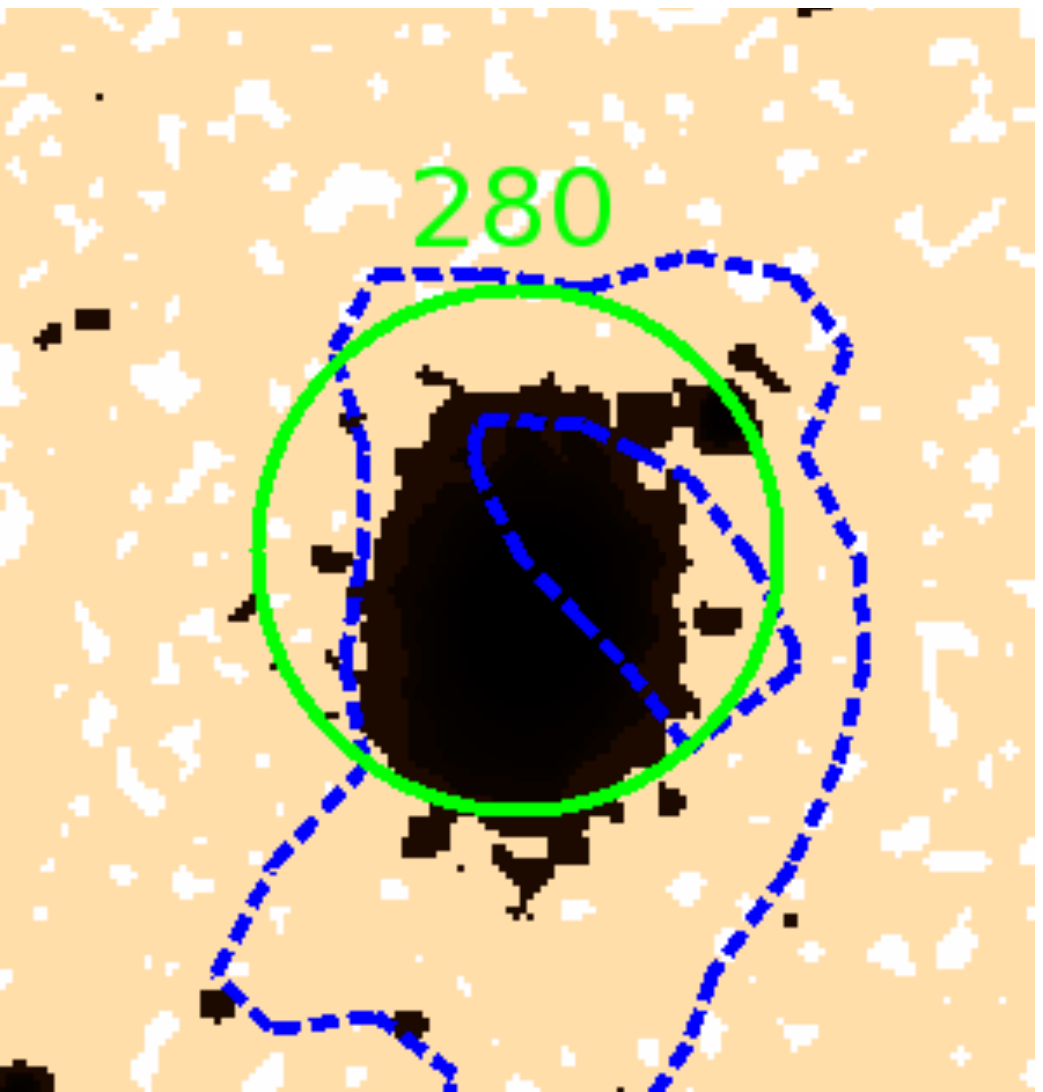}
  &\includegraphics[height=.19\textwidth,bb=0 0 300 315,clip]{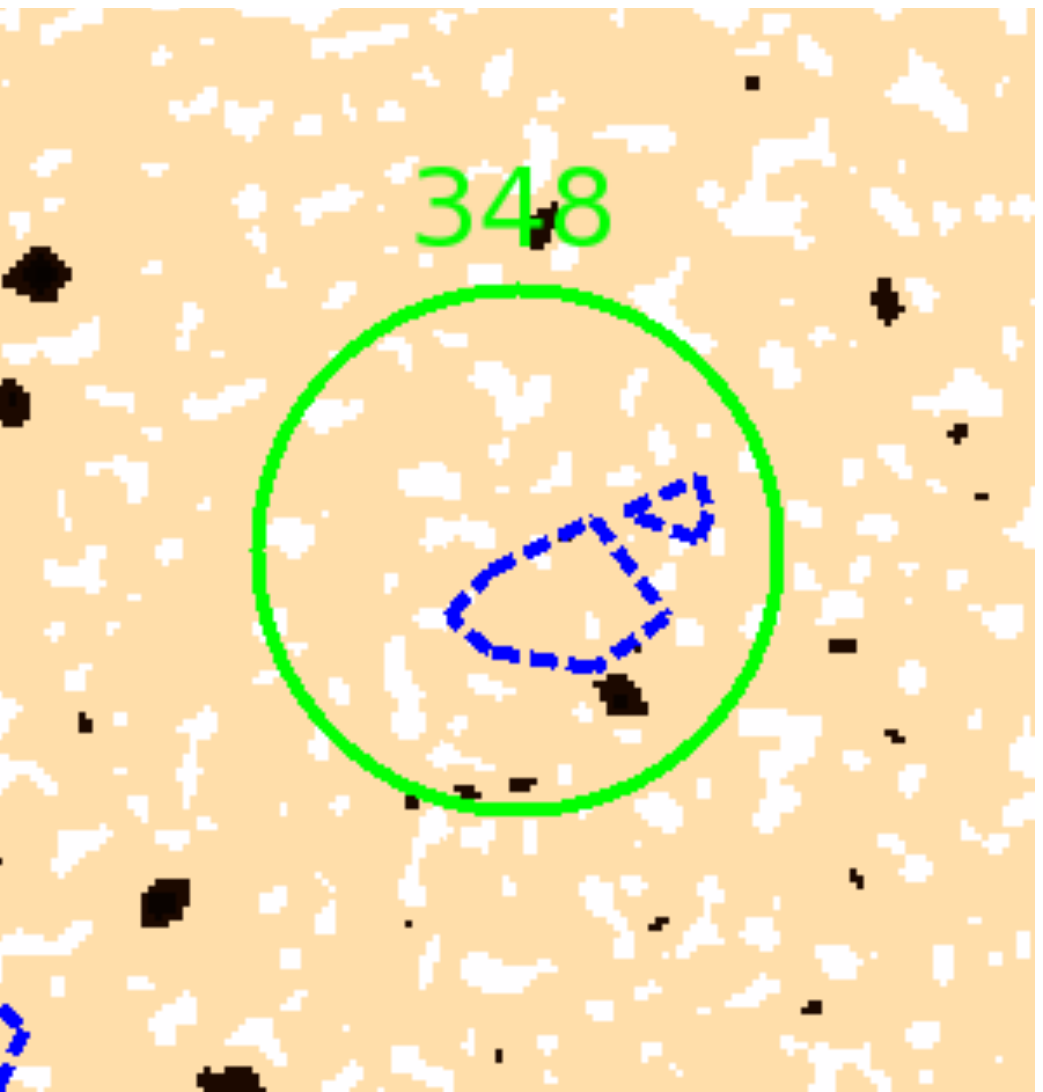}
  &\includegraphics[height=.19\textwidth,bb=0 0 300 315,clip]{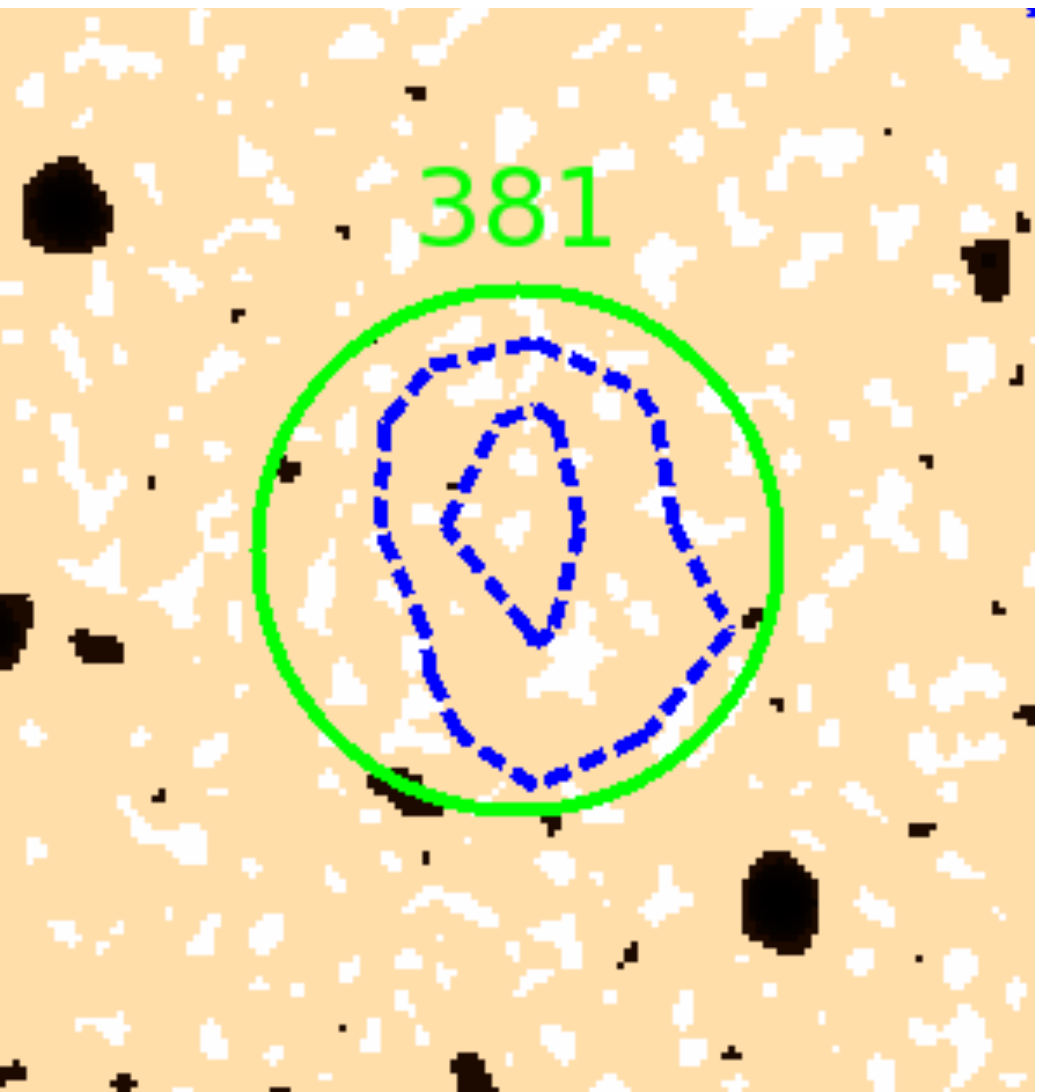} \\
  \includegraphics[height=.19\textwidth,bb=0 0 300 315,clip]{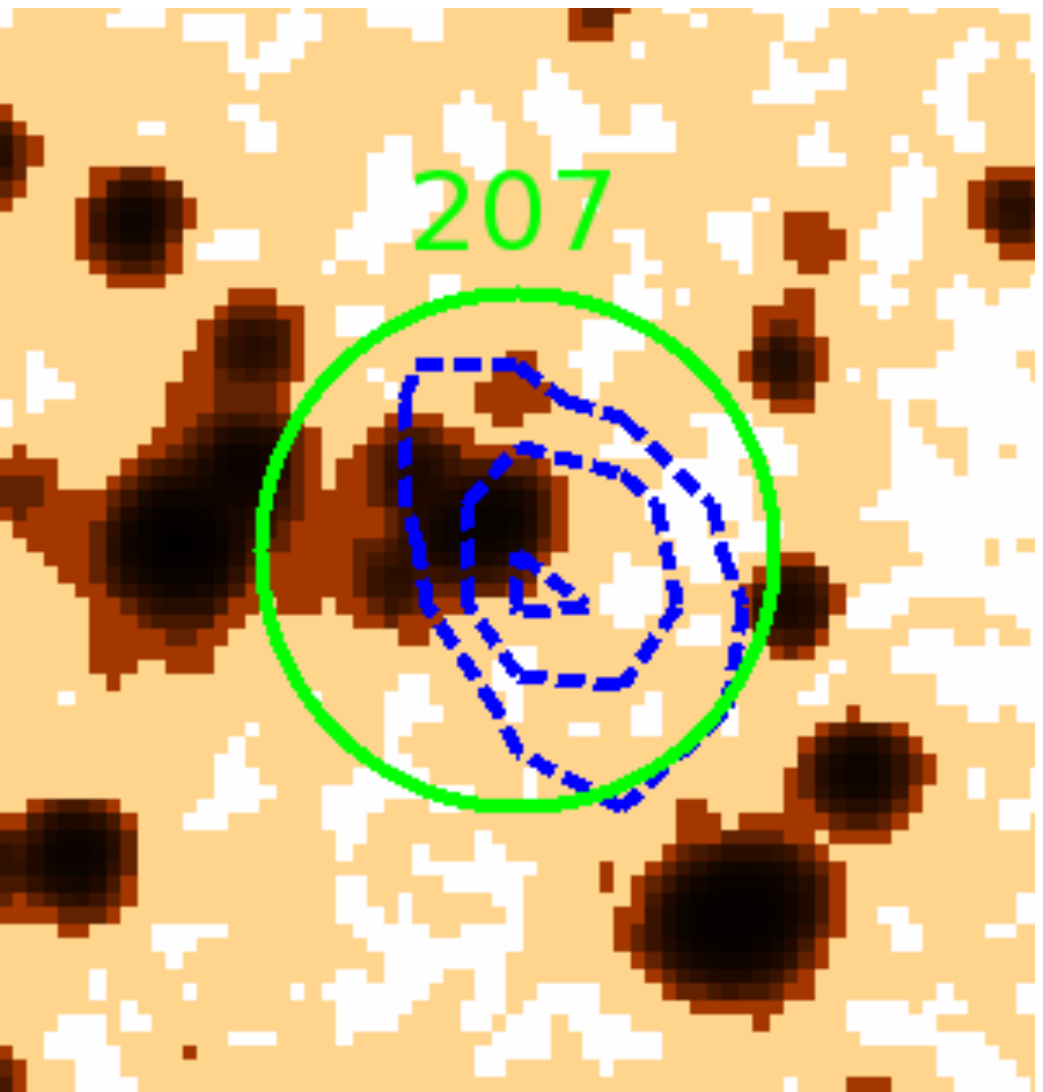}
  &\includegraphics[height=.19\textwidth,bb=0 0 300 315,clip]{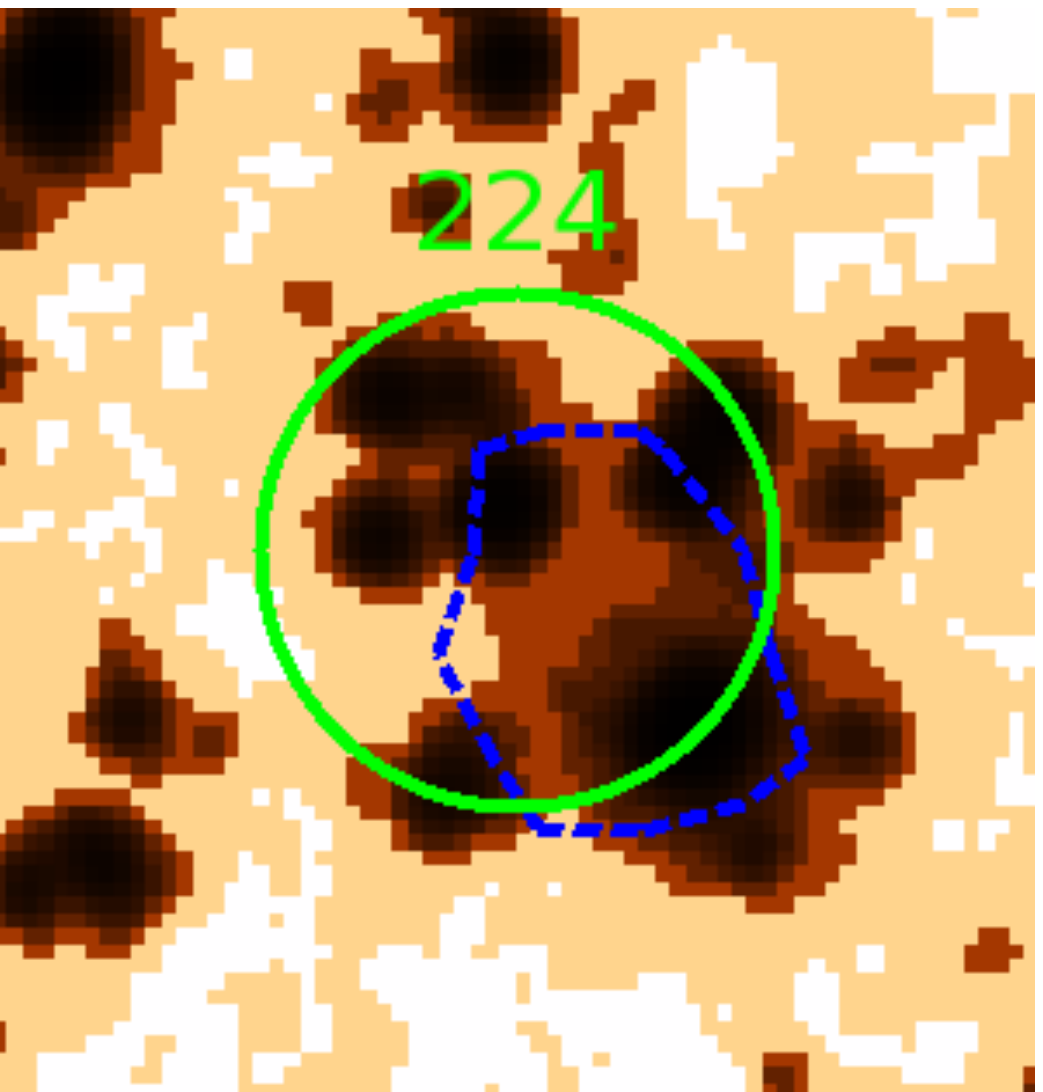}
  &\includegraphics[height=.19\textwidth,bb=0 0 300 315,clip]{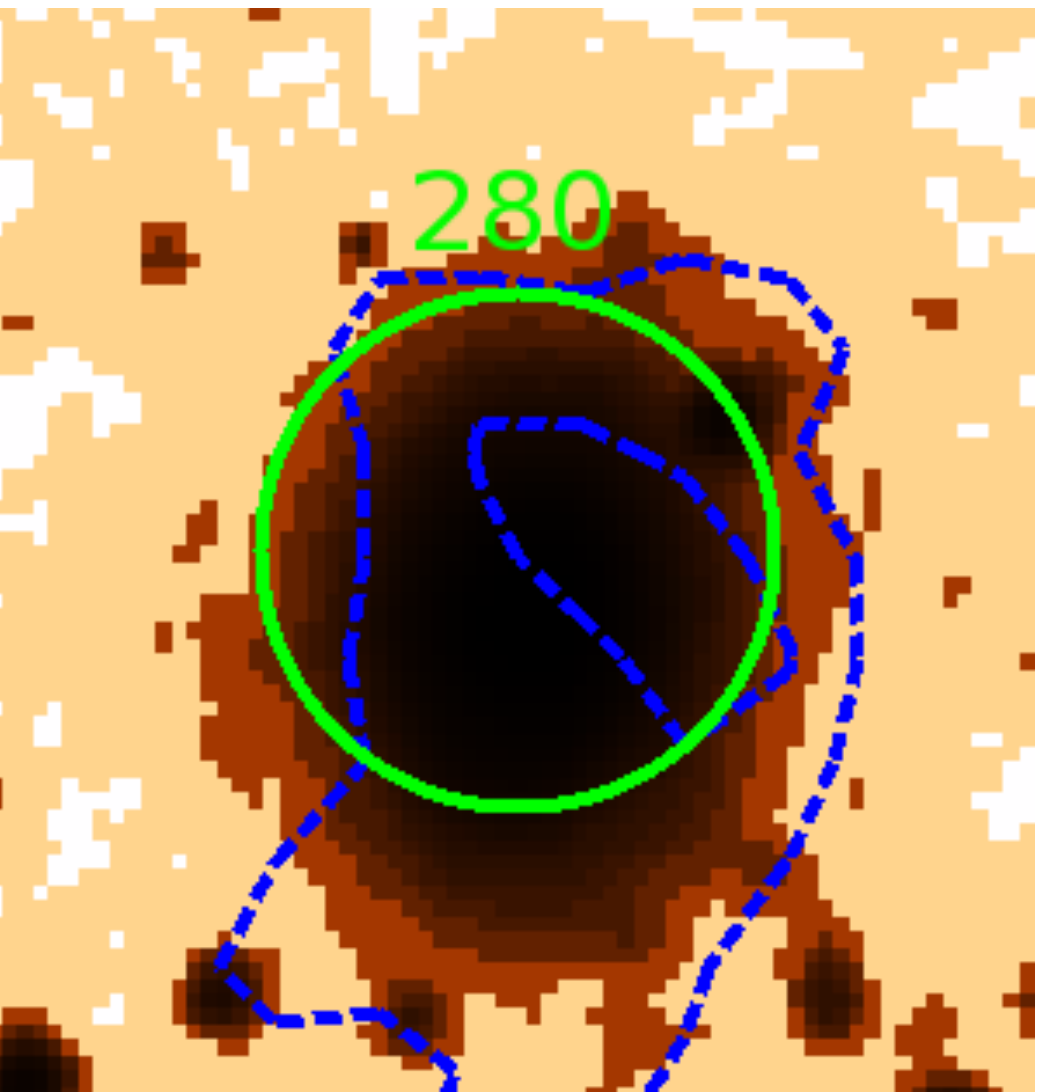}
  &\includegraphics[height=.19\textwidth,bb=0 0 300 315,clip]{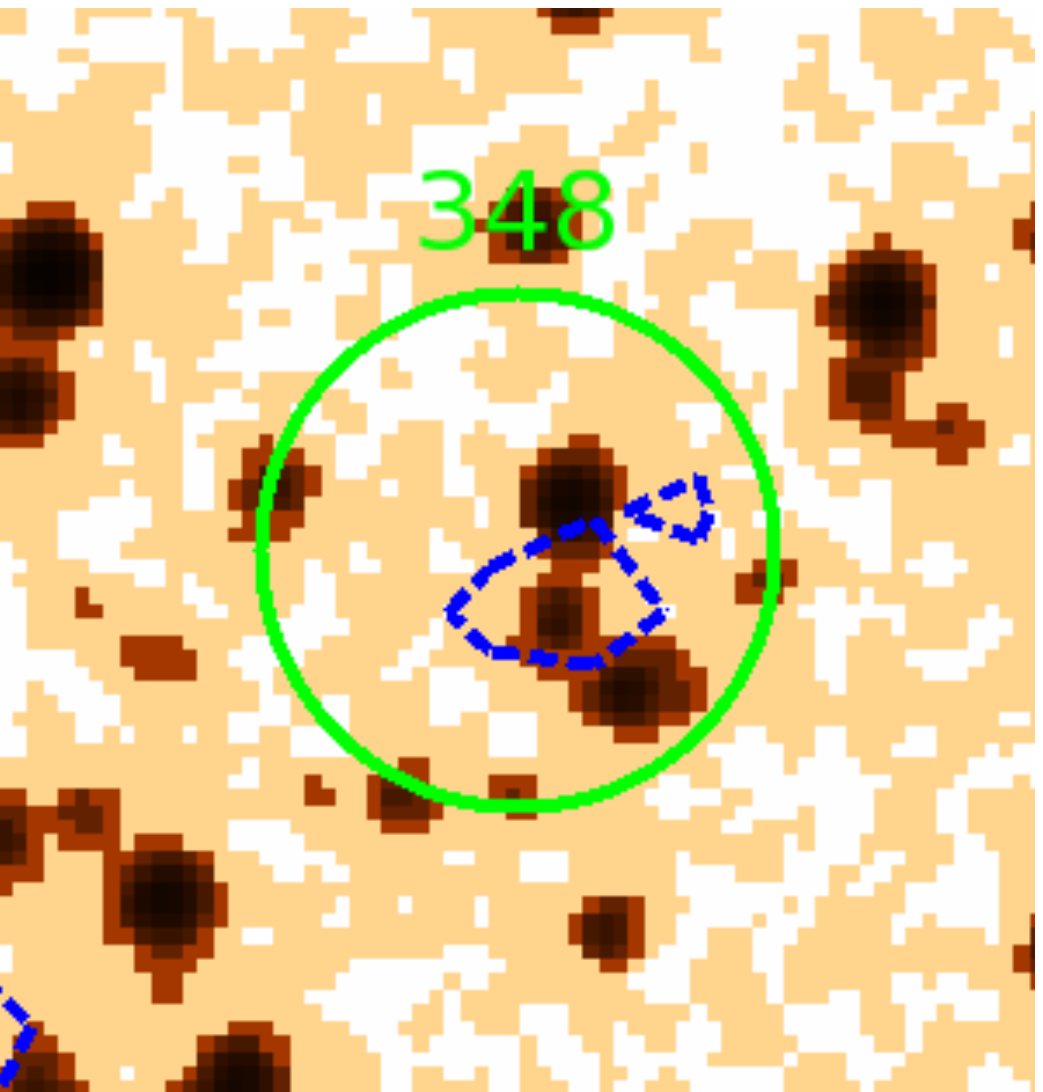}
  &\includegraphics[height=.19\textwidth,bb=0 0 300 315,clip]{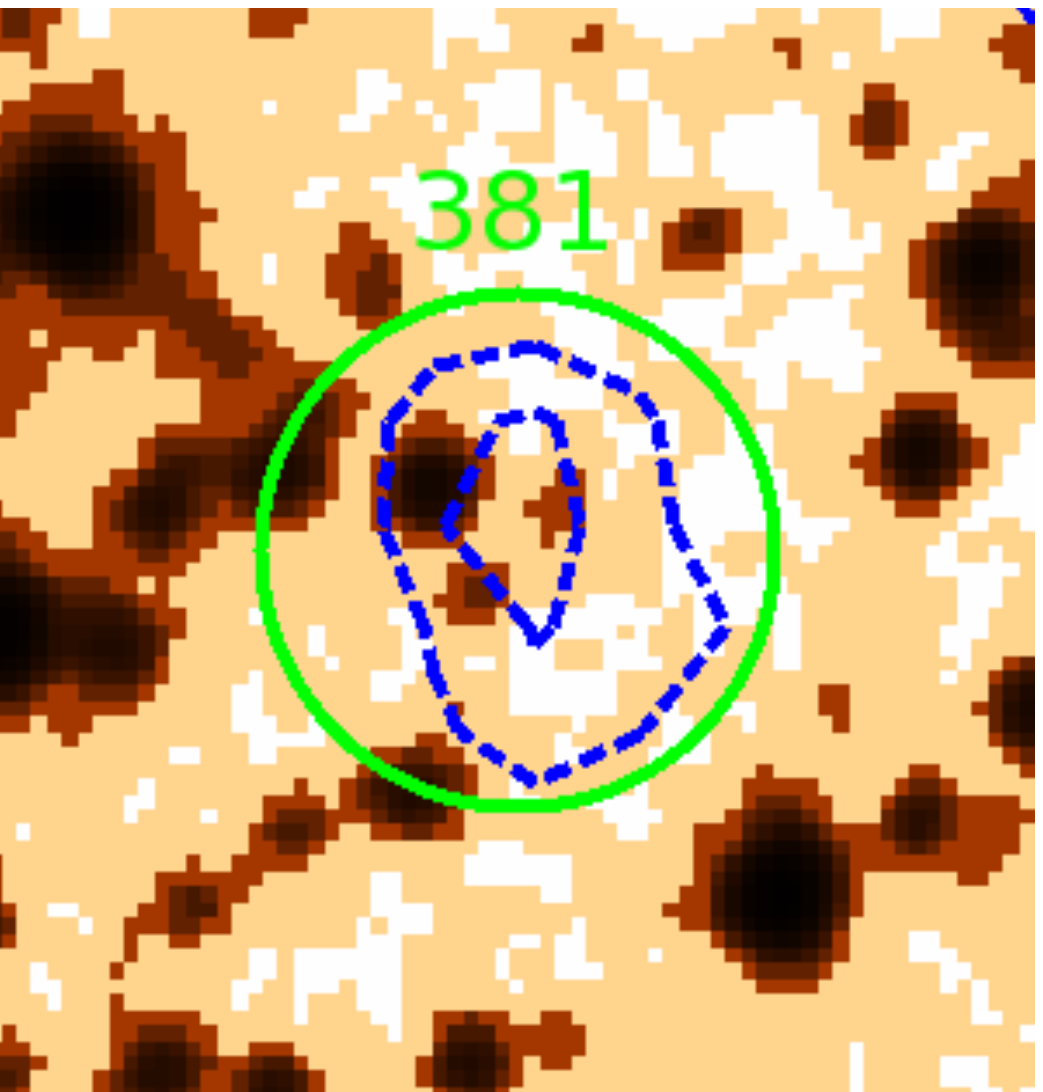} \\
  \includegraphics[height=.19\textwidth,bb=0 0 300 315,clip]{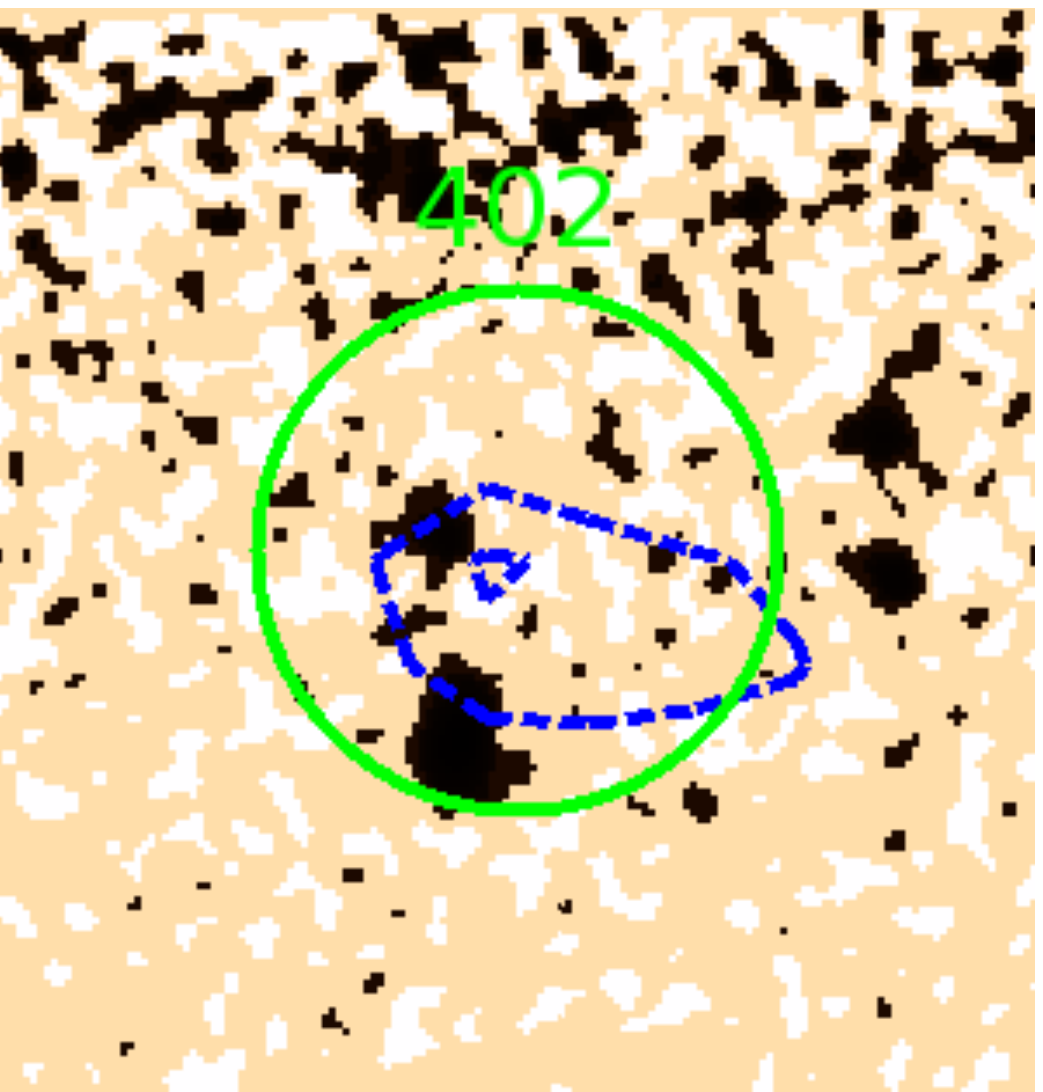}
  &
  &
  &\includegraphics[height=.19\textwidth,bb=0 0 300 315,clip]{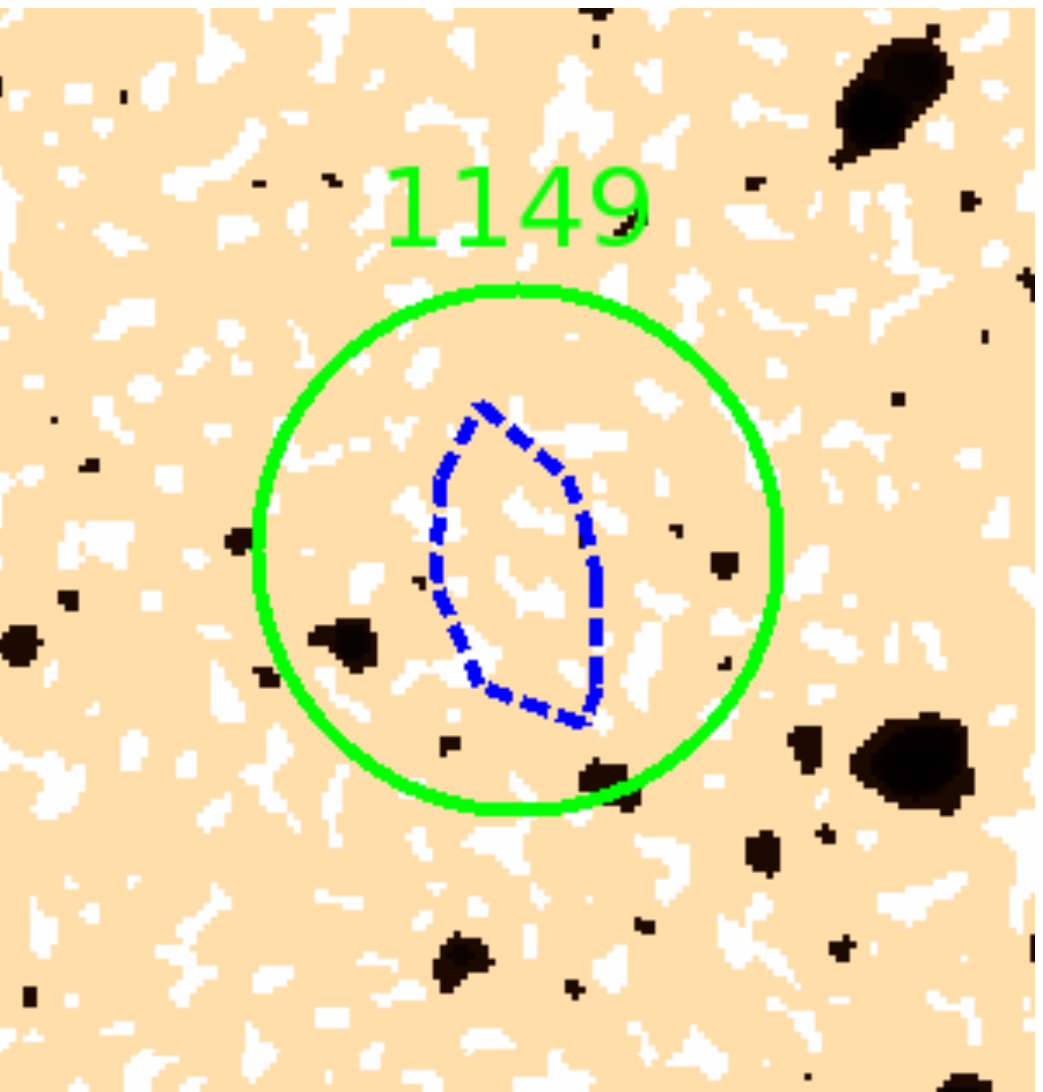}
  &\includegraphics[height=.19\textwidth,bb=0 0 300 315,clip]{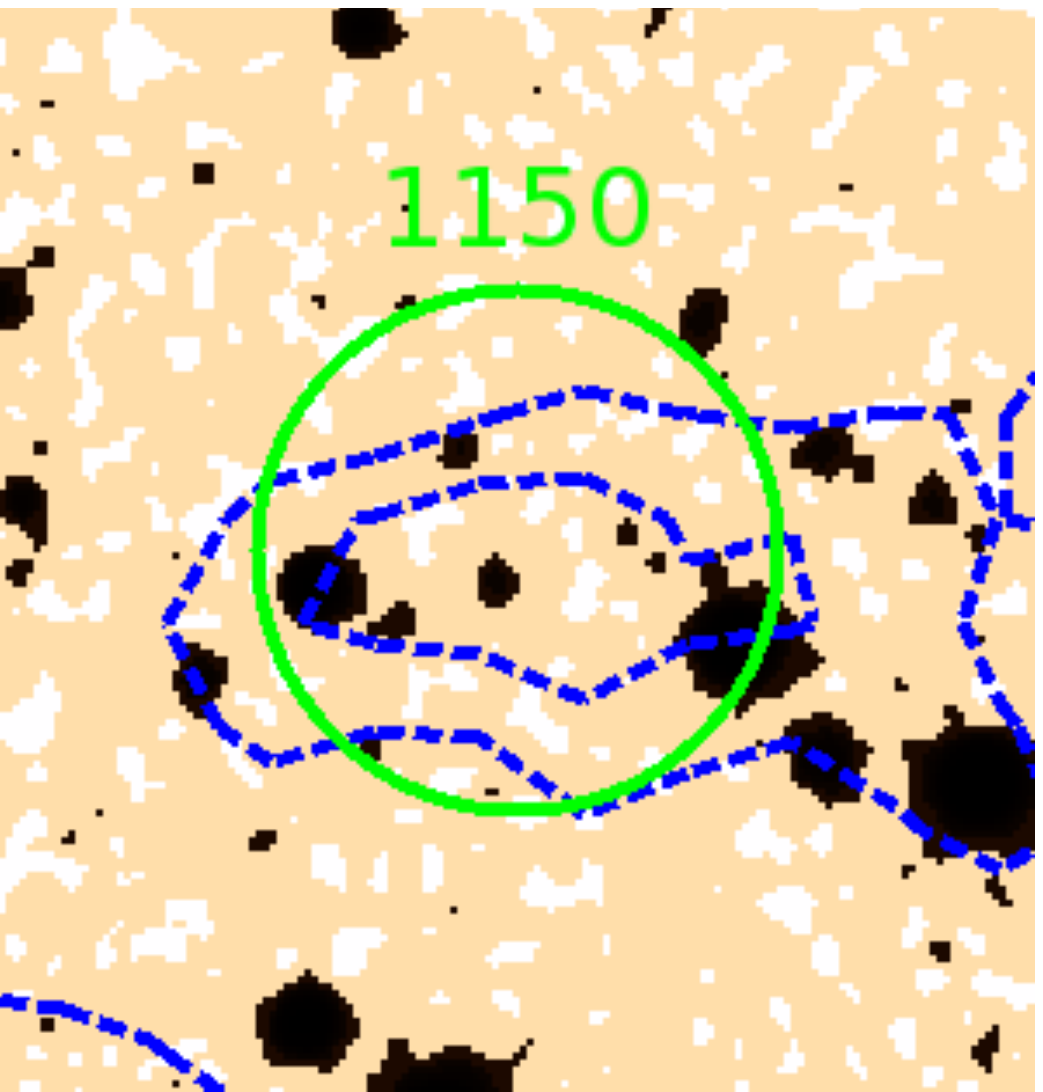}\\
  \includegraphics[height=.19\textwidth,bb=0 0 300 315,clip]{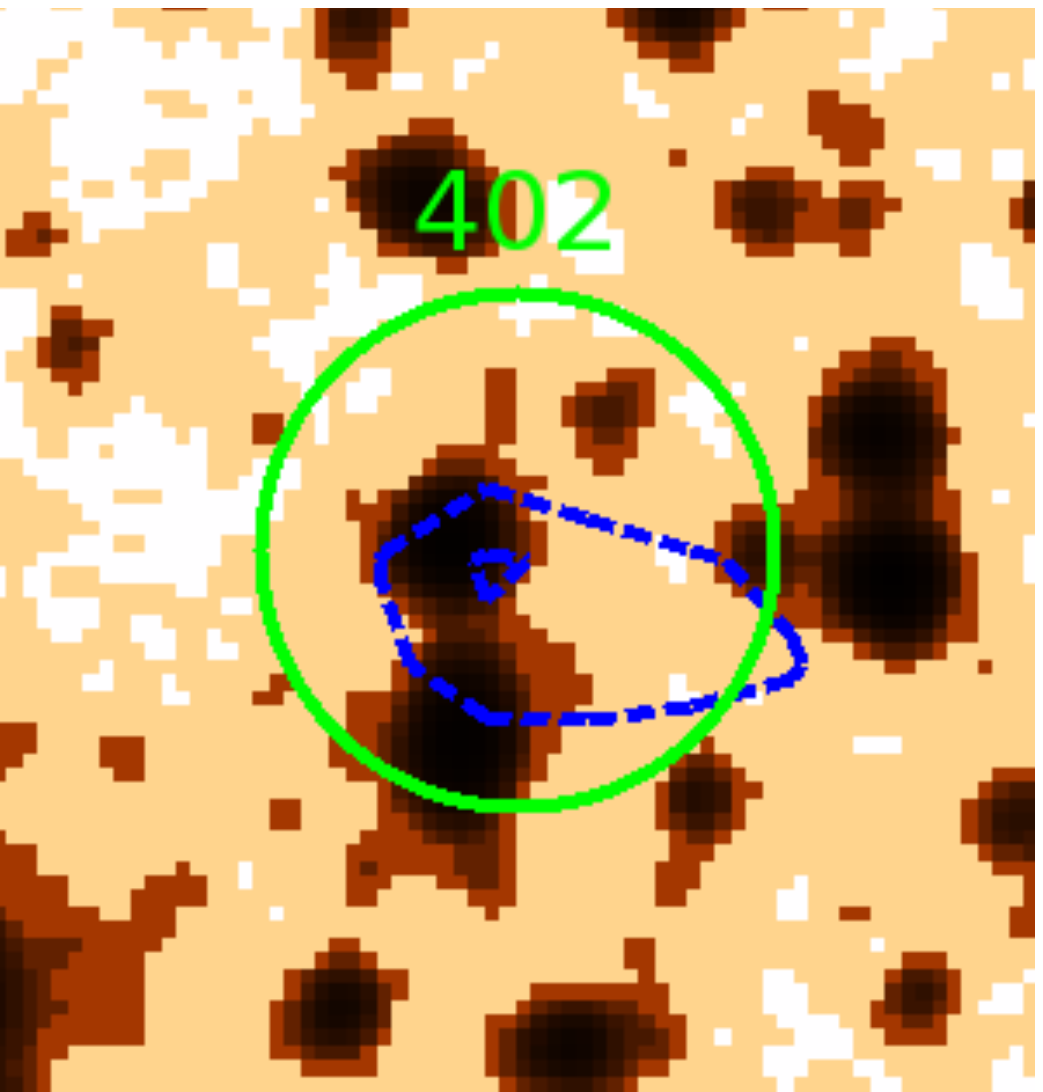}
  &\includegraphics[height=.19\textwidth,bb=0 0 300 315,clip]{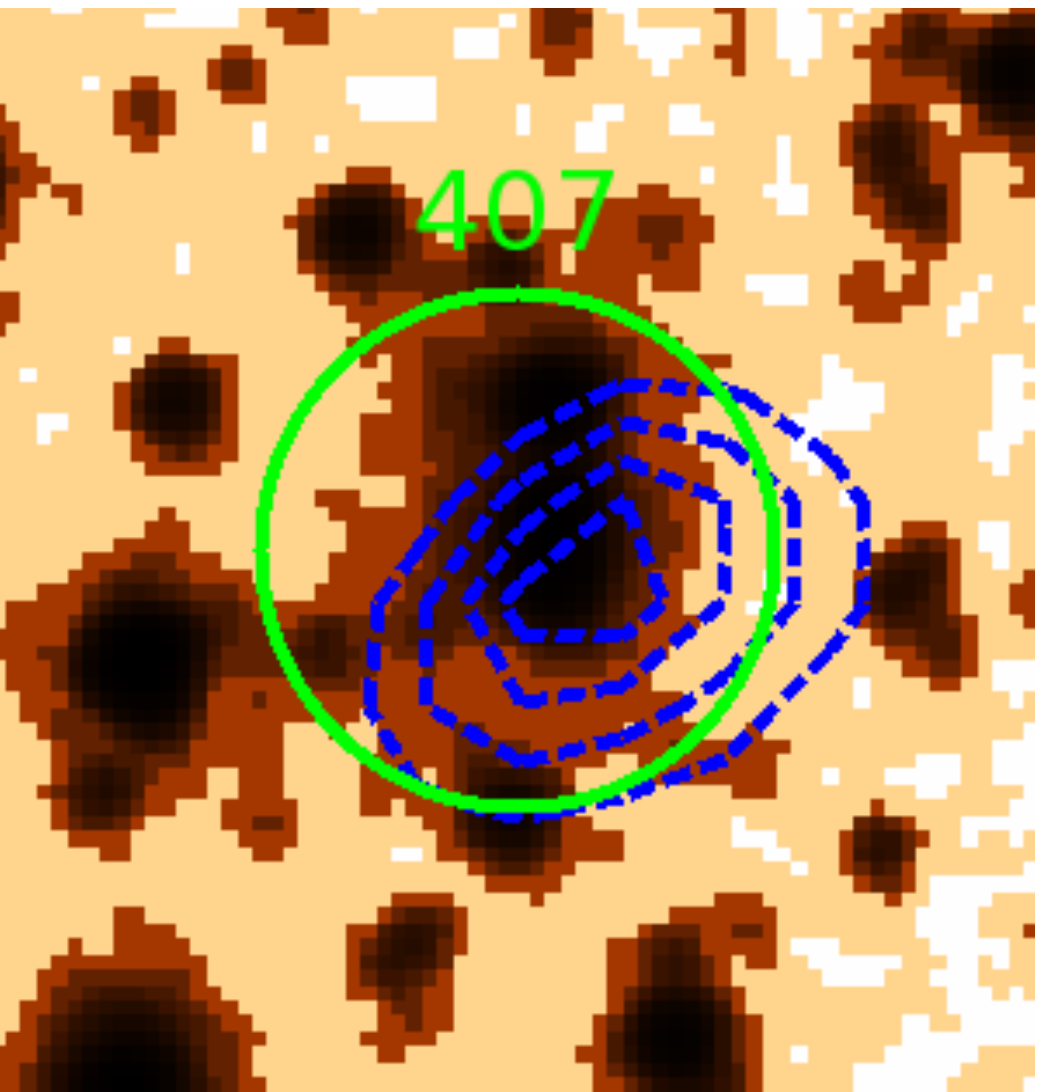}
  &\includegraphics[height=.19\textwidth,bb=0 0 300 315,clip]{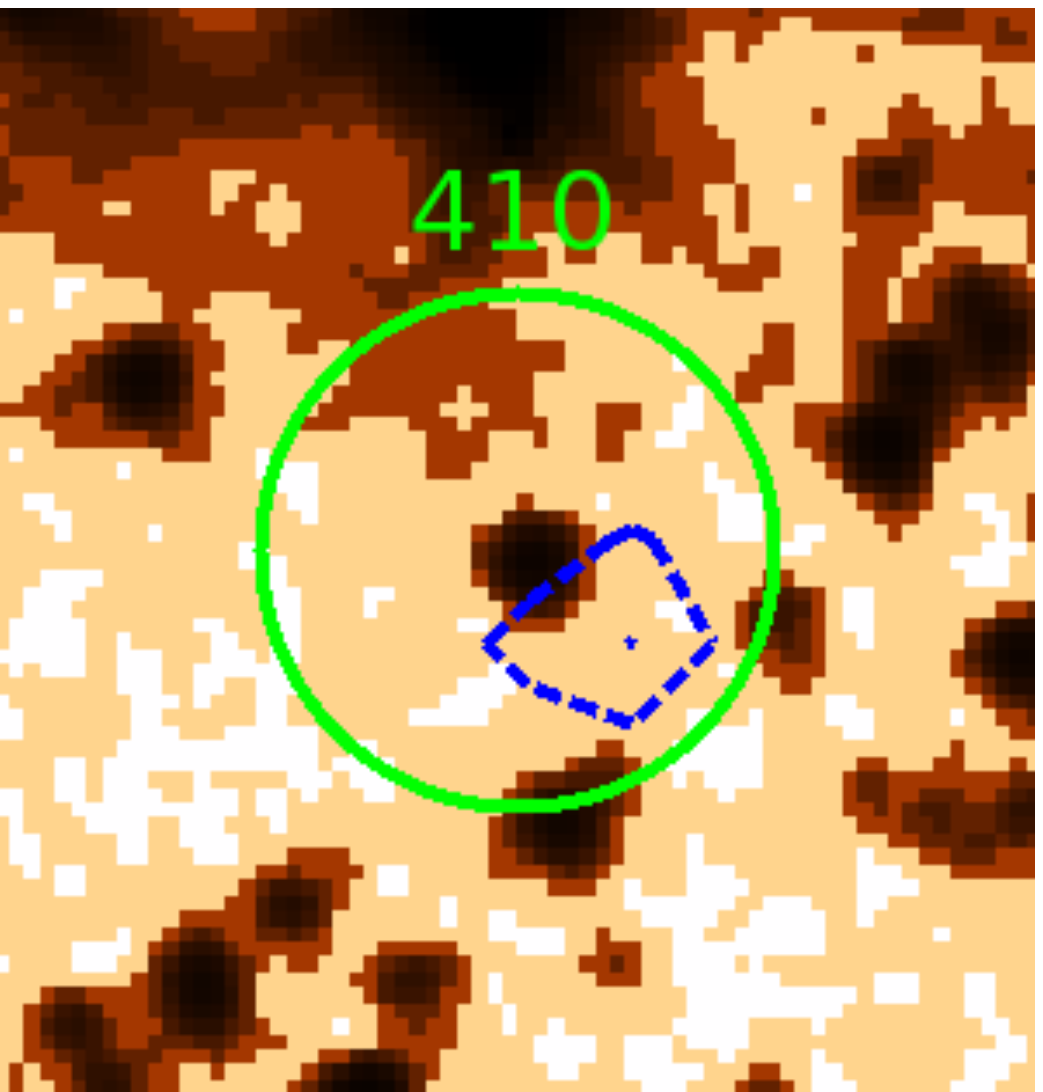}
  &\includegraphics[height=.19\textwidth,bb=0 0 300 315,clip]{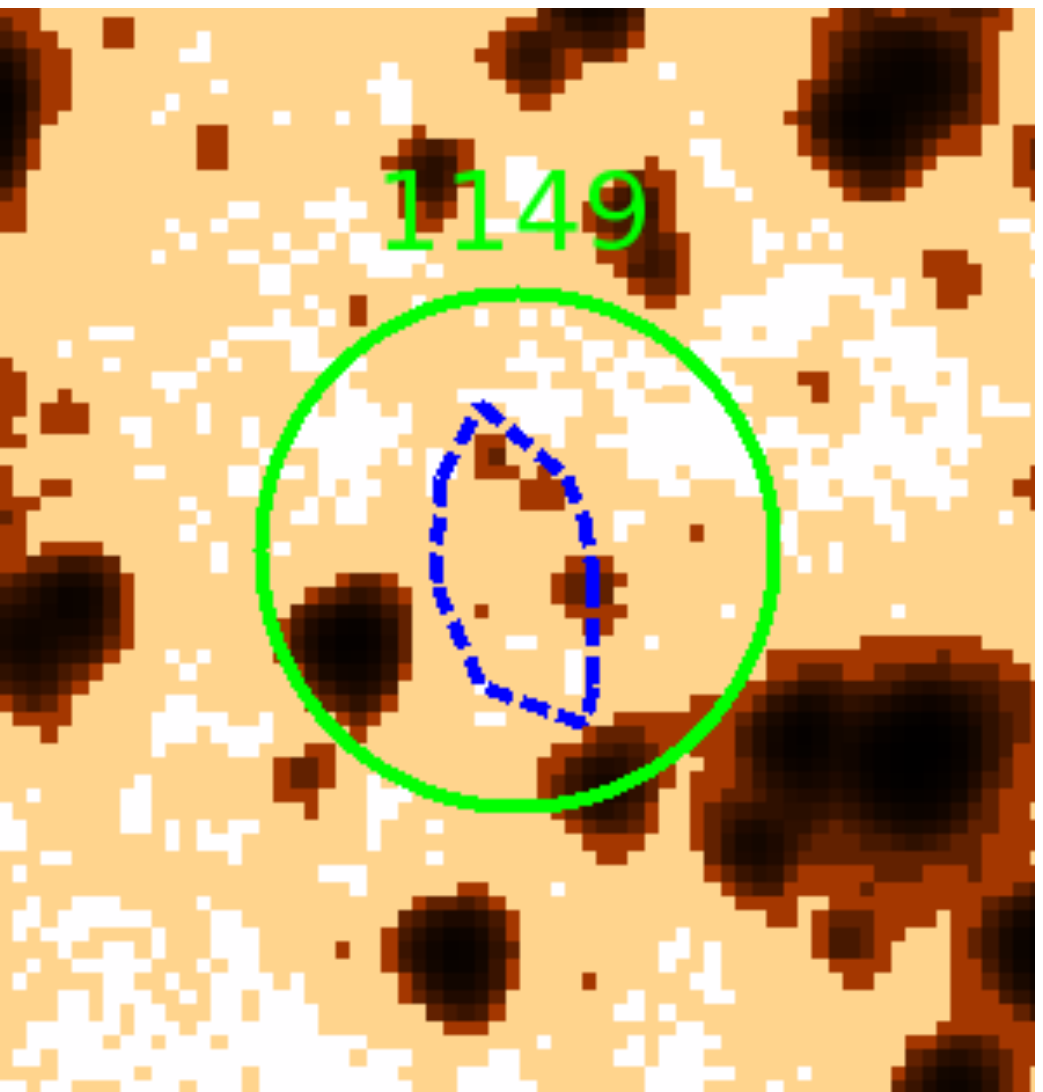}
  &\includegraphics[height=.19\textwidth,bb=0 0 300 315,clip]{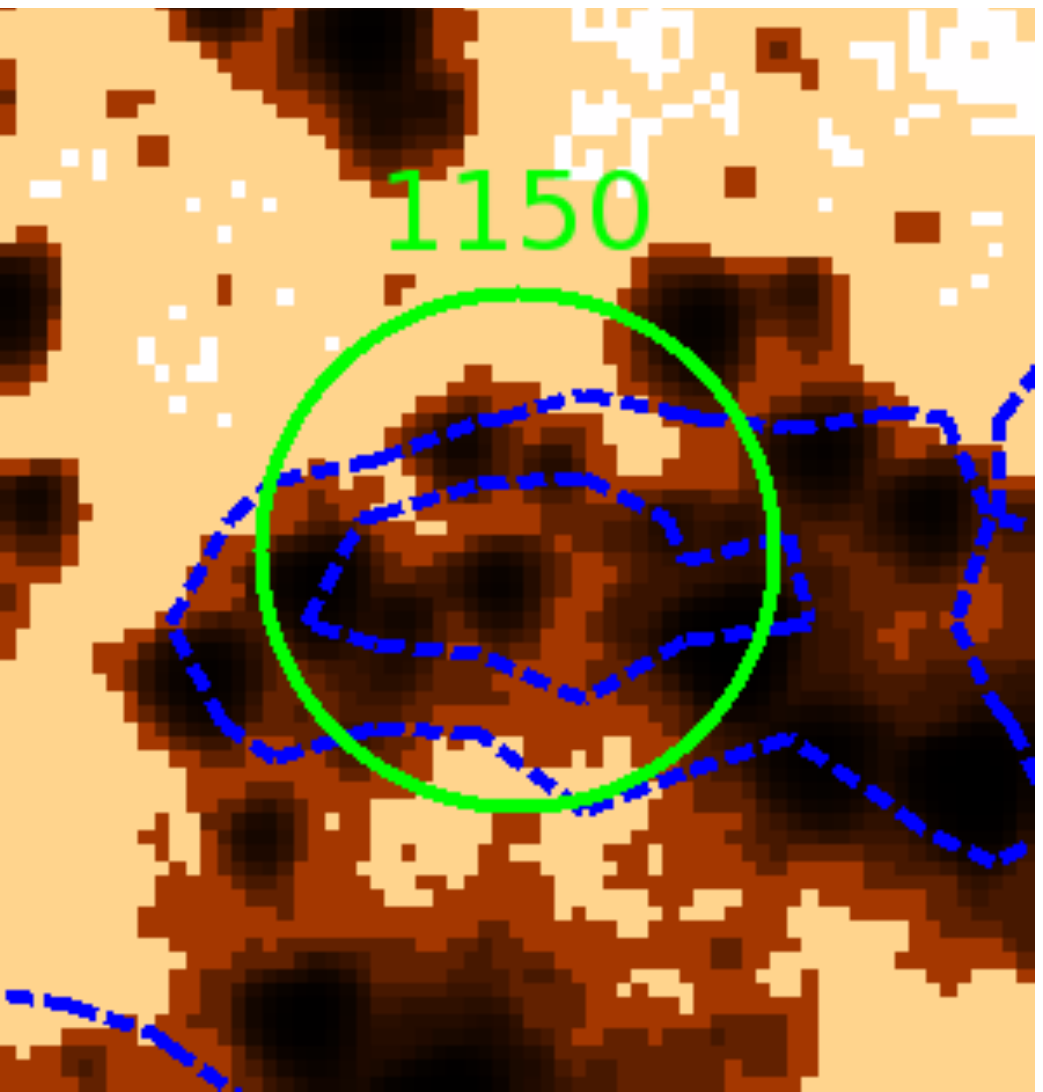}
  \end{array}$
  \caption{K-band and 3.6$\mu$m cutouts of the 15 XMM-CDFS sources not
    previously detected by \chandra\ (images from the MUSYC and SIMPLE
    surveys). The \xmm\ contours from the 2--10 keV band (5--10 keV
    for sources 1098 and 1149) are superimposed in blue. A green
    circle with radius $10\arcsec$ is drawn around the \xmm\ position. The first,
    third and fifth rows show K-band images; the second, fourth and
    sixth rows show 3.6$\mu$m images.  From left to right, and from
    top to bottom: ID210 5, 85, 176, 186, 189, 207, 224, 280, 348,
    381, 402, 407, 410, and ID510 1149, 1150. K-band data are not
    available for ID210 407 and 410. Each cutout covers an
    area of $40\arcsec\times 42\arcsec$. The K-band images have been
    smoothed with a Gaussian kernel of size 3 pixel
    ($0.75\arcsec$). The images are shown in histogram-equalized
    scaling, ranging from light (fainter fluxes) to dark brown
    (brighter fluxes).  }
  \label{fig:newsrc}
\end{figure*}

\subsection{\chandra\ sensitivity limits and optical/IR counterparts for the new sources}
\label{sec:newsrc-chandra}

\begin{figure}
  \centering
  \resizebox{\hsize}{!}{\includegraphics[width=\columnwidth,bb=28 116 566 722,clip]{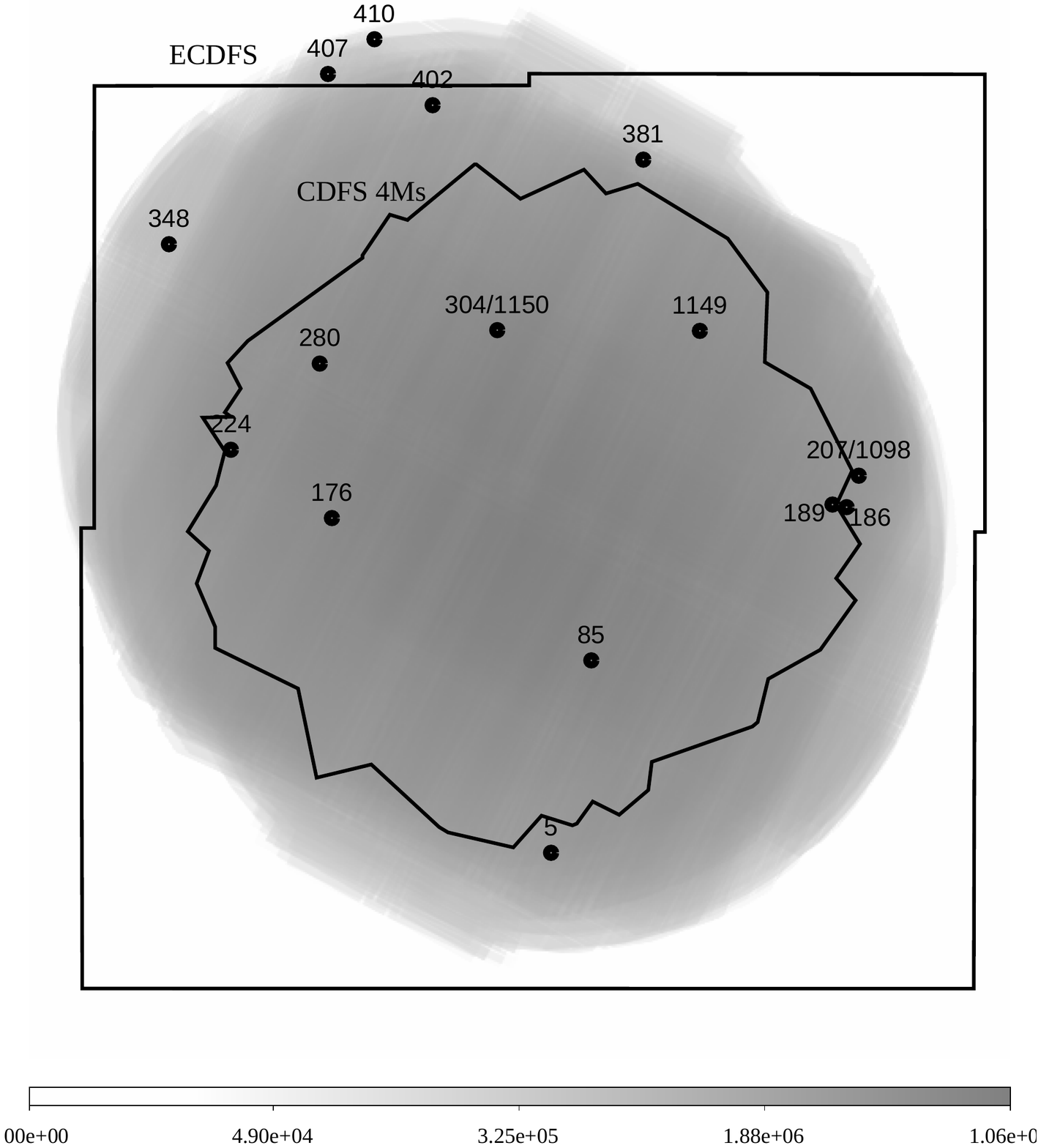}}
  \caption{Positions and ID210/ID510 numbers of the 15 candidate real
    sources not previously detected by \chandra, superimposed on the
    \xmm\ 2--10 keV exposure map. The exposure map is drawn in
    logarithmic scale to show also the field border where the exposure
    is very low. The areas previously surveyed by \chandra\ are also
    shown.  }
  \label{fig:newsrc-on-expmap}
\end{figure}

Among the 15 candidate new sources, 2 are outside the 4~Ms and ECDFS
areas (ID210: 407, 410) and were not observed by \chandra. Of the
remaining 13, 8 are only covered by the ECDFS (ID210/ID510: 5, 186,
189, 207/1098, 224, 348, 381, 402), and 5 have also data from the 4 Ms
survey (ID210/ID510: 85, 176, 280, 304/1150, 1149; though source 280
falls in an area with a strong exposure gradient). The 4~Ms flux
limits are fainter than the ECDFS ones by about two order of
magnitudes, therefore we will focus on the 5 sources falling in the 4
Ms area. The source positions are shown in
Fig.~\ref{fig:newsrc-on-expmap} on an \xmm\ exposure map with the
\chandra\ survey boundaries superimposed.

Visual inspection of \chandra\ 4~Ms data showed low-significance local
enhancements in the positions of the new sources.  The \xmm\ fluxes
are compared to the \chandra\ sensitivity in the soft and hard bands
in Table~\ref{tab:sensitivity}.  The \chandra\ limits in 2--8 keV band
for sources 85, 280 and 119 inside the 4~Ms area are a factor 2--3
lower than the flux observed by \xmm\ in the 2--10 keV band, while for
source 176 the \chandra\ limit is 15 times lower than the
\xmm\ observed flux. Only for source 304/1150 the \chandra\ limit is
above the \xmm\ flux. Therefore, sources 85, 176, 280 and 119 should,
in principle, have been detected by \chandra.

All the sources inside the \chandra-covered areas are hard sources,
because no signal is present in \xmm\ data in the 0.5--2 keV band.
Also, the hardness ratios\footnote{Calculated following \citet{behr}
  as the median of the posterior porability distribution of the ratio
  $(H-S)/(H+S)$ where $H$ and $S$ are the counts in the 2--10 and
  0.5--2 keV band, respectively.}  for the 5 sources inside the 4~Ms
area are $\ge 0.3$ (Table~\ref{tab:sensitivity}) and consistent with
an obscured spectrum with $N_H\gtrsim 10^{22}$ cm$^{-2}$.  The
\chandra\ flux limits were calculated for an unabsorbed spectrum with
$\Gamma=1.4$; for obscured sources the limits would be higher
depending on the column density.  Thus it is possible that the
differences between the \chandra\ and \xmm\ effective area curves
(\chandra\ having a higher drop at high energies than \xmm) may favour
the detection by \xmm\ of very hard sources close to the nominal
\chandra\ flux limit in the hard band. Consistently source 176, which
has the largest ratio between observed \xmm\ flux and \chandra\ limit,
is also the hardest source with a $HR\sim 0.88$ (this HR may be
produced by a power-law with $\Gamma=1.7$ and $N_H\sim 5\e{22}$
cm$^{-2}$; source 176 is not detected in the 5--10 keV band probably
because of its faintness, though an excess is still visible in the
5--10 keV image).  If the hardness of these sources is due to
obscuration, then the non-detection by \chandra\ in the 0.5--2 keV
band would also be justified.  None of these sources was included in
the \citet{iwasawa2012} paper because their \pwxd\ significance is
below the threshold set in that paper (except for source 407 which has
however a low number of counts).  Further study of these sources may
help shed light on the obscured AGN population responsible for the
missing fraction of the cosmic X-ray background, and will be presented
in a further paper of this series (Ranalli et al., in prep.).

For the 15 new sources, we have searched for identifications in the K
and 3.6$\mu$m bands from the MUSYC \citep{taylor2009} and SIMPLE
\citep{damen2011} surveys. The closest possible matches are presented
in Table~\ref{tab:newsrc-counterparts}; in a few case we list more
than one possibility due to the crowdness of the field.  Finally, in
Fig.~\ref{fig:newsrc} we present K and 3.6$\mu$m image cutouts in the
with \xmm\ contours sumperimposed.

\begin{table}
\caption{\xmm\ fluxes and \chandra\ sensitivities for the 5 candidate
  new sources falling in the 4~Ms area.}
\centering
\begin{tabular}{lrrrr}
\hline\hline
ID210/ID510 & XMM  & XMM  & \multicolumn{2}{c}{\chandra\ limits}                     \\
            & flux & HR   & 0.5--2 keV                           & 2--8 keV                 \\
\hline
85          & 1.2  & 0.22 & 0.080                                & 0.40  \\
176         & 8.3  & 0.88 & 0.089                                & 0.53  \\
280         & 3.0  & 0.31 & 0.22                                 & 1.1   \\
304/1150    & 0.21 & 0.30 & 0.073                                & 0.42  \\
1149        & 2.4  & 1    & 0.21                                 & 1.2   \\ 
\hline
\end{tabular}
\tablefoot{\xmm\ fluxes are for the 2--10 keV band (except for source ID510
  1149, which is in the 5--10 keV band). \chandra\ sensitivity limits
  in the 2--8 and 0.5--2 keV bands from the 4~Ms survey. All fluxes in
  units of $10^{-15}$ \ergscmq. Hardness ratio are from PN data.}
\label{tab:sensitivity}
\end{table}

\begin{table*}
\caption{K-band and 3.6$\mu$m counterparts to new XMM-CDFS sources.}
\label{tab:newsrc-counterparts}
\centering
\begin{tabular}{rrrrrcccccc} 
\hline\hline             
ID210/   & Location & RA$_{\rm NIR}$ & DEC$_{\rm NIR}$ & dist   & ID    & K     & ID     & 3.6$\mu$m & redshift & notes                               \\ 
ID510    &          & J2000          & J2000           & arcsec & MUSYC & AB    & SIMPLE & AB        &          &                                     \\
\hline
407      & out      & 53.25516       & -27.54088       & 1.46   & --    & --    & 54179  & 18.58     & 1.13     & \tablefootmark{a}$\!^,$\tablefootmark{b} \\
410      & out      & 53.22289       & -27.51963       & 0.83   & --    & --    & 56022  & 21.33     & --       & \tablefootmark{c}                   \\
\hline                                                                    
5        & ECDFS    & 53.09970       & -28.02297       & 1.92   & 1613  & 20.56 & 11010  & 20.31     & 0.68     &                                     \\
186      & ECDFS    & 53.89292       & -27.80909       & 1.09   & 8374  & 19.71 & 30699  & 19.67     & --       &                                     \\
189      & ECDFS    & 52.90288       & -27.80730       & 1.60   & 8413  & 21.35 & 30653  & 20.02     & 1.43     &                                     \\
207/1098 & ECDFS    & 52.88460       & -27.78947       & 1.81   & 8998  & 20.94 & 32579  & 20.91     & 0.62     &                                     \\
224      & ECDFS    & 53.32407       & -27.77321       & 1.86   & 9603  & 20.51 & 34104  & 19.80     & 1.38     & \tablefootmark{d}                   \\ 
348      & ECDFS    & 53.36649       & -27.64710       & 3.31   & --    & --    & 46367  & 23.17     & --       & \tablefootmark{e}                   \\ 
         &          & 53.36634       & -27.64582       & 2.71   & --    & --    & 46310  & 21.71     & --       & \tablefootmark{e}                   \\ 
         &          & 53.36575       & -27.64786       & 6.89   & 13769 & 21.81 & 46252  & 22.63     & --       &                                     \\ 
381      & ECDFS    & 53.03626       & -27.59340       & 4.2    & --    & --    & 50549  & 21.56     & --       & \tablefootmark{e}                   \\
         &          & 53.03574       & -27.59456       & 2.40   & --    & --    & 50687  & 23.64     & --       & \tablefootmark{e}                   \\
         &          & 53.03471       & -27.59362       & 2.32   & --    & --    & 50766  & 23.69     & --       & \tablefootmark{e}                   \\
402      & ECDFS    & 53.18346       & -27.56021       & 3.14   & 15482 & 20.51 & 52934  & 19.79     & --       & \tablefootmark{f}                   \\
         &          & 53.18321       & -27.56246       & 7.74   & 16344 & 18.95 & 52602  & 19.53     & --       &                                     \\
\hline                                                                    
85       & 4~Ms     & 53.07197       & -27.90447       & 1.93   & 4907  & 21.37 & 21475  & 20.53     & 1.1125   & \tablefootmark{c}                   \\ %
176      & 4~Ms     & 53.25321       & -27.81607       & 0.12   & 8087  & 21.49 & 30251  & 21.17     & 1.75     &                                     \\
280      & 4~Ms     & 53.26146       & -27.72112       & 2.64   & 10952 & 15.63 & 37523  & 16.42     & --       & \tablefootmark{d}                   \\
1149     & 4~Ms     & 52.99465       & -27.70056       & 3.2    & --    & --    & 41386  & 23.93     & --       & \tablefootmark{e}                   \\
         &          & 52.99755       & -27.70113       & 7.4    & 12085 & 21.23 & 41190  & 20.53     & --       &                                     \\
304/1150 & 4~Ms     & 53.13756       & -27.70008       & 2.49   & 12132 & 21.54 & 41087  & 20.57     & 1.2357   & \tablefootmark{c}                   \\ %
\hline
\end{tabular}
\tablefoot{The columns show: ID210 (ID510 if the number is
    $>1000$); if the source falls inside or outside the 4~Ms and ECDFS
  surveys; J2000 positions from the K band, unless in case of a missing K counterpart, in which case
we list the IRAC $3.6\mu$m position; distance in arcsec between the
\xmm\ and NIR position; ID number and K magnitude from the MUSYC
survey; ID number and $3.6\mu$m magnitude from the SIMPLE survey;
redshift; notes. Magnitudes are in the AB scale.\\
\tablefoottext{a}{No K-band imaging.}
\tablefoottext{b}{Bright \xmm\ source.}
\tablefoottext{c}{Possible extended NIR source, only closest \xmm/NIR match listed.}
\tablefoottext{d}{K-band empty field.}
\tablefoottext{e}{Another source is present at 7.7$\arcsec$.}
\tablefoottext{f}{Bright galaxy.}
}
\end{table*}

\subsection{Number of chance associations}

We have used the simulations detailed in Sect.~\ref{sec:simulations}
to assess how many of these XMM-\chandra\ associations are due to chance:
i.e., we check the reliability of the match, not that of the
individual sources.
We took the output XMM source lists from five of those simulations
(for each band) and cross-correlated them with the CDFS 4~Ms and the
ECDFS catalogues following the procedure outlined above
(cross-correlation with CDFS 4~Ms and ECDFS independently, removal of
the duplicated sources, merging, subselection out to $r\le 5$, removal
of further counterparts with low relative probability and selection of
associations with high probability). We then merged the five resulting
catalogues and calculated the fraction of simulated XMM sources that
had one or more counterparts with probability above 90~per cent: out
of the 1793 (685) sources in those five simulations, 71 (30) had one
``Good'' \chandra\ counterpart and 6 (3) had two, there were none with
three or more. These correspond to a total of chance associations with
one or more \chandra\ sources of $4.3\pm0.5$~per cent\footnote{We have
  estimated the uncertainties in those fractions using a Bayesian
  approach and the binomial distribution \citep{walljenkins} with the
  shortest 68 per cent condence interval (S. Andreon, priv.  comm.).}
($4.8_{-0.8}^{+0.9}$).  Therefore, for the 2--10 keV main catalogue of
339 sources, we would expect between about 13 and 16 chance
associations of sources, while for the 5--10 keV main catalogue of 136
sources there would be between about 5 and 8 spurious pairs.

\section{Conclusions}
\label{sec:conclusion}

We have presented the observations, data reduction, catalogues and
number counts of the XMM-CDFS survey. Currently the deepest \xmm\
survey, the XMM-CDFS observations pose several challenges in their
reduction. The large number of independent exposures (33 obsids times
3 cameras) centred on the same field was unprecedented for
\xmm.

\begin{itemize}

\item The very large time span of the observations allowed us to find
  an increase by a factor $\sim 2$ in the instrumental background in
  the years 2008--2010 with respect to the years 2001--2002, and
  estimate the variations of the various background components.

\item A careful study of the instrumental background of \xmm\ during
  the XMM-CDFS observations was done to produce simulated
  observations, whose properties were designed to be as close as
  possible to the real XMM-CDFS. By using the count ratio between
  in-FOV and out-of-FOV data, and the Filter Wheel Closed (FWC)
  observations, the background was decomposed in cosmic, particle, and
  residual soft proton. The simulated observations reproduce the
  spatial details of the background (chip gaps, a missing MOS1 CCD,
  vignetting), and include detected sources distributed according to
  the \citet{gilli07} and \citet{rcs05} model \lognlogs$\!$. Mock
  catalogues were produced by running the source detection procedure
  on the simulated observations, and were used to calculate the
  coverage and the number of chance \xmm-\chandra\ associations.

\item
We derived the catalogues in the 2--10 and 5--10 keV bands with a
two-step procedure. First, the \pwxd\ wavelet software was used to
identify candidate sources with a significance equivalent to
$4\sigma$, and to find their coordinates. Next, we used the SAS \emld\
tool to further check the significance of the sources, and obtain
counts, count rates and fluxes. The final catalogues contain 339 and
137 sources in the 2--10 and 5--10 keV bands respectively, and have a
significance equivalent to $4.8\sigma$. 

The faintest sources have fluxes of $6.6\e{-16}$ and $9.5\e{-16}$
\ergscmq, respectively. The flux limits at 50\% of the sky coverage
are $1.8\e{-15}$ and $4.0\e{-15}$ \ergscmq, respectively.  The
catalogues were cross-correlated with the \chandra\ 4~Ms (X11) and
ECDFS \citep{lehmer05}, using a likelihood-ratio method.  Simulations
provided an upper limit to the number of spurious sources, whose
number is better estimated with comparison with \chandra\ data and
inspection of the \xmm\ signal/noise ratio. The spurious fraction is
thus $13/339\sim 3.8\%$ in the 2--10 keV band, and $4/137\sim 2.9\%$
in the 5--10 keV.

\item Despite the high background level, \xmm\ was able to detect a
  few extremely hard sources. Further study of these objects will help
  understanding the most obscured AGN population responsible for the
  missing fraction cosmic X-ray background (Ranalli et al., in prep.).

\item The number counts where derived in both the 2--10 keV and 5--10
  keV bands, and extend from $6.6\e{-16}$ and $9.5\e{-16}$ to
  $1.1\e{-13}$ and $6.7\e{-14}$ \ergscmq, respectively. They are in
  agreement with previous derivations with the \chandra\ and \xmm\
  telescopes, and in different fields (the CDFS, the Lockman Hole, and
  the Hellas2XMM and 2XMM surveys). The XMM-CDFS currently
  reaches the faintest flux limit in the 5--10 keV band ($9.5\e{-16}$
  \ergscmq), $\sim 3$ times fainter than the Lockman Hole (the
  \chandra\ 4~Ms survey reaches fainter fluxes, though the band is
  formally softer, 4--8 keV, \citealt{lehmer2012}).

\item Finally, the simulations were used to study the source confusion
  in the XMM-CDFS relative to the 2--10 keV band. An average of 14
  confused sources per simulation (i.e.\ detected as a single source,
  but corresponding to two input sources separated by $<15\arcsec$)
  was found; when comparing to the XMM-CDFS, this number should be
  considered as a lower limit because the simulations do not include
  source clustering. The cross-correlation with \chandra\ yielded 20 \xmm\
  sources associated with two or three \chandra\ counterparts.

\end{itemize}

\begin{appendix} 
\section{The XMM-CDFS simulator}
\label{sec:simulator}

We have developed a general purpose simulator of X-ray CCD
observations, tuned for the \xmm\ EPIC camera, but easily extensible
to other missions. Although its primary use has been the validation of
the XMM-CDFS catalogue, the simulator has in fact been written with
future missions in mind (from eROSITA to concepts such as the Wide
Field X-ray Telescope, and the project formerly known as
XEUS/IXO/Athena). The simulator reproduces both point-like sources and
the (cosmic, instrumental, and soft proton) background. As to sources
and cosmic background, the only assumptions are that a library of
position- and energy-dependent PSFs is available, that the effective
area of the telescope and detector is known, and that an exposure map
is provided to account for vignetting and eventual chip gaps.  As to
the instrumental background, the levels for the particle and soft
proton contributions are needed, which in our case have been estimated
by analysing the existing XMM-CDFS observations.

The main features are:
\begin{itemize}
\item it produces event files, to be analyzed with common X-ray data
  analysis software;
\item it supports an arbitrary number of sources;
\item arbitrary spectra can be assigned to the sources;
\item it calculates the most appropriate PSF for each source,
  according to position and energy, by interpolating from the PSF
  library.
\end{itemize}

The simulator is written in the Perl language, making use of the Perl
Data Language\footnote{Available at http://pdl.perl.org} libraries
\citep[PDL;][]{pdl} for numerical computation and FITS
input/output. The Ftools package is used to process the simulated
event file headers and make it readable by the SAS; the SAS tool
\texttt{merge} is used to join the source and background event files.

The computing time for a single-camera, single-\obsid\ simulation is
around 4--5 min on a 2.80 GHz Xeon processor running Linux; the time
needed to simulate the whole XMM-CDFS on a single CPU is therefore
around 7.5 hours. The simulations of different cameras/\obsid\ can
however be run in parallel.

The simulator is released under the terms of the General Public
License (GPL) as published by the Free Software Foundation, and is
available on the XMM-CDFS
website\footnote{http://www.bo.astro.it/xmmcdfs/deeprime/index.html}.

\subsection{Point-like sources and cosmic X-ray background}

The goal is to produce an event list with the same properties (camera,
pointing, exposure) of a real observation.  This stage of the
simulator acts basically as a PSF sampler. Given a list of coordinates
and count rates, for each source it calculates the correct PSF
according to the off-axis angle and the spectrum, and samples it for a
number of photons depending on the count rate and exposure time. The
PSF library is provided, for \xmm, by three calibration files (CCF),
one for each of the MOS1, MOS2 and PN, which contain normalized images
of the PSFs at 11 energies (from 0.1 keV to 15 keV) times 6 off-axis
angles (from 0$\arcmin$ to 15$\arcmin$).
Photon times and energies are also assigned to the events, to
obtain an output file which looks like a real XMM-Newton event file.

The cosmic X-ray background is simulated by generating many faint
sources according to the number counts models of AGN \citep{gilli07}
and galaxies \citep{rcs05}.

The simulator accepts the following input:
%\begin{enumerate}
a list of RA, DEC, and count rates of the sources to simulate;
 a model spectrum formatted as an ASCII table; % format as output from XSPEC (see Sect.\ref{sec:model-spec});
 an event file, which is used to get the information about the boresight
and camera and (optional) an exposure map. %If the exposure map is
%not specified, than the source list should include counts instead
%of rates.
%\item the filename for the output event list.
%\end{enumerate}

The output is an event list in FITS format, which is readable by both
PWXDetect and the SAS.

The algorithm works as follows. For each source, the expected counts
are calculated, and rounded if they are fractional. The rounding needs
to conserve the sum of the counts from all sources, in order to
correctly reproduce the level of the cosmic X-ray background.
The model spectrum is sampled, obtaining the event energies, which are
binned according to the same sampling scheme of the PSF.
Finally, the PSF is sampled, according to the event energies; if
needed, PSFs at different off-axis angles are interpolated. The
positions of the sampled events are placed at the RA and DEC
coordinates specified for the source, with the correct position angle.

\subsection{Particle and soft proton background}

\begin{figure}
  \centering \includegraphics[width=\columnwidth,bb=15 160 571 712,clip]{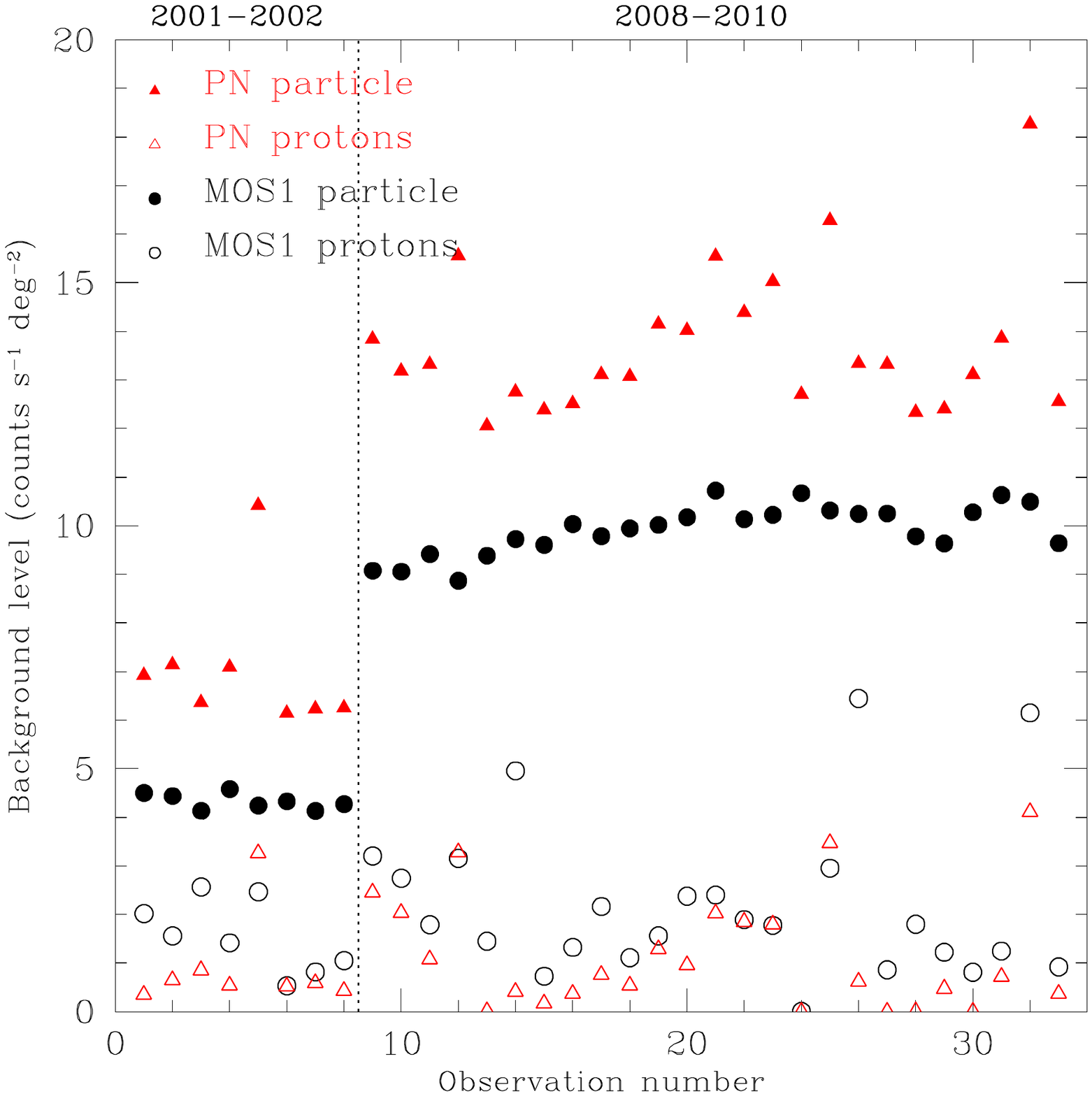} \caption{Intensity
  of the particle and proton components of the quiescent background in
  the 2--10 keV band for the 33 \obsid s of the XMM-CDFS. The \obsid s
  are numbered from 1 to 33 according to observation date; the
  vertical dotted line marks the separation between the 2001--2002 and
  2008--2010 observations. Filled triangles: PN particle background;
  empty triangles: PN residual soft protons (after light curve
  screening; see Sect.~\ref{sec:protons}); filled circles: MOS1
  particle; empty circles: MOS1 protons. The MOS2 camera has values
  very similar to the MOS1.}
   %Squares: MOS1; Triangles: MOS2; Circles: PN.}
  \label{fig:secularbkg}
\end{figure}

The input for the simulation of the background are: the levels of the
particle and soft-proton components (as an example, we show in
Fig.~\ref{fig:secularbkg} the levels for the XMM-CDFS), which can be
estimated from the in/out of FOV count ratio (see
Sects.\ref{sec:particle} and
\ref{sec:protons}); an energy inteval; and an event file used to
derive the pointing coordinates, the position angle and the exposure
time.

The FWC observations, which consist of event files, are filtered for
the energy interval resampled with the bootstrap method \citep{bootstrap}
to obtain the simulated particle background event file.

The soft proton are simulated using a series of images, part of the
ESAS CALDB files. These were originally obtained from severely flared
\xmm\ observations, by filtering out the low-background periods and
retaining only the flares. The appropriate image is sampled, producing
a simulated soft proton event file.

The two event files of the background components are finally merged
with that for the source and cosmic X-ray background.

\end{appendix}

\begin{acknowledgements}
  We thank an anonymous referee whose comments have contributed to
  improve the presentation of this paper.
  We thank F. Damiani for support and help with the \pwxd\ software,
  and K. Kuntz, S. Molendi, and S. Sciortino for useful discussions.
  PR thanks R. Di Luca for very valuable assistance in recovering data
  from a problematic disc; and A. Longinotti and all the \xmm\
  helpdesk team for their kind, focused and prompt support of the XMM-SAS
  software.

We acknowledge financial contribution from the agreement ASI-INAF
I/009/10/0 and from the PRIN-INAF-2011.  PR acknowledges a grant from
the Greek General Secretariat of Research and Technology in the
framework of the program Support of Postdoctoral Researchers. The
XMM-CDFS simulator was developed by PR as per contract with the
Astronomy Department of the University of Bologna. FJC acknowledges
support by the Spanish ministry of Economy and Competitiveness through
the grant AYA2010-21490-C02-01. WNB acknowledges the NASA ADP grant
NNX11AJ59G. 

This research has made use of the Perl Data Language (PDL)
which provides a high-level numerical functionality for the Perl
programming language (\citealt{pdl}; http://pdl.perl.org).

\end{acknowledgements}

\bibliographystyle{aa}
\bibliography{../fullbiblio}

\end{document}